\documentclass{article}

\usepackage{microtype}
\usepackage{graphicx}
\usepackage{subcaption}
\usepackage{booktabs} %

\usepackage{hyperref}

\usepackage[preprint]{icml2026}

\usepackage{amsmath}
\usepackage{amssymb}
\usepackage{mathtools}
\usepackage{amsthm}

\usepackage{xurl} %
\usepackage{amsthm} %
\usepackage{textcomp} %
\usepackage{mathrsfs} %

\usepackage{algorithm} %
\makeatletter
\@ifpackageloaded{algorithmic}{%

}{}
\makeatother

\usepackage{algpseudocode} %

\usepackage{amsmath}
\usepackage{amssymb}
\usepackage{svg}
\usepackage{tcolorbox}
\usepackage[table]{xcolor} %
\usepackage{lscape}
\usepackage{graphicx}
\usepackage{rotating}
\usepackage{caption} %
\usepackage{subcaption} %

\usepackage{booktabs,tabularx} %
\usepackage{graphicx}         %
\usepackage{tabularx}         %
\usepackage[capitalize,noabbrev]{cleveref}

\theoremstyle{plain}
\newtheorem{theorem}{Theorem}[section]

\theoremstyle{definition}
\newtheorem{definition}[theorem]{Definition}

\theoremstyle{remark}

\usepackage[textsize=tiny]{todonotes}

\icmltitlerunning{Algorithmic Approaches to Opinion Selection for Online Deliberation: A Comparative Study}

\begin{document}

\twocolumn[
  \icmltitle{Algorithmic Approaches to Opinion Selection for Online Deliberation: A Comparative Study}

  \icmlsetsymbol{equal}{*}

  \begin{icmlauthorlist}
    \icmlauthor{Salim Hafid}{yyy}%
    \icmlauthor{Manon Berriche}{yyy}%
    \icmlauthor{Jean-Philippe Cointet}{yyy}%
  \end{icmlauthorlist}

  \icmlaffiliation{yyy}{médialab, Sciences Po Paris, France}

  \icmlcorrespondingauthor{Salim Hafid}{salim.hafid@sciencespo.fr}

  \icmlkeywords{Machine Learning, ICML}

  \vskip 0.3in
]

\printAffiliationsAndNotice{}  %

\begin{abstract}
During deliberation processes, mediators and facilitators typically need to select a small and representative set of opinions later used to produce digestible reports for stakeholders. In online deliberation platforms, algorithmic selection is increasingly used to automate this process. However, such automation is not without consequences. For instance, enforcing consensus-seeking algorithmic strategies can imply ignoring or flattening conflicting preferences, which may lead to erasing minority voices and reducing content diversity. More generally, across the variety of existing selection strategies (e.g., consensus, diversity), it remains unclear how each approach influences desired democratic criteria such as proportional representation. To address this gap, we benchmark several algorithmic approaches in this context. We also build on social choice theory to propose a novel algorithm that incorporates both diversity and a balanced notion of representation in the selection strategy. We find empirically that while no single strategy dominates across all democratic desiderata, our social-choice-inspired selection rule achieves the strongest trade-off between proportional representation and diversity.
\end{abstract}

\section{Introduction}
Deliberative democracy argues that proper deliberation yields rational collective outcomes \cite{habermas2015between, rawls1997idea}. The essence of democratic legitimacy is seen as the "capacity of those affected by a collective decision to deliberate in the production of that decision" \cite{dryzek2003social}. To bring such collective decision-making to the masses, platforms for participatory democracy run public online consultations and deliberations for various use-cases such as peacebuilding, participatory budgeting, and public policy design (e.g., Polis \cite{small2021polis}, Remesh \cite{konya2022elicitation}, make\footnote{\url{https://make.org/en}}). However, such processes increasingly involve large numbers of free-form contributions (e.g., opinions, critiques, proposals) from diverse participants \cite{zhu2025can}. It thus becomes impossible for participants to process all inputs during deliberation. In this context, a crucial task is to select a small set of opinions that faithfully represents participants, and that can serve as a digestible input. This representative subset is typically selected before, during, and after the process by human facilitators, and is shared with participants or experts in order to prepare the next discussion cycle, or used to produce reports for stakeholders, decision-makers, or for the general public. However, while essential for all stakeholders, producing such representative subsets is resource-intensive and therefore costly to scale-up \cite{landemore2020open, fishkin2019deliberative}. 

\noindent
To address this problem, a vast literature from recommender systems \cite{deldjoo2024fairness} and collective response systems \cite{ovadya2023generative} has developed automated selection strategies that optimize for various desiderata such as popularity \cite{klimashevskaia2024survey} and fairness \cite{zhao2025fairness}. In the case of online deliberation, designing an algorithmic mechanism for selecting a subset of representative opinions requires making an explicit choice with regard to the type of opinions that the algorithm ought to favour. In other words, it requires defining what makes an opinion representative or desirable for selection: is it the most popular? the most consensual? the most disruptive? In deliberation contexts where multiple, often incompatible perspectives coexist \cite{crawford2016can}, answering these questions is not trivial. Nonetheless, their corresponding answers are embedded in algorithmic choices which directly influence deliberation results and yet are typically not made transparent to participants \cite{revel2025ai}. 

In parallel, a large body of literature from social choice theory has developed axioms whose goal is to formalize when collective outcomes adequately reflect the preferences of significant minorities. %
Such axioms offer precise and well-studied fairness guarantees against the systematic erasure of minority viewpoints, and are thus directly relevant for deliberation contexts. Recent efforts towards incorporating such guarantees into algorithmic systems for participatory democracy are promising but remain scattered \cite{fish2024generative,revel2025representative}. To our best knowledge, research has yet to comparatively evaluate the variety of existing algorithmic strategies for opinion selection using a comprehensive set of metrics specifically tailored to democratic desiderata for online deliberation. To address this gap, this work contributes the following:

\textbf{(1) A systematic study of how algorithmic mechanisms for opinion selection impact specific democratic desiderata in online deliberation}. Our analysis using real-world deliberation data from Remesh\footnote{Remesh is a "collective response system" \cite{ovadya2023generative} that enables participants to discuss divisive issues and surface shared insights.} reveals that existing algorithms exhibit clear trade-offs, and that enforcing constraints based on social choice theory leads to strong gains in representation while reasonably covering the opinion space and limiting redundancy among selected opinions. We make our code publicly available for both researchers and democracy practitioners to facilitate future usage and benchmarking of algorithms in this context.\footnote{To reproduce the paper's results or benchmark algorithms on any deliberation dataset, see \url{https://github.com/SalimHFX/Algorithms-for-Opinion-Selection-in-Deliberation}}%

\textbf{(2) A novel selection algorithm based on social choice theory to promote diversity without breaking proportional representation constraints}. We find empirically that, when the selected subset is small ($k\approx1-5$), our algorithm offers the best compromise between proportional representation and diversity, making it well suited to deliberation settings where (i) the subset must represent the diversity of expressed opinions, (ii) the subset must remain small and digestible.

\noindent
With this work, we aim to strengthen the systematic study of algorithmic impacts in contexts of digital democracy, and to contribute to existing efforts towards incorporating fairness guarantees into algorithmic systems in such contexts.

\section{Related Work}\label{sec:rel-work}

\noindent
\textbf{Deliberative democracy platforms and bridging systems:} Platforms for online participatory democracy rely on automated methods for condensing large-scale participations (comments, opinions, proposals) into small-size, digestible inputs. One such example is Remesh \cite{konya2023deliberative}, a platform for collective dialogues previously used by the UN in peacebuilding contexts \cite{bilich2023faster}. On Remesh, each participant expresses their personal opinion in free-form text, and votes on the opinions of a few other participants. This results in a sparse approval matrix, which is computationally made dense by inference of missing votes through a model combining language-model embeddings and latent-factors \cite{konya2022elicitation}. Once the approval matrix is full, the platform selects a subset of representative statements through a bridging-based algorithm \cite{ovadya2023bridging}. In short, bridging algorithms aim at highlighting items which receive approval across groups that typically disagree. The underlying assumption is that focusing on items which bridge across existing divides can reduce unproductive conflict and promote cooperation. Various online platforms have shown benefits of bridging-based selection. For instance, Community Notes, the community-based fact-checking system on X (ex-Twitter) relies on a bridging algorithm.\footnote{\url{https://communitynotes.x.com/guide/en/under-the-hood/ranking-notes}}. Another example is the popular Pol.is platform which has used bridging for large-scale online deliberations \cite{small2021polis}. While bridging-based algorithms have proven useful for opinion selection in online deliberation, they have, to our best knowledge, yet to be systematically compared to other algorithmic strategies to quantify how they rank on democratic desiderata such as proportional representation, coverage, and diversity.

\noindent
\textbf{Diversity in algorithmic selection:} Diversity has long been the interest of a large body of work in recommender systems \cite{zhao2025fairness, zhou2025calibrated} and algorithmic fairness \cite{pessach2022review, shrestha2019fairness}. This is in part due to popularity bias \cite{abdollahpouri2019managing}, a phenomenon whereby systems (e.g., online press, streaming services) tend to select popular items by default and ignore or rarely select less popular, niche items in the "long tail". In our work, niche items might represent minority voices or novel disruptive ideas whose inclusion in the selection is arguably required for the deliberation process to be seen as democratic \cite{shortall2021inclusion}. To solve this, recommender systems have mainly focused on developing (i) measures to assess diversity, for instance through distance-based metrics such as Intra-List Diversity \cite{ziegler2005improving}, Jaccard similarity, Hamming distance, or cosine similarity, (ii) algorithms and learning objectives that promote diversity, where many involve optimizing maximin-based objectives for learning, clustering, or re-ranking, and are typically operationalized via greedy approximation procedures. In this work, we build on this literature to (1) develop diversity-related metrics that best serve the use-case of online deliberation and incorporate them in our comparative analysis, (2) develop a novel algorithm that promotes diversity while also guaranteeing criteria of proportional representation based on social choice theory.

\noindent
\textbf{Theoretical guarantees from social choice theory:} 
Despite the proven empirical efficiency of bridging-based and diversity-based algorithmic selection, the outcomes in such approaches typically lack theoretical guarantees. For example, can a selection algorithm guarantee that minority opinions will not be ignored? Social choice theory aims precisely at answering such questions, by studying the formal aspects of aggregating  human preferences \cite{brandt2016handbook}. As observed by \cite{halpern2023representation}, selecting a subset of opinions based on agreements
and disagreements is "equivalent to electing a committee
based on approval votes", a process known as approval-based
committee elections. In this context, a vast and robust literature has defined various desired properties for election processes. One popular property previously used in contexts of online deliberation \cite{revel2025representative,halpern2023representation} is Justified Representation (JR) \cite{aziz2017justified}. In simple terms, the satisfaction of JR requires that every large enough group of cohesive voters has an approved candidate in the elected committee.\footnote{Explanations of notions such as "large enough group" and "cohesive voters", as well as definitions of all relevant properties—including JR—will be given in Section \ref{sec:diversebjr}.} In our context, JR would require that every large enough group of cohesive voters has an approved opinion in the selected subset. However, while an algorithm satisfying JR would ensure that opinions supported by enough participants end up in the selected set, it does not necessarily guarantee that the matching between participants and opinions is balanced in the final set. For example, in a scenario where the task is to select 7 representative opinions, selecting very few highly popular opinions (e.g., 2) might be enough to satisfy JR, leaving the remaining 5 opinions to be selected without any constraints of proportional representation. To solve this, many stronger variants of JR have been proposed \cite{sanchez2017proportional,kalayci2025full}. Perhaps the most intuitive is Balanced Justified Representation (BJR) \cite{fish2024generative}. In the same example, under BJR constraints, it becomes impossible for a small amount of popular opinions to satisfy the property. Instead, representation must be equally balanced across all selected opinions, meaning that every selected opinion should represent approximately the same amount of participants.\footnote{We discuss more in detail the limitations of JR and how JR and BJR differ with a concrete example in Appendix \ref{appendix:jr-vs-bjr-vs-diversebj}.} Assuming a BJR-satisfying subset of opinions, one limitation remains: the subset is not necessarily maximally diverse, i.e., it might include redundant opinions. In the context of online deliberation where the size of the subset must be small, selecting redundant opinions is not desired. While existing works from approval-based committee election have investigated combining the quality of the elected committee with diversity constraints \cite{bredereck2018multiwinner, beal2025multiwinner, ijcai2022p714}, such works have focused on specific diversity attributes (e.g., gender, age) and have assumed their availability in data. In this work, we build on social choice theory to (1) operationalize existing properties (JR and BJR) into algorithms for opinion subset selection; (2) define and operationalize a novel selection property (DiverseBJR) that accounts for both balanced representation and diversity, where diversity is based exclusively on voting data and does not require access to any additional attributes; (3) evaluate the resulting algorithms against strategies which lack theoretical guarantees (e.g. bridging) in the context of online deliberation. 

\noindent
\textbf{Scope:} A related line of research has used LLMs to produce abstractive summaries during or after deliberation sessions \cite{tessler2024ai, zhu2025can, fish2024generative}. This use-case is outside the scope of our study. While arguably easier to streamline, the LLM-based summary of a deliberation session comes with its own set of biases: beyond general reformulation and trust biases in LLMs \cite{akpinar2026llm, bouyzourn2025shapes}, recent work has revealed persistent weaknesses in LLM-based deliberation summarization, including the under-representation of minority voices \cite{zhu2025can}. In this work, we focus exclusively on algorithmically selecting representative opinions, not on reformulating them into a single cohesive summary.

\section{Methodology}
\subsection{Theoretical framework}\label{sec:prob-formal}
We adopt the \cite{elkind2022price} framework for multiwinner approval voting , adapted to online deliberation by \cite{revel2025representative}. In this setting, the goal is to select $k$ representative microposts by maximizing a given scoring rule $f$ (e.g., engagement, consensus). In our case, the microposts are opinions expressed during a deliberation process.%

\noindent
\textbf{Problem formalization: }Let $[m] = \{1,...,m\}$ be the set of all opinions and $[n]$ the corresponding set of users. The task is to select a subset $S \subseteq [m]$ of $k$ opinions. %
$\mathscr{A}_n = \{A_1,...,A_n\}$ is the preference profile across all users, where each user $u \in [n]$ approves a set of opinions $A_u \subseteq [m]$. The setting is then described by the tuple $I_{m,\mathscr{A}_n,k} = <m, \mathscr{A}_n,k>$, where $k$ opinions must be selected based on $n$ approval profiles over $m$ opinions.

\noindent
\textbf{Scoring rule: } To select a subset $S$ of $k$ opinions from $[m]$, a scoring rule is used, whereby each opinion $i \in [m]$ is a assigned a score $f(i,\mathscr{A_n})$. Scoring rules are additive: $f(S,\mathscr{A}_n) = \Sigma_{i\in S} f(i, \mathscr{A}_n)$, and may or may not depend on the approval profile $\mathscr{A}_n$. For instance, random selection as a scoring rule is not approval-dependent, whereas examples of approval-dependent rules include engagement (also described as the utilitarian or social welfare rule \cite{elkind2022price}): $f_{\mathrm{engagement}}(i, \mathscr{A}_n) = \bigl|\{u : i \in A_u \}\bigr|$, and coverage: $f_{\mathrm{coverage}}(i, \mathscr{A}_n, V) = \bigl|\{ u \in V : i \in A_u \}\bigr|,$ where $V$ is the set of uncovered users.

\subsection{DiverseBJR: Diversity-aware Balanced Justified Representation}\label{sec:diversebjr}

To investigate the effects of directly enforcing guarantees of proportional representation into our task, we use two existing axioms from social choice theory: Justified Representation \cite{revel2025representative, aziz2017justified}, and Balanced Justified Representation \cite{fish2024generative}, which we define below. We adapt their formulation from approval-based committee election to our problem setting. We also define a novel criterion: Diversity-aware Balanced Justified Representation (DiverseBJR). DiverseBJR aims to preserve the balanced and justified representation properties while additionally promoting diversity in the selected opinions, an aspect not explicitly accounted for by either JR or BJR.  

\begin{definition}[Justified Representation (JR) \cite{revel2025representative, aziz2017justified}]
Let $[m]$ be the set of all opinions expressed by a set of users $[n]$. A subset of opinions $S\subseteq[m]$ with $|S|=k$ satisfies \emph{JR} if, for every group $G$ of at least $n/k$ users, where all users in $G$ approve at least one common opinion in $[m]$, there is at least one user from $G$ who approves at least one item in $S$ .
\end{definition}

\noindent
In words, a selected subset of opinions satisfies JR if every large enough group of users that has shared preferences (approves of at least one opinion in common) is allocated at least one opinion in this subset.

\begin{definition}[Balanced Justified Representation (BJR) \cite{fish2024generative}, adapted to binary approvals]
Let $[m]$ be the set of all opinions expressed by a set of users $[n]$. A subset of opinions $S \subseteq [m]$ with $|S| = k$ satisfies \emph{Balanced Justified Representation (BJR)} if there exists a mapping $\omega : [n] \to S$ such that (i) each opinion $q \in S$ is assigned either $\lfloor n/k \rfloor$ or $\lceil n/k \rceil$ users, and (ii) there is no group $T \subseteq [n]$ with $|T| \ge n/k$ and opinion $q \in [m]$ such that every user $u \in T$ approves $q$ while $q \notin \omega(T)$, where $\omega(T) := \{\omega(u) \mid u \in T\}$.
\end{definition}
\noindent
In words, a selected set of opinions satisfies BJR if there exists a way to assign every user to one of the selected opinions so that the users are split as evenly as possible across the selected opinions, and there can be no group of at least $n/k$ users which unanimously approves some opinion without at least one member of that group being assigned to that opinion.

\begin{definition}[$\varepsilon$-neighbor]\label{def:epsilon-neighbor}
Let $A\in\{0,1\}^{n\times m}$ be the approval matrix and let $\mathbf a_i\in\{0,1\}^n$ denote the approval vector of opinion $i$. Let $d:\{0,1\}^n\times\{0,1\}^n\to[0,1]$ be any distance function. For $\varepsilon\in[0,1]$, an opinion $j\neq i$ is an \emph{$\varepsilon$-neighbor} of opinion $i$ (w.r.t.\ $d$) iff $d(\mathbf a_i,\mathbf a_j)\le\varepsilon$.
\end{definition}

\begin{definition}[DiverseBJR]
Let $[m]$ be the set of all opinions expressed by a set of users $[n]$. Let $\mathrm{dist}:[m]\times[m]\to \mathbb{R}_{\ge 0}$ be a distance function and $\varepsilon>0$ a diversity threshold. A subset of opinions $S \subseteq [m]$ with $|S| = k$ satisfies \emph{DiverseBJR} if there exists a mapping $\omega : [n] \to S$ such that (i) each opinion $q \in S$ is assigned either $\lfloor n/k \rfloor$ or $\lceil n/k \rceil$ users, and (ii) there is no group $T \subseteq [n]$ with $|T| \ge n/k$ and opinion $q \in [m]$ such that every user $u \in T$ approves $q$ while $q \notin \omega(T)$, where $\omega(T) := \{\omega(u) \mid u \in T\}$ \emph{(BJR constraints)}. Additionally, (iii) S is locally $\varepsilon$-diverse among sets satisfying the BJR constraints, i.e., there do not exist $p\in S$ and $q\in [m]\setminus S$ such that $q$ is not an $\varepsilon$-neighbor of any opinion in $S\setminus\{p\}$ and $(S\setminus\{p\})\cup\{q\}$ also satisfies the BJR constraints. 
\end{definition}

\vspace{-0.5em}

In words, a selected set of opinions satisfies DiverseBJR if it satisfies BJR and the selected opinions are $\varepsilon$-diverse, i.e., there is no way to replace one selected opinion by an unselected opinion that is not an $\varepsilon$-neighbor of any of the other selected opinions while still satisfying the BJR constraints. The notion of $\varepsilon$-diversity relates to the notion of approximate clones from social choice theory \cite{procaccia2025clone}, whereby the selected outcome should be approximately invariant to duplicating an item within $\varepsilon$ distance. %

\noindent
Note that in this work, we \textbf{do not attempt to prove existence of DiverseBJR for every instance where BJR holds}. We instead (1) precisely define DiverseBJR, (2) extend the existing GreedyCC procedure \cite{elkind2022price} to construct a diversity-aware approximation of BJR, and (3) use a separate greedy simulator as a \emph{sufficient} feasibility test, ensuring that the imposed diversity constraints do not violate BJR under the greedy policy.\footnote{Alternatively, BJR constraints could be checked exactly via ILP, but ILP solving is NP-hard and doesn't scale. In large online democracy initiatives (e.g., Pol.is consultations with tens of thousands of votes \cite{hsiao2018vtaiwan}), exact verification is impractical, making approximations necessary in real-world settings.}

\section{Experiments}\label{sec:experiments}

\subsection{Evaluation metrics}\label{sec:metrics}
Let $A \in \{0,1\}^{n \times m}$ be the binary approval matrix across all users, where $A_{u,s} = 1$ iff user $u$ approves opinion $s$. Let $\mathscr{G}=\{G_1,\dots,G_\gamma\}$ be a partition of users into political groups, with $G_g\subseteq[n]$ and $|G_g|$ the size of a given political group $g$. Let $S\subseteq[m]$ be the selected set of opinions with $|S|=k$. We operationalize the following democratic criteria as quantitative metrics:

{\textbf{(i) Individual- and Group-level Representation.}}
Let $\bar r_u(S)$ be the "unrepresentation" indicator for a user $u$, where $u$ is unrepresented if they approve none of the selected opinions in $S$ : $\bar r_u(S)=\mathbf{1}\left\{\sum_{s\in S} A_{u,s}=0\right\}$. Following \cite{revel2025representative}, we measure overall "unrepresentation" across users as
\begin{equation}
U_{\mathrm{all}}(S)=100\cdot \frac{1}{n}\sum_{u=1}^n \bar r_u(S).
\label{eq:unrepr}
\end{equation}
$U_{\mathrm{all}}(S)$ relates to classic social welfare, it accounts for total approvals among selected opinions. %
However, as shown by \cite{aziz2017justified}, it does not assess proportionality of representation (a single highly-approved opinion can dominate the sum). Therefore, we also measure the median-voter welfare across political groups. This metric relates to the Rawlsian max-min (maximizing the utility of the worst-off user) used in fairness-aware recommendation systems \cite{deldjoo2024fairness}. In comparison to Rawlsian max-min, our median measure is robust to outliers (a single highly disadvantaged group can dominate the Rawlsian objective, forcing the algorithm to sacrifice overall quality to "protect" the worst-case), and is faster to compute. Let the political group-level unrepresentation be defined as $U_g(S)=100\cdot \frac{1}{|G_g|}\sum_{u\in G_g}\bar r_u(S)$, we measure "unrepresentation" across political groups as
\begin{equation}
\mathrm{median\_}U(S)=\mathrm{median}\bigl\{U_g(S): g=1,\dots,\gamma\bigr\}.
\label{eq:unrepr-pol-groups}
\end{equation}
\paragraph{(ii) Consensus.}
Following recent work in consensus-building via AI-assisted collective dialogues \cite{konya2025using}, we define consensus as the maximin cross-political-group agreement. It is the maximum, over selected opinions, of the minimum approval share across groups. More intuitively, the measured consensus can be seen as an answer to the question "Is there at least one selected opinion which satisfies all political groups?”
\begin{equation}
\mathrm{Consensus}(S)=\max_{s\in S}\min_{g\in[\gamma]}\frac{1}{|G_g|}\sum_{u\in G_g} A_{u,s}.
\label{eq:consensus}
\end{equation}
\textbf{(iii) Diversity: Coverage gap (CG).}
CG quantifies how well the selected $k$ opinions cover the full opinion space.  It is defined as the largest distance between any non-selected opinion and its most similar selected opinion, and is closely related to maximin coverage objectives in recommender systems \cite{zhao2025fairness}. More intuitively, the least-covered opinion differs from every selected opinion by at least the coverage gap (as a fraction of votes).

\begin{equation}
    \mathrm{CG}(S)=\max_{o\in[m]\setminus S}\min_{s\in S} d(o,s).
\label{eq:coverage-gap}
\end{equation}
Note that JR/BJR do not guarantee a low CG.\footnote{Conversely, a small CG does not imply JR/BJR.} For instance, a very novel but highly impopular opinion could be ignored under JR/BJR. Measuring CG enables surfacing precisely such scenarios (identified by a large CG). In our task, the goal is both to guarantee proportional representation in the selected subset (JR/BJR), but also to surface distinct perspectives by covering a large spectrum of expressed views (small CG) \cite{small2021polis, hsiao2018vtaiwan}. %

\textbf{(iv) Diversity: Opinion Redundancy.} While CG measures how well the selected subset $S$ covers the full opinion space, the opinion redundancy metric measures how many distinct viewpoints exist \emph{within} $S$ itself. As discussed in Section \ref{sec:rel-work}, redundancy is not desired in $S$. We define opinion redundancy as the fraction of redundant opinions in $S$. This metric is inspired by outcome-level diversity measures from recommender systems \cite{deldjoo2024fairness, zhao2025fairness} which we tailor to discrete opinion similarity to match our data. For \(|S|=k\), and given a distance metric $d$ and a diversity threshold $\varepsilon$, let \(C_1,\dots,C_m\) denote the clone groups in $S$ formed by grouping opinions that are pairwise linked by the relation \(d(i,j)\le \varepsilon\) (where chains are allowed). We define opinion redundancy as
\begin{equation}
\mathrm{Redundancy\_score}(S)
\;=\;
\frac{\sum_{g=1}^{m} (|C_g|-1)}{k}.
\label{eq:redundancy-score}
\end{equation}

\noindent
In all metrics and baselines, we measure distances using normalized Hamming. In our data, 0 encodes explicit disapproval (not a missing vote), so shared zeros are as informative as shared ones. Hamming therefore captures similarity in both support and opposition. By contrast, Jaccard ignores shared zeros, which is suitable for missingness but not for meaningful disapproval. We thus use $d_H(\mathbf a_i,\mathbf a_j)=\frac{1}{n}\sum_{u=1}^n \mathbf 1[a_{ui}\ne a_{uj}]$.

\noindent
Together, these five metrics operationalize democratic desiderata that are relevant for online deliberation: social welfare, proportional representation, consensus, coverage, and low redundancy. Moreover, no single strategy can optimize them all: work in information retrieval shows that jointly maximizing relevance and diversity is NP-hard \cite{gollapudi2009axiomatic}, while work from social choice shows that optimizing welfare or coverage under Justified Representation is also NP-hard \cite{bredereck2019experimental}. We therefore use them as a comprehensive evaluation tool for existing selection algorithms.

\subsection{Baselines}
To assess adherence to democratic desiderata, we define for each metric a baseline algorithm that optimizes it by construction (e.g., bridging targets consensus). We then evaluate every algorithm on all metrics to characterize which principles it promotes and what it sacrifices (e.g., engagement does not target low redundancy). For notational simplicity, we denote implementations approximating each axiom as JR, BJR, and DiverseBJR. We consider three families of baseline selection algorithms:

\noindent
\textbf{Consensus-based baselines}: We include (i) \emph{Engagement}, a utilitarian rule scoring each item by its number of approvals:
$f_{\text{eng}}(i, \mathscr{A}_n) = \left| \{\, u : i \in A_u \,\} \right|$, and (ii) \emph{Bridging}, for which we use maximin cross-group agreement (CGA, also called Diverse Approval \cite{revel2025representative}). CGA ranks items by their minimum approval rate across political groups, and has been used in online deliberation \cite{konya2025using, konya2024chain, konya2023democratic, revel2025representative}:

$f_{\text{CGA}}(i, \mathscr{A}_n) = \min_{g \in [\gamma]} \frac{1}{|G_g|} \left| \{\, u \in G_g : i \in A_u \,\} \right|$.

\noindent
\textbf{Diversity-based baselines}: We operationalize this baseline as greedy minimization of the coverage gap defined in Equation \ref{eq:coverage-gap}.%

\noindent\textbf{Representation-constraint baselines:}
We implement approximations of JR and BJR as two distinct baselines using a GreedyCC-style \cite{elkind2022price} two-stage selection procedure.
For JR, we use GreedyCC (Algorithm D.1 in \cite{revel2025ai}) with threshold $r=\lceil n/k\rceil$: in Stage~1, while there exists an opinion approved by at least $r$ currently unrepresented users, we repeatedly add the opinion maximizing coverage among those users; once no such opinion exists, Stage~2 fills the remaining slots up to $k$ using a scoring rule $f$ (we use random scoring to minimize dependence on additional scoring rules).
For BJR, we add an explicit greedy matching for balance: we fix per-slot capacities $r_j\in\{\lfloor n/k\rfloor,\lceil n/k\rceil\}$ whose sum is $n$.
For each slot $j$, if some opinion is approved by at least $r_j$ currently unmatched users, we select the opinion with maximum approval among them and assign (i.e., remove) $r_j$ of its approvers; otherwise, we select an opinion by $f$ and assign up to $r_j$ of its approvers.
When choosing which users to assign/remove, we prioritize users with fewer remaining approved opinions (residual degree).
As with JR, we use random scoring for $f$.

\noindent
\textbf{DiverseBJR (ours)}: Building on the BJR-style balanced selection described above, we additionally enforce local $\varepsilon$-diversity using an opinion-distance. We first precompute, for each opinion, its set of direct $\varepsilon$-neighbors (under Hamming distance on approval vectors).
Selection proceeds in $k$ rounds with the same per-slot capacities $r_j\in\{\lfloor n/k\rfloor,\lceil n/k\rceil\}$ and Stage~1/Stage~2 structure as our BJR baseline.
In Stage~1, when there exists an opinion approved by at least $r_j$ currently unmatched users, we pick a maximum-coverage opinion, breaking ties to favor opinions with fewer $\varepsilon$-neighbors and more ``unique'' approvers (users who do not approve any neighbor), and then assign $r_j$ approvers using the residual-degree rule.
In Stage~2, we select by the scoring rule $f$ (random), and then assign up to $r_j$ of its approvers using the residual-degree rule, while avoiding opinions marked ineligible for Stage~2 because they are $\varepsilon$-neighbors of already selected opinions.
After each selection, we attempt to mark the newly selected opinion's $\varepsilon$-neighbors as ineligible for Stage~2; we do so only if a greedy feasibility check indicates that the remaining rounds can still satisfy the BJR constraints given the remaining voters and candidate opinions. Full algorithm descriptions are given in Algorithm~\ref{alg:diversebjr-simple} (DiverseBJR) and Algorithm \ref{alg:greedy-bjr-simulator} (BJR feasibility checker). In experiments, we set $\varepsilon=0.8$. A sensitivity analysis over $\varepsilon$ is reported in Appendix \ref{appendix:epsilon-radius}.

We compare all baselines to each other and to \textbf{DiverseBJR}, our proposed method. We also include (i) a random-selection sanity check (N=100 seeds) to ensure results are non-trivial, and (ii) a zero-shot ChatGPT baseline to probe LLM adherence (details in Appendix \ref{appendix:llm-baseline}).

\subsection{Data}
\label{sec:data}
\begin{figure}[htbp]
\setlength{\abovecaptionskip}{1pt}
\centerline{\includeinkscape[width=\columnwidth]{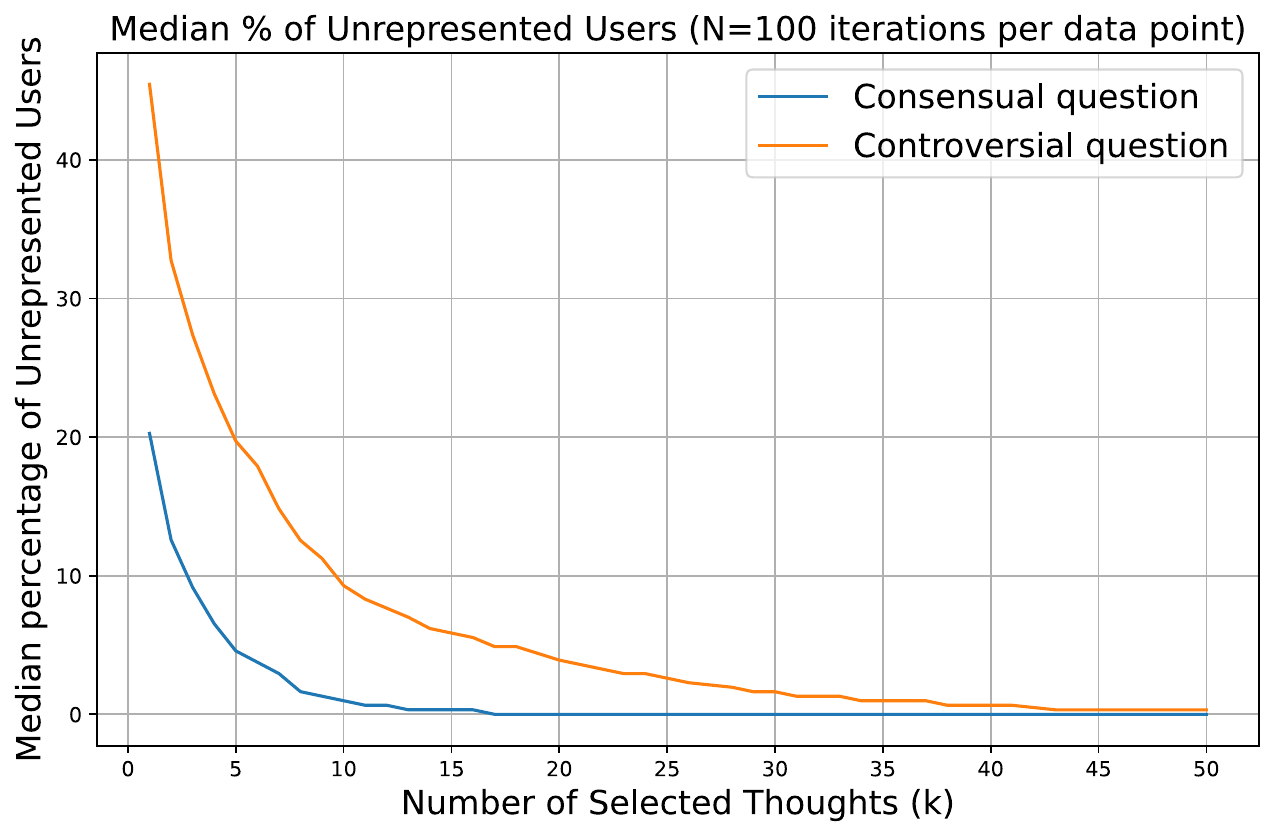}}
\caption{Random selection for a consensual (blue) and controversial (orange) question. At $k=5$, the consensual question leaves $\leq$5\% unrepresented, whereas the controversial question leaves $\geq$20\% unrepresented, motivating more sophisticated selection.} 
\label{fig:consensual_vs_controversial}
\vspace{-2em}
\end{figure}

We use the "Polarized Issues Dataset" by Remesh.\footnote{The data is publicly available at \url{https://github.com/akonya/polarized-issues-data}. We preprocess it only to remove duplicate or empty opinions.} The data contains real-world collective dialogues conducted with N300 representative samples of the US public, focusing on polarized topics such as the right to protest, and includes participants’ political-leaning attributes, captured via self-reported choices drawn from \textit{\{Moderate, Slightly conservative, Very conservative, Slightly liberal, Very liberal\}}. %
Across various questions (e.g., \textit{"What are features or characteristics that make a protest appropriate?"}), each participant expresses their personal opinion in free-form text, and votes on the opinions of a few other participants. Following Remesh's process (see Section \ref{sec:rel-work}), missing votes are inferred probabilistically. Following \cite{revel2025representative}, we threshold probabilities by 0.5 to get a binary approval matrix, thus matching our problem setting. To control task difficulty, we split questions by how well a random set of five opinions represents users: \emph{consensual} if it leaves $\leq$5\% unrepresented (near-trivial), and \emph{controversial} if it leaves $\geq$20\% unrepresented (Fig. \ref{fig:consensual_vs_controversial}). We choose these empirical thresholds to increase difficulty while retaining sufficient data for analysis (Table \ref{tab:data}). We also replicate all experiments on a different dataset with fully observed approvals.\footnote{This validates the robustness of our findings to Remesh’s vote inference and to the 0.5 threshold, see Appendix \ref{appendix:validation-habermas}.}

\subsection{Results}\label{sec:results}
For each metric, we plot performance with respect to the number of selected opinions across baselines. We focus on controversial questions, where selection is more challenging (Fig. \ref{fig:consensual_vs_controversial}). Results on consensual questions are presented in Appendix \ref{appendix:full-results}. We omit the zero-shot ChatGPT baseline from the main results because it occasionally returns results with fewer than 
k indices (i.e., selects less than required), which makes the metrics not directly comparable. We report it separately in Appendix \ref{appendix:llm-baseline}.

\begin{figure*}[htbp]%
\centering
    \begin{subfigure}[c]{.49\textwidth}
      \centering
      \includeinkscape[height=5cm]{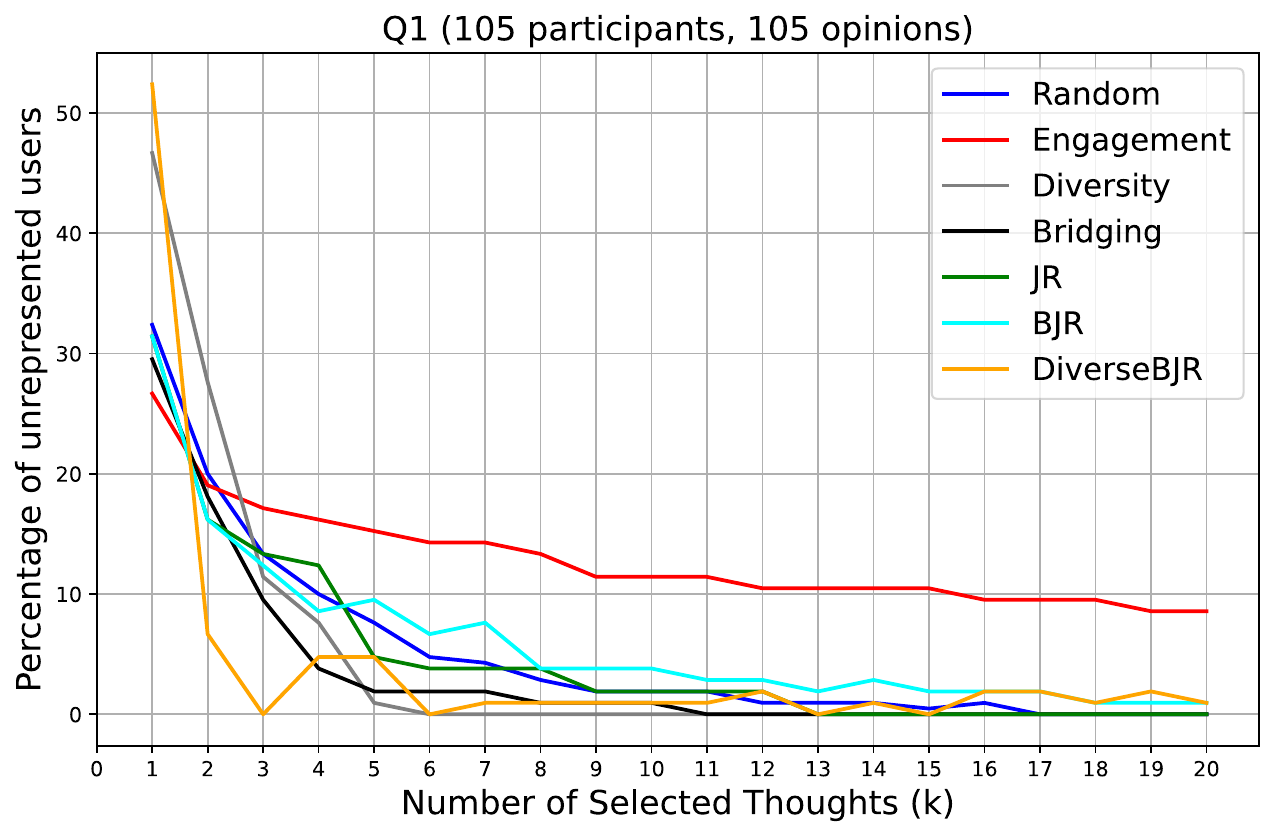}
    \end{subfigure}\hfill
  \begin{subfigure}[c]{.49\textwidth}
    \centering
    \includeinkscape[height=5cm]{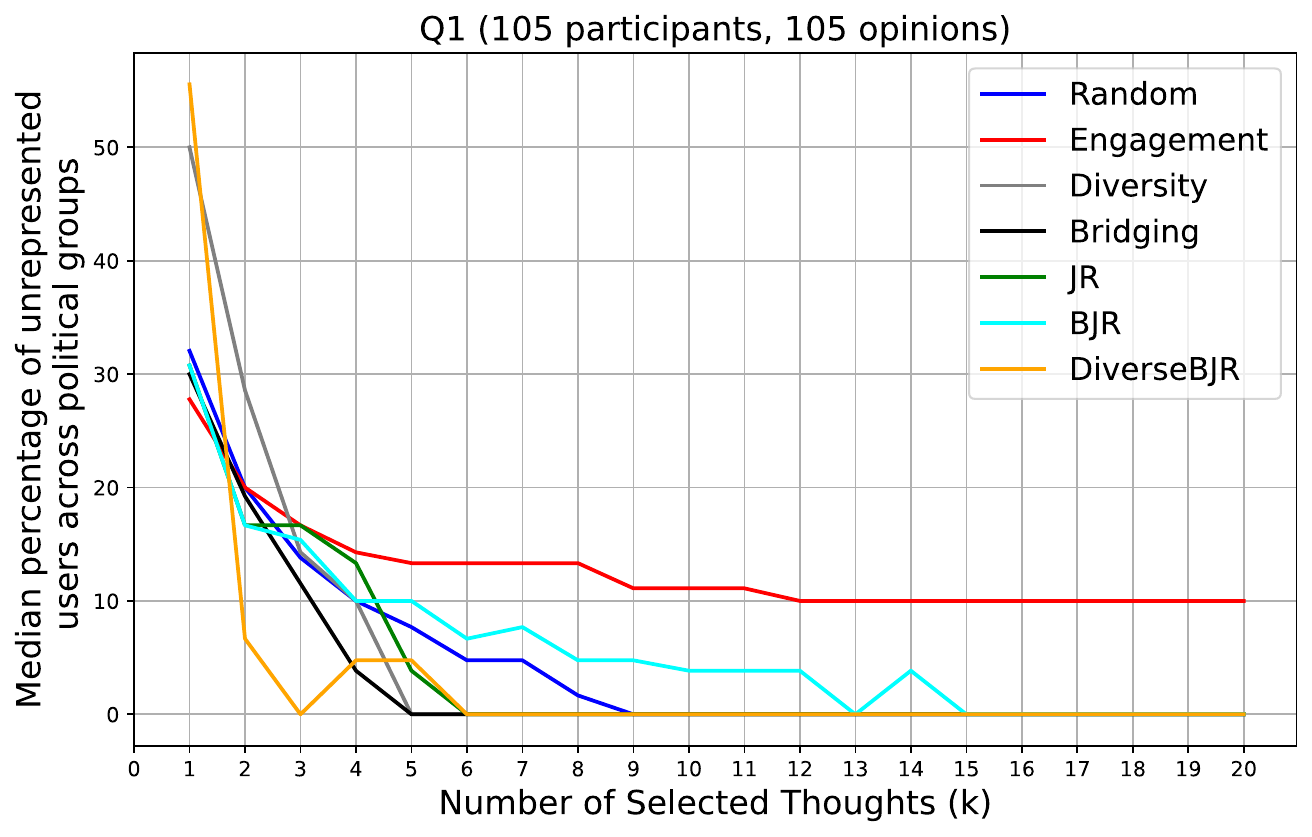}%
  \end{subfigure}\\
    \begin{subfigure}[c]{.49\textwidth}
      \centering
      \includeinkscape[height=5cm]{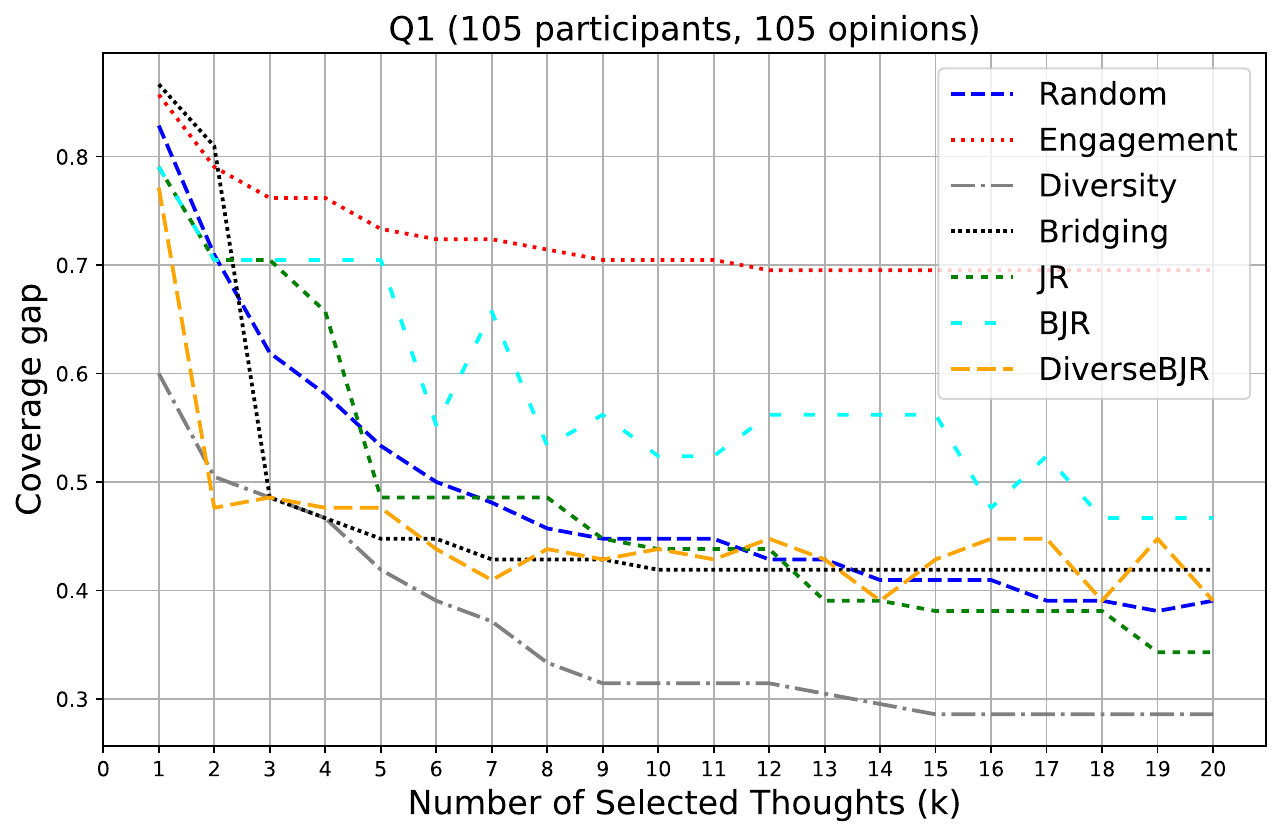}
    \end{subfigure}\hfill
  \begin{subfigure}[c]{.49\textwidth}
    \centering
    \includeinkscape[height=5.5cm]{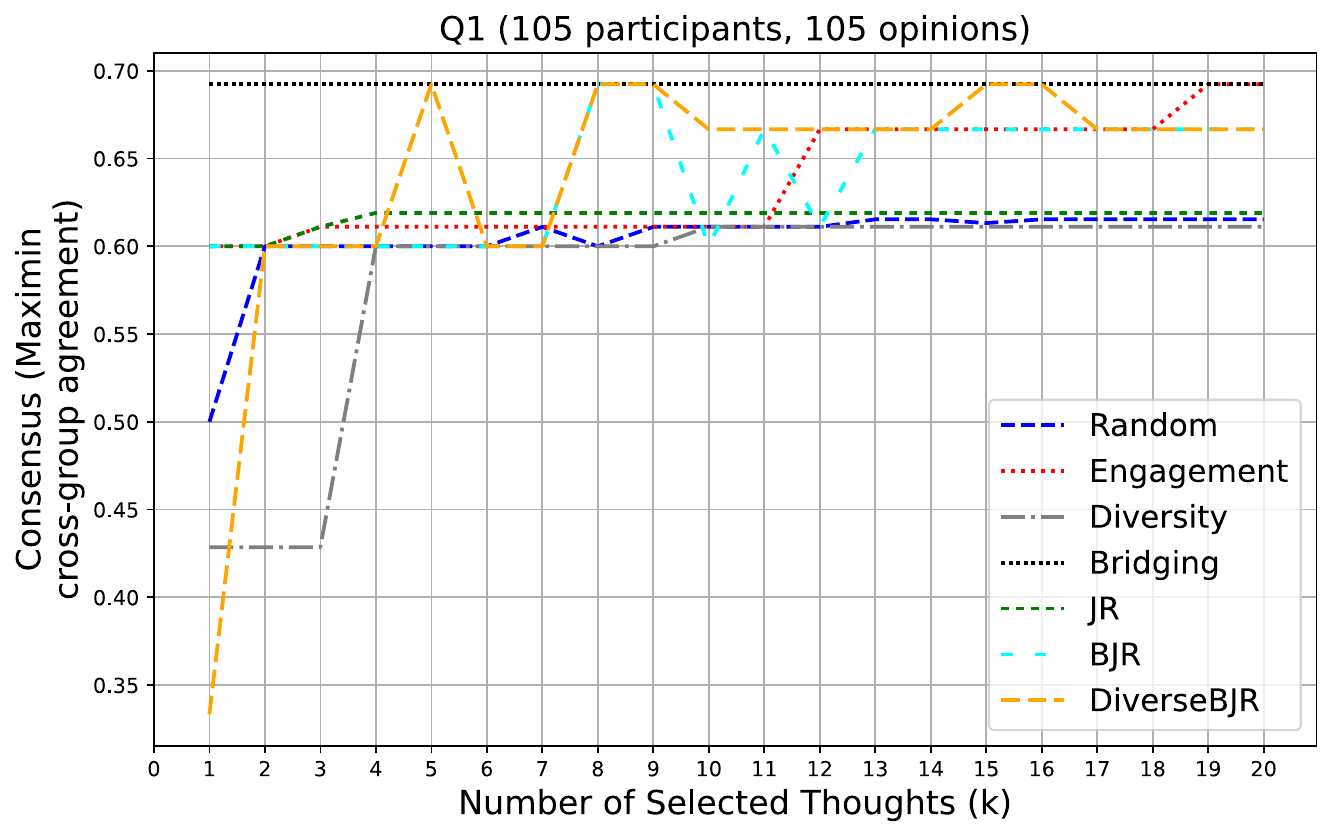}%
  \end{subfigure}
  \begin{subfigure}[c]{.49\textwidth}
    \centering
    \includeinkscape[height=5cm]{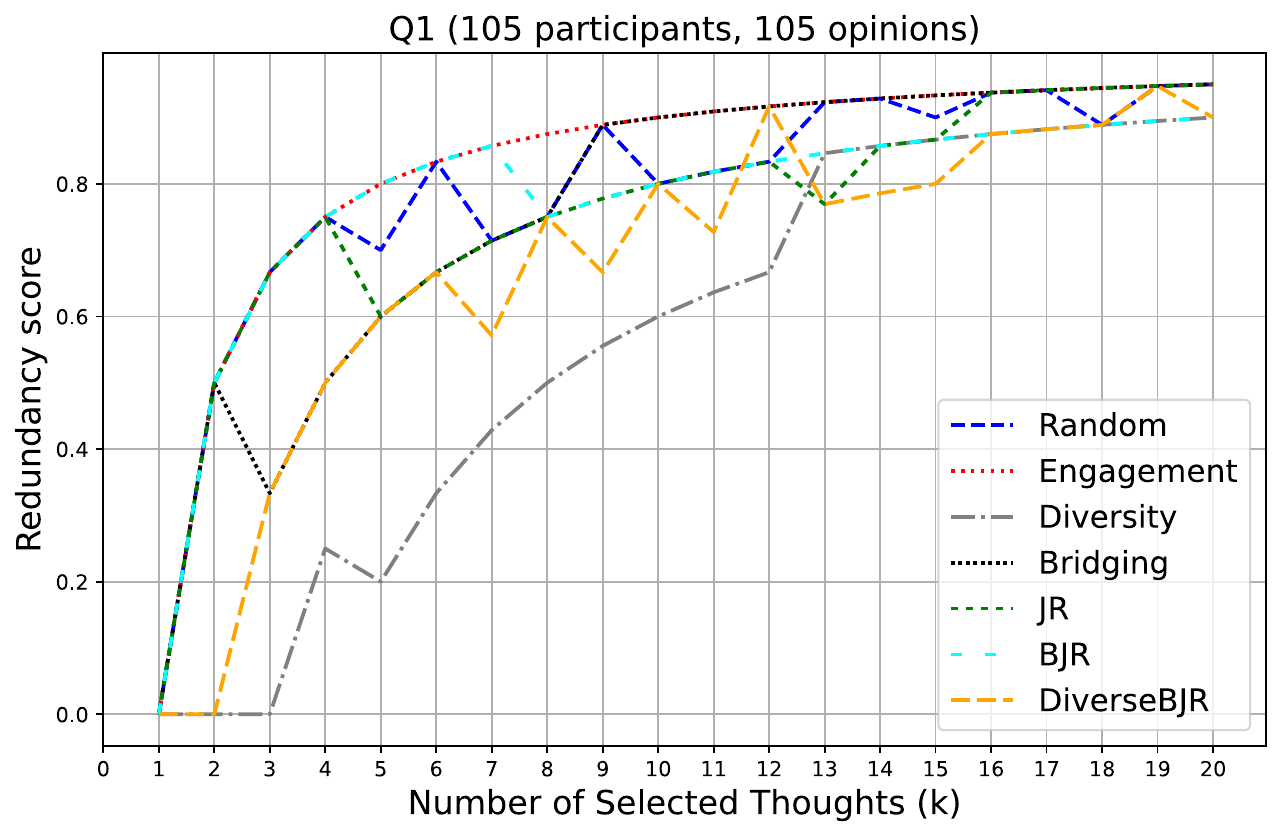}%
  \end{subfigure}\\
  \caption{Evolution of various democratic desiderata with the number of selected opinions (dotted lines are used for better distinction of overlapping zones).%
      \label{fig:q1-all}}
\end{figure*}

\textbf{(1) DiverseBJR typically gives the largest gains in representation (overall and across political groups) across all baselines when selecting a small subset $(k\approx2-3)$, making it most efficient for deliberation where subset size matters}. More generally, across both overall representation (Fig. \ref{fig:q1-all} - upper-left) and group representation (Fig. \ref{fig:q1-all} - upper-right), representation-constraint algorithms (JR, BJR, DiverseBJR) substantially reduce unrepresented users and achieve the strongest early gains, particularly at very small k, consistent with prior work \cite{revel2025representative, fish2024generative, boehmer2025generative}. However, they do not systematically dominate for all $k$. At moderate k, Diversity can outperform JR and BJR . Consensus-oriented algorithms perform less favorably: Engagement is comparable to or worse than random, while Bridging consistently improves on Engagement with strong early gains but typically remains behind representation-constraint methods. %
Results on consensual questions are reported in Appendix \ref{appendix:full-results}: trends for representation-constraint algorithms are consistent, while all baselines improve, especially Diversity and Bridging, an expected result since most opinions are widely approved in this set (Fig. \ref{fig:consensual_vs_controversial}), making representation easier to achieve without constraints. Note also that some baselines are non-monotonic because they re-select from scratch for each $k$. In particular, BJR/DiverseBJR need not have the $k+1$ solution contain the $k$ solution.

\textbf{(2) Coverage-wise, representation-constraint algorithms (JR, BJR, DiverseBJR) and Bridging generally outperform Engagement, particularly at small $k$.} (Fig. \ref{fig:q1-all} - center-left) Diversity typically achieves the lowest coverage gap, as expected since it minimizes CG by design. %
Engagement has the largest gaps for most $k$ values, yielding the worst opinion-space coverage. On consensual questions (Appendix \ref{appendix:full-results}), BJR and DiverseBJR generally perform worse for moderate-to-high $k$, typically trailing JR and Bridging. Overall, this indicates that although representation-constraint methods target proportional representation, they still deliver broad opinion-space coverage (particularly on controversial questions), matching the goal in online deliberation of guaranteeing proportional representation (JR/BJR) while surfacing diverse perspectives (low CG) \cite{small2021polis, hsiao2018vtaiwan}.  %

\begin{table*}[htbp]%
    \centering
    \resizebox{.8\textwidth}{!}{
    \begin{tabular}{llp{8cm}cc}
        \hline
        \textbf{ID} & \textbf{Split} & \textbf{Questions} & \textbf{Participants(n)}& \textbf{Opinions(m)}  \\
        \hline
        Q1& Controversial&  How has your personal experience with protests influenced your viewpoint on the right to assemble?& 105& 105\\
        Q2& Controversial& What are features or characteristics that make a protest appropriate?& 307& 306\\
        \hline
        Q3& Consensual& What characteristics or actions, in your view, deem a protest inappropriate?& 306& 299\\
        Q4& Consensual& What are some of the reasons you have not participated in or attended a protest?& 201& 201\\
        Q5& Consensual& What measures (if any) could be taken to restrict or limit inappropriate protests?& 303& 299\\
        Q6& Consensual& What measures (if any) could be taken to ensure appropriate protests are protected?& 305& 302\\
        \hline
    \end{tabular}}
    \caption{Data summary}
    \label{tab:data}
\end{table*}

\textbf{(3) JR/BJR (and DiverseBJR for $k\geq2$) typically select opinions with consensus levels similar to Engagement, but less consensual than Bridging}, which maximizes consensus by design (Fig. \ref{fig:q1-all} - center-right). This aligns with prior work showing that while JR-style selection can in theory reduce engagement, in practice it yields little engagement loss \cite{revel2025representative}. Crucially, it suggests that enforcing proportional representation and diversity still allows selecting broadly acceptable opinions when they exist. Results are consistent on consensual questions (Appendix \ref{appendix:full-results}).

\textbf{(4) For small k, representation-constraint algorithms typically select less redundant opinions than consensus-oriented algorithms}. On controversial questions (Fig. \ref{fig:q1-all} - bottom), DiverseBJR achieves low redundancy at small k, second only to Diversity (which minimizes redundancy by design). To illustrate the differing selection dynamics, Fig. \ref{appendix:viz-diversebjr-vs-engagement}.\ref{fig:diversity-viz} visualizes Engagement versus DiverseBJR. Engagement selects closely clustered opinions, whereas DiverseBJR spreads selection across the opinion space, reducing redundancy. Results are consistent on consensual questions (Appendix \ref{appendix:full-results}), where redundancy rises quickly for all baselines, as many opinions are similar.

\textbf{Takeaways:} DiverseBJR provides the strongest trade-off at small $k$: it improves proportional representation relative to diversity baselines while preserving high coverage and low redundancy. More generally, representation-constraint methods are beneficial for deliberation settings, where only a few opinions can be surfaced. Diversity achieves high coverage and low redundancy but offers no proportional-representation guarantees, whereas consensus-oriented methods can miss minority views and typically underperform on representation. Bridging is a strong compromise but requires sensitive attributes (e.g., political leanings) and depends on their quality. Findings replicate on a separate dataset from a virtual UK citizen assembly (Appendix \ref{appendix:validation-habermas}). %

\textbf{Limitations and Future work:} 
Algorithms can significantly improve the efficiency of online deliberation by helping participants, moderators and facilitators process large volumes of inputs. This work takes a step in this direction by incorporating notions of diversity and balanced representation into the algorithmic selection process. While our results clearly show how different strategies for algorithmic selection perform on democratic desiderata, they are still offline results, and may differ from those observed in real-world deployments. In future work, it would be interesting to go beyond observational data by deploying these algorithms during an online deliberation, similar to what existing works have done for the related use-case of algorithmic intervention on social media \cite{tucker2023,bail2018exposure}.

\section*{Reproducibility Statement}
We provide, in full details, our algorithms' pseudocode and corresponding empirical parameters, as well as our evaluation protocols and metrics. We also release the code and evaluation scripts at \url{https://github.com/SalimHFX/Algorithms-for-Opinion-Selection-in-Deliberation} to enable independent replication of experiments and extension of our evaluation framework. Special attention was given to making code easy-to-use, with the goal of disseminating it to both researchers and civil-society practitioners.

\section*{Impact Statement}
This work has direct societal consequences on the efficiency and safety of online democracy initiatives. It presents work whose goal is to advance fairness in algorithmic selection in contexts of online deliberation. It assesses how various existing algorithmic strategies perform on specific metrics reflecting democratic desiderata, and proposes a novel algorithm that incorporates notions of diversity and balanced representation into the algorithmic selection process. Given the increasingly large amount of online participatory democracy platforms and initiatives across various use-cases (e.g., peacebuilding, participatory budgeting, public policy design), it becomes crucial to develop publicly available (i) principled evaluation benchmarks that can be used by practitioners and researchers, (ii) fairness-aware algorithms that guarantee desired democratic properties and can be safely deployed in such high stakes online contexts.

\bibliography{custom}

\begin{thebibliography}{47}
\providecommand{\natexlab}[1]{#1}
\providecommand{\url}[1]{\texttt{#1}}
\expandafter\ifx\csname urlstyle\endcsname\relax
  \providecommand{\doi}[1]{doi: #1}\else
  \providecommand{\doi}{doi: \begingroup \urlstyle{rm}\Url}\fi

\bibitem[Abdollahpouri et~al.(2019)Abdollahpouri, Burke, and
  Mobasher]{abdollahpouri2019managing}
Abdollahpouri, H., Burke, R., and Mobasher, B.
\newblock Managing popularity bias in recommender systems with personalized
  re-ranking.
\newblock \emph{arXiv preprint arXiv:1901.07555}, 2019.

\bibitem[Akpinar et~al.(2026)Akpinar, Avula, Lee, Dang, Razat, and
  Murdock]{akpinar2026llm}
Akpinar, N.-J., Avula, S., Lee, C., Dang, B., Razat, K., and Murdock, V.
\newblock Llm or human? perceptions of trust and information quality in
  research summaries.
\newblock \emph{arXiv preprint arXiv:2601.15556}, 2026.

\bibitem[Aziz et~al.(2017)Aziz, Brill, Conitzer, Elkind, Freeman, and
  Walsh]{aziz2017justified}
Aziz, H., Brill, M., Conitzer, V., Elkind, E., Freeman, R., and Walsh, T.
\newblock Justified representation in approval-based committee voting.
\newblock \emph{Social Choice and Welfare}, 48\penalty0 (2):\penalty0 461--485,
  2017.

\bibitem[Bail et~al.(2018)Bail, Argyle, Brown, Bumpus, Chen, Hunzaker, Lee,
  Mann, Merhout, and Volfovsky]{bail2018exposure}
Bail, C.~A., Argyle, L.~P., Brown, T.~W., Bumpus, J.~P., Chen, H., Hunzaker,
  M.~F., Lee, J., Mann, M., Merhout, F., and Volfovsky, A.
\newblock Exposure to opposing views on social media can increase political
  polarization.
\newblock \emph{Proceedings of the National Academy of Sciences}, 115\penalty0
  (37):\penalty0 9216--9221, 2018.

\bibitem[B{\'e}al et~al.(2025)B{\'e}al, Deschamps, Diss, and
  Takengd]{beal2025multiwinner}
B{\'e}al, S., Deschamps, M., Diss, M., and Takengd, R.~T.
\newblock Multiwinner elections with diversity constraints on individual
  preferences.
\newblock \emph{Revue d'{\'e}conomie politique}, 135\penalty0 (1):\penalty0
  169--203, 2025.

\bibitem[Bilich et~al.(2023)Bilich, Varga, Masood, and Konya]{bilich2023faster}
Bilich, J., Varga, M., Masood, D., and Konya, A.
\newblock Faster peace via inclusivity: An efficient paradigm to understand
  populations in conflict zones.
\newblock \emph{arXiv preprint arXiv:2311.00816}, 2023.

\bibitem[Boehmer et~al.(2025)Boehmer, Fish, and
  Procaccia]{boehmer2025generative}
Boehmer, N., Fish, S., and Procaccia, A.~D.
\newblock Generative social choice: The next generation.
\newblock In \emph{Forty-second International Conference on Machine Learning},
  2025.

\bibitem[Bouyzourn \& Birch(2025)Bouyzourn and Birch]{bouyzourn2025shapes}
Bouyzourn, K. and Birch, A.
\newblock What shapes user trust in chatgpt? a mixed-methods study of user
  attributes, trust dimensions, task context, and societal perceptions among
  university students.
\newblock \emph{arXiv preprint arXiv:2507.05046}, 2025.

\bibitem[Brandt et~al.(2016)Brandt, Conitzer, Endriss, Lang, and
  Procaccia]{brandt2016handbook}
Brandt, F., Conitzer, V., Endriss, U., Lang, J., and Procaccia, A.~D.
\newblock \emph{Handbook of computational social choice}.
\newblock Cambridge University Press, 2016.

\bibitem[Bredereck et~al.(2018)Bredereck, Faliszewski, Igarashi, Lackner, and
  Skowron]{bredereck2018multiwinner}
Bredereck, R., Faliszewski, P., Igarashi, A., Lackner, M., and Skowron, P.
\newblock Multiwinner elections with diversity constraints.
\newblock In \emph{Proceedings of the AAAI Conference on Artificial
  Intelligence}, volume~32, 2018.

\bibitem[Bredereck et~al.(2019)Bredereck, Faliszewski, Kaczmarczyk, and
  Niedermeier]{bredereck2019experimental}
Bredereck, R., Faliszewski, P., Kaczmarczyk, A., and Niedermeier, R.
\newblock An experimental view on committees providing justified
  representation.
\newblock In \emph{IJCAI}, pp.\  109--115, 2019.

\bibitem[Crawford(2016)]{crawford2016can}
Crawford, K.
\newblock Can an algorithm be agonistic? ten scenes from life in calculated
  publics.
\newblock \emph{Science, Technology, \& Human Values}, 41\penalty0
  (1):\penalty0 77--92, 2016.

\bibitem[Deldjoo et~al.(2024)Deldjoo, Jannach, Bellogin, Difonzo, and
  Zanzonelli]{deldjoo2024fairness}
Deldjoo, Y., Jannach, D., Bellogin, A., Difonzo, A., and Zanzonelli, D.
\newblock Fairness in recommender systems: research landscape and future
  directions.
\newblock \emph{User Modeling and User-Adapted Interaction}, 34\penalty0
  (1):\penalty0 59--108, 2024.

\bibitem[Dryzek \& List(2003)Dryzek and List]{dryzek2003social}
Dryzek, J.~S. and List, C.
\newblock Social choice theory and deliberative democracy: A reconciliation.
\newblock \emph{British journal of political science}, 33\penalty0
  (1):\penalty0 1--28, 2003.

\bibitem[Elkind et~al.(2022)Elkind, Faliszewski, Igarashi, Manurangsi,
  Schmidt-Kraepelin, and Suksompong]{elkind2022price}
Elkind, E., Faliszewski, P., Igarashi, A., Manurangsi, P., Schmidt-Kraepelin,
  U., and Suksompong, W.
\newblock The price of justified representation.
\newblock In \emph{Proceedings of the AAAI Conference on Artificial
  Intelligence}, volume~36, pp.\  4983--4990, 2022.

\bibitem[Fish et~al.(2024)Fish, G{\"o}lz, Parkes, Procaccia, Rusak, Shapira,
  and W{\"u}thrich]{fish2024generative}
Fish, S., G{\"o}lz, P., Parkes, D.~C., Procaccia, A.~D., Rusak, G., Shapira,
  I., and W{\"u}thrich, M.
\newblock Generative social choice.
\newblock In \emph{Proceedings of the 25th ACM Conference on Economics and
  Computation}, pp.\  985--985, 2024.

\bibitem[Fishkin et~al.(2019)Fishkin, Garg, Gelauff, Goel, Munagala,
  Sakshuwong, Siu, and Yandamuri]{fishkin2019deliberative}
Fishkin, J., Garg, N., Gelauff, L., Goel, A., Munagala, K., Sakshuwong, S.,
  Siu, A., and Yandamuri, S.
\newblock Deliberative democracy with the online deliberation platform.
\newblock In \emph{The 7th AAAI Conference on Human Computation and
  Crowdsourcing (HCOMP 2019)}, pp.\  1--2, 2019.

\bibitem[Gollapudi \& Sharma(2009)Gollapudi and Sharma]{gollapudi2009axiomatic}
Gollapudi, S. and Sharma, A.
\newblock An axiomatic approach for result diversification.
\newblock In \emph{Proceedings of the 18th international conference on World
  wide web}, pp.\  381--390, 2009.

\bibitem[Guess et~al.(2023)Guess, Malhotra, Pan, Barberá, Allcott, Brown,
  Crespo-Tenorio, Dimmery, Freelon, Gentzkow, González-Bailón, Kennedy, Kim,
  Lazer, Moehler, Nyhan, Rivera, Settle, Thomas, Thorson, Tromble, Wilkins,
  Wojcieszak, Xiong, de~Jonge, Franco, Mason, Stroud, and Tucker]{tucker2023}
Guess, A.~M., Malhotra, N., Pan, J., Barberá, P., Allcott, H., Brown, T.,
  Crespo-Tenorio, A., Dimmery, D., Freelon, D., Gentzkow, M.,
  González-Bailón, S., Kennedy, E., Kim, Y.~M., Lazer, D., Moehler, D.,
  Nyhan, B., Rivera, C.~V., Settle, J., Thomas, D.~R., Thorson, E., Tromble,
  R., Wilkins, A., Wojcieszak, M., Xiong, B., de~Jonge, C.~K., Franco, A.,
  Mason, W., Stroud, N.~J., and Tucker, J.~A.
\newblock How do social media feed algorithms affect attitudes and behavior in
  an election campaign?
\newblock \emph{Science}, 381\penalty0 (6656):\penalty0 398--404, 2023.
\newblock \doi{10.1126/science.abp9364}.
\newblock URL \url{https://www.science.org/doi/abs/10.1126/science.abp9364}.

\bibitem[Habermas(2015)]{habermas2015between}
Habermas, J.
\newblock \emph{Between facts and norms: Contributions to a discourse theory of
  law and democracy}.
\newblock John Wiley \& Sons, 2015.

\bibitem[Halpern et~al.(2023)Halpern, Kehne, Procaccia, Tucker-Foltz, and
  W{\"u}thrich]{halpern2023representation}
Halpern, D., Kehne, G., Procaccia, A.~D., Tucker-Foltz, J., and W{\"u}thrich,
  M.
\newblock Representation with incomplete votes.
\newblock In \emph{Proceedings of the AAAI Conference on Artificial
  Intelligence}, volume~37, pp.\  5657--5664, 2023.

\bibitem[Hsiao et~al.(2018)Hsiao, Lin, Tang, Narayanan, and
  Sarahe]{hsiao2018vtaiwan}
Hsiao, Y.-T., Lin, S.-Y., Tang, A., Narayanan, D., and Sarahe, C.
\newblock vtaiwan: An empirical study of open consultation process in taiwan.
\newblock \emph{Taiwan: Center for Open Science}, 2018.

\bibitem[Kalayci et~al.(2025)Kalayci, Liu, and Kempe]{kalayci2025full}
Kalayci, Y.~H., Liu, J., and Kempe, D.
\newblock Full proportional justified representation.
\newblock In \emph{Proceedings of the 24th International Conference on
  Autonomous Agents and Multiagent Systems}, pp.\  1070--1078, 2025.

\bibitem[Klimashevskaia et~al.(2024)Klimashevskaia, Jannach, Elahi, and
  Trattner]{klimashevskaia2024survey}
Klimashevskaia, A., Jannach, D., Elahi, M., and Trattner, C.
\newblock A survey on popularity bias in recommender systems.
\newblock \emph{User Modeling and User-Adapted Interaction}, 34\penalty0
  (5):\penalty0 1777--1834, 2024.

\bibitem[Konya et~al.(2022)Konya, Qiu, Varga, and Ovadya]{konya2022elicitation}
Konya, A., Qiu, Y.~L., Varga, M.~P., and Ovadya, A.
\newblock Elicitation inference optimization for multi-principal-agent
  alignment.
\newblock 2022.

\bibitem[Konya et~al.(2023{\natexlab{a}})Konya, Schirch, Irwin, and
  Ovadya]{konya2023democratic}
Konya, A., Schirch, L., Irwin, C., and Ovadya, A.
\newblock Democratic policy development using collective dialogues and ai.
\newblock \emph{arXiv preprint arXiv:2311.02242}, 2023{\natexlab{a}}.

\bibitem[Konya et~al.(2023{\natexlab{b}})Konya, Turan, Ovadya, Qui, Masood,
  Devine, Schirch, Roberts, and Forum]{konya2023deliberative}
Konya, A., Turan, D., Ovadya, A., Qui, L., Masood, D., Devine, F., Schirch, L.,
  Roberts, I., and Forum, D.~A.
\newblock Deliberative technology for alignment.
\newblock \emph{arXiv preprint arXiv:2312.03893}, 2023{\natexlab{b}}.

\bibitem[Konya et~al.(2024)Konya, Ovadya, Feng, Chen, Schirch, Irwin, and
  Zhang]{konya2024chain}
Konya, A., Ovadya, A., Feng, K., Chen, Q.~Z., Schirch, L., Irwin, C., and
  Zhang, A.~X.
\newblock Chain of alignment: Integrating public will with expert intelligence
  for language model alignment.
\newblock In \emph{Pluralistic Alignment Workshop at NeurIPS 2024}, 2024.

\bibitem[Konya et~al.(2025)Konya, Thorburn, Almasri, Leshem, Procaccia,
  Schirch, and Bakker]{konya2025using}
Konya, A., Thorburn, L., Almasri, W., Leshem, O.~A., Procaccia, A., Schirch,
  L., and Bakker, M.
\newblock Using collective dialogues and ai to find common ground between
  israeli and palestinian peacebuilders.
\newblock In \emph{Proceedings of the 2025 ACM Conference on Fairness,
  Accountability, and Transparency}, pp.\  312--333, 2025.

\bibitem[Landemore(2020)]{landemore2020open}
Landemore, H.
\newblock Open democracy: Reinventing popular rule for the twenty-first
  century.
\newblock 2020.

\bibitem[Ovadya(2023)]{ovadya2023generative}
Ovadya, A.
\newblock 'generative ci'through collective response systems.
\newblock \emph{arXiv preprint arXiv:2302.00672}, 2023.

\bibitem[Ovadya \& Thorburn(2023)Ovadya and Thorburn]{ovadya2023bridging}
Ovadya, A. and Thorburn, L.
\newblock Bridging systems: open problems for countering destructive
  divisiveness across ranking, recommenders, and governance.
\newblock \emph{arXiv preprint arXiv:2301.09976}, 2023.

\bibitem[Pessach \& Shmueli(2022)Pessach and Shmueli]{pessach2022review}
Pessach, D. and Shmueli, E.
\newblock A review on fairness in machine learning.
\newblock \emph{ACM Computing Surveys (CSUR)}, 55\penalty0 (3):\penalty0 1--44,
  2022.

\bibitem[Procaccia et~al.(2025)Procaccia, Schiffer, and
  Zhang]{procaccia2025clone}
Procaccia, A.~D., Schiffer, B., and Zhang, S.
\newblock Clone-robust ai alignment.
\newblock \emph{arXiv preprint arXiv:2501.09254}, 2025.

\bibitem[Rawls(1997)]{rawls1997idea}
Rawls, J.
\newblock The idea of public reason revisited.
\newblock \emph{The university of Chicago law review}, 64\penalty0
  (3):\penalty0 765--807, 1997.

\bibitem[Relia(2022)]{ijcai2022p714}
Relia, K.
\newblock Dire committee : Diversity and representation constraints in
  multiwinner elections.
\newblock In Raedt, L.~D. (ed.), \emph{Proceedings of the Thirty-First
  International Joint Conference on Artificial Intelligence, {IJCAI-22}}, pp.\
  5143--5149. International Joint Conferences on Artificial Intelligence
  Organization, 7 2022.
\newblock \doi{10.24963/ijcai.2022/714}.
\newblock URL \url{https://doi.org/10.24963/ijcai.2022/714}.
\newblock AI for Good.

\bibitem[Revel \& P{\'e}nigaud(2025)Revel and P{\'e}nigaud]{revel2025ai}
Revel, M. and P{\'e}nigaud, T.
\newblock Ai-facilitated collective judgements.
\newblock \emph{arXiv preprint arXiv:2503.05830}, 2025.

\bibitem[Revel et~al.(2025)Revel, Milli, Lu, Watson-Daniels, and
  Nickel]{revel2025representative}
Revel, M., Milli, S., Lu, T., Watson-Daniels, J., and Nickel, M.
\newblock Representative ranking for deliberation in the public sphere.
\newblock \emph{arXiv preprint arXiv:2503.18962}, 2025.

\bibitem[S{\'a}nchez-Fern{\'a}ndez et~al.(2017)S{\'a}nchez-Fern{\'a}ndez,
  Elkind, Lackner, Fern{\'a}ndez, Fisteus, Val, and
  Skowron]{sanchez2017proportional}
S{\'a}nchez-Fern{\'a}ndez, L., Elkind, E., Lackner, M., Fern{\'a}ndez, N.,
  Fisteus, J., Val, P.~B., and Skowron, P.
\newblock Proportional justified representation.
\newblock In \emph{Proceedings of the AAAI Conference on Artificial
  Intelligence}, volume~31, 2017.

\bibitem[Shortall et~al.(2021)Shortall, Itten, van~der Meer, Murukannaiah, and
  Jonker]{shortall2021inclusion}
Shortall, R., Itten, A., van~der Meer, M., Murukannaiah, P.~K., and Jonker,
  C.~M.
\newblock Inclusion, equality and bias in designing online mass deliberative
  platforms.
\newblock \emph{arXiv preprint arXiv:2107.12711}, 2021.

\bibitem[Shrestha \& Yang(2019)Shrestha and Yang]{shrestha2019fairness}
Shrestha, Y.~R. and Yang, Y.
\newblock Fairness in algorithmic decision-making: Applications in multi-winner
  voting, machine learning, and recommender systems.
\newblock \emph{Algorithms}, 12\penalty0 (9):\penalty0 199, 2019.

\bibitem[Small et~al.(2021)Small, Bjorkegren, Erkkil{\"a}, Shaw, and
  Megill]{small2021polis}
Small, C., Bjorkegren, M., Erkkil{\"a}, T., Shaw, L., and Megill, C.
\newblock Polis: Scaling deliberation by mapping high dimensional opinion
  spaces.
\newblock \emph{Recerca: revista de pensament i an{\`a}lisi}, 26\penalty0 (2),
  2021.

\bibitem[Tessler et~al.(2024)Tessler, Bakker, Jarrett, Sheahan, Chadwick,
  Koster, Evans, Campbell-Gillingham, Collins, Parkes, et~al.]{tessler2024ai}
Tessler, M.~H., Bakker, M.~A., Jarrett, D., Sheahan, H., Chadwick, M.~J.,
  Koster, R., Evans, G., Campbell-Gillingham, L., Collins, T., Parkes, D.~C.,
  et~al.
\newblock Ai can help humans find common ground in democratic deliberation.
\newblock \emph{Science}, 386\penalty0 (6719):\penalty0 eadq2852, 2024.

\bibitem[Zhao et~al.(2025)Zhao, Wang, Liu, Cheng, Aggarwal, and
  Derr]{zhao2025fairness}
Zhao, Y., Wang, Y., Liu, Y., Cheng, X., Aggarwal, C.~C., and Derr, T.
\newblock Fairness and diversity in recommender systems: a survey.
\newblock \emph{ACM Transactions on Intelligent Systems and Technology},
  16\penalty0 (1):\penalty0 1--28, 2025.

\bibitem[Zhou et~al.(2025)Zhou, Neumann, Garimella, and
  Gionis]{zhou2025calibrated}
Zhou, T., Neumann, S., Garimella, K., and Gionis, A.
\newblock Calibrated and diverse news coverage.
\newblock In \emph{Proceedings of the 34th ACM International Conference on
  Information and Knowledge Management}, pp.\  4509--4518, 2025.

\bibitem[Zhu et~al.(2025)Zhu, Yang, Bakker, Pentland, and Pei]{zhu2025can}
Zhu, S., Yang, S., Bakker, M.~A., Pentland, A., and Pei, J.
\newblock Can ai truly represent your voice in deliberations? a comprehensive
  study of large-scale opinion aggregation with llms.
\newblock \emph{arXiv preprint arXiv:2510.05154}, 2025.

\bibitem[Ziegler et~al.(2005)Ziegler, McNee, Konstan, and
  Lausen]{ziegler2005improving}
Ziegler, C.-N., McNee, S.~M., Konstan, J.~A., and Lausen, G.
\newblock Improving recommendation lists through topic diversification.
\newblock In \emph{Proceedings of the 14th international conference on World
  Wide Web}, pp.\  22--32, 2005.

\end{thebibliography}
\bibliographystyle{icml2026}

\newpage
\appendix
\onecolumn

\section{Illustrative differences in selection between DiverseBJR and Engagement}
\label{appendix:viz-diversebjr-vs-engagement}

\textbf{Differences in the selected opinions' content:}

\begin{figure}[H]%
\setlength{\abovecaptionskip}{1pt}
\centerline{\includeinkscape[width=\columnwidth]{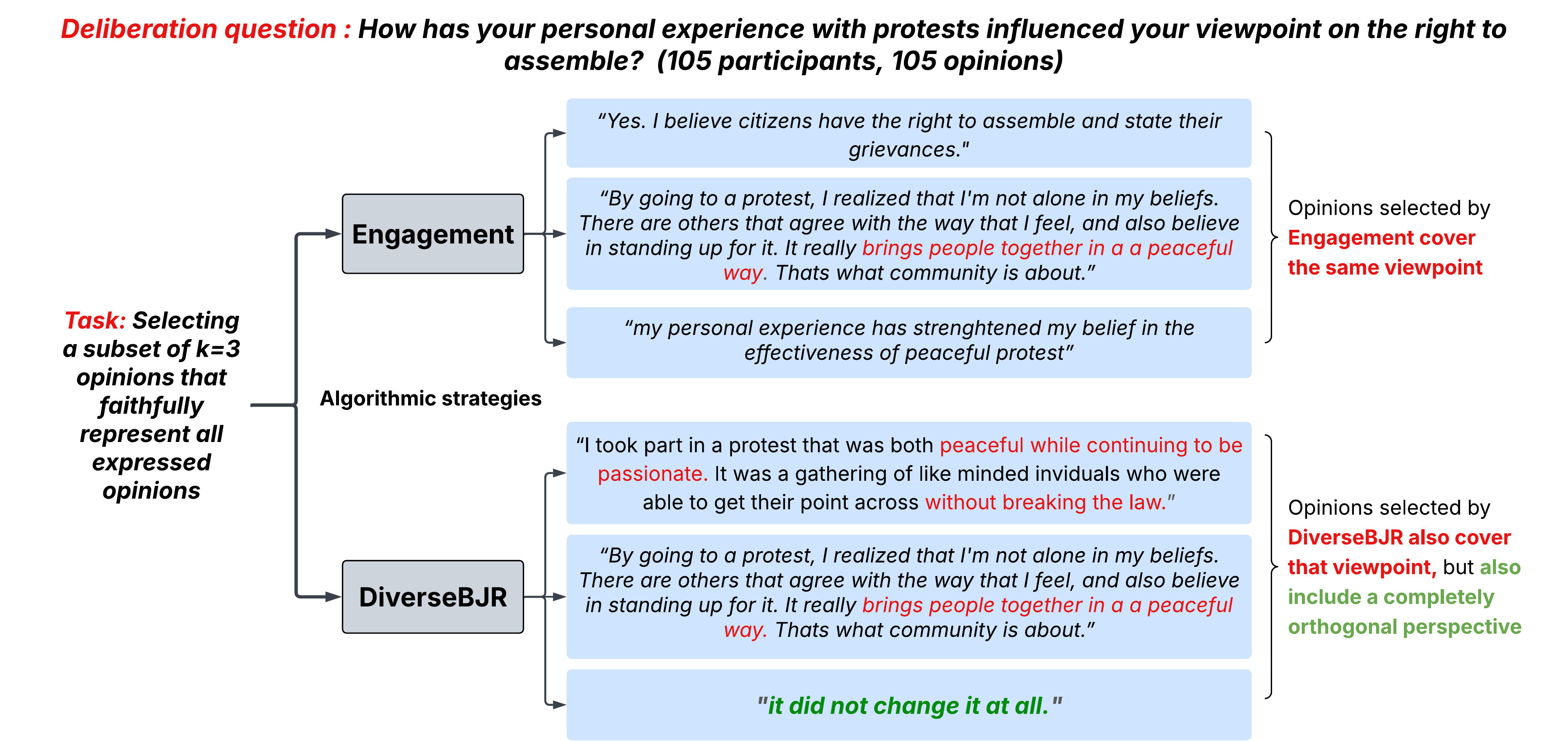}}
\label{fig:diversity-viz-2}
\end{figure}

\textbf{Differences in the selected opinions' distribution:}

\begin{figure*}[htbp]%
\setlength{\abovecaptionskip}{1pt}
\centerline{\includeinkscape[width=\textwidth]{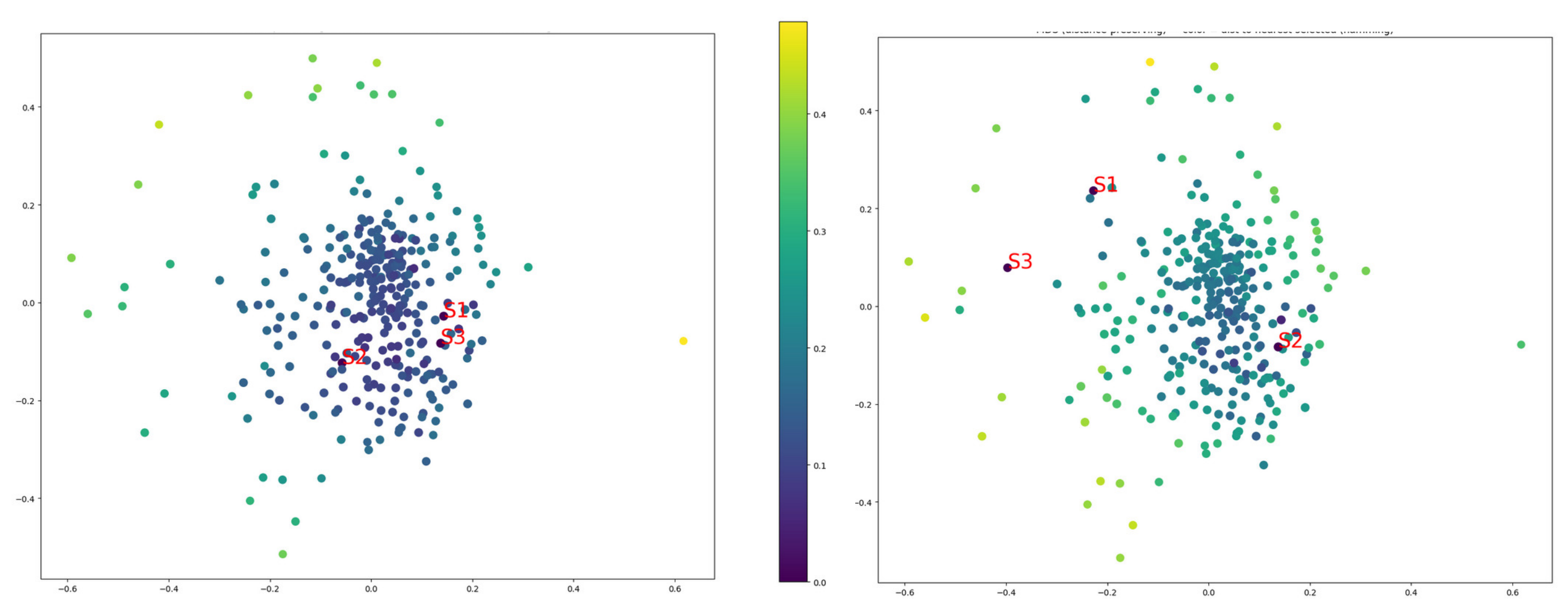}}
\caption{Visualization of the opinion space - Comparison of \textbf{selecting 3 representative opinions} based on \textbf{Engagement (left)} versus \textbf{DiverseBJR (right)}. Embeddings are based on MDS (distance-preserving) - Color = distance to nearest selected opinion (normalized Hamming). DiverseBJR selects more distant (i.e., diverse) thoughts, whereas Engagement selects based on close clusters, therefore potentially selecting redundant opinions.
} 
\label{fig:diversity-viz}
\end{figure*}

\section{Relationship between JR, BJR and DiverseBJR}
\label{appendix:jr-vs-bjr-vs-diversebj}
Previous work has shown that BJR is different from JR and does not imply it \cite{fish2024generative}. To illustrate the limitations of JR as well as the need for a balanced-matching-based notion of justified representation (BJR), we cite the following standard example previously used by multiple works \cite{fish2024generative, aziz2017justified}:

\[
\begin{array}{ccccc}
 & \alpha & \alpha' & \beta & \beta' \\
\hline
u_0 & 1 & 1 & 0 & 0\\
u_1 & 1 & 1 & 0 & 0\\
u_2 & 0 & 0 & 1 & 1\\
\hline
\end{array}
\]
\noindent
In this example, as was previously observed (e.g., \cite{fish2024generative} Appendix A), JR is satisfied by the slate $\{\alpha, \beta, \beta'\}$. However, the slate is unproportional as it \textit{"represents two-thirds of the population by one-third of the slate, and vice versa"} \cite{fish2024generative}. In simpler terms, the selection of very few highly popular items is enough to satisfy JR and can in some scenarios give a false sense of representation in the selected set. For instance, in a task similar to ours, \cite{revel2025representative} find, when selecting $k=8$ opinions, that the size of JR-sets is typically $\approx$1-2, thus giving no representation guarantees for $\approx75\%$ of the final slate. To solve this issue, BJR introduces the notion of budgets in selection, whereby the users represented by the selected items must be balanced across items, thus avoiding a scenario where the inclusion of a single or a few very popular items suffices to satisfy desired representation guarantees. 

\noindent
Next, we illustrate the benefits of enforcing diversity on top of BJR. DiverseBJR enables the selection of more diverse opinions without breaking BJR guarantees under the greedy policy. It does so through two distinct mechanisms: tie-breaking by diversity in Stage 1, and flagging $\varepsilon$-neighbors as ineligible for future picks in Stage 2 iff flagging does not jeopardize future BJR budgets (see Algorithm \ref{alg:diversebjr-simple}). To illustrate how the diversity-based tie-breaking mechanism works, let us consider the following approval matrix with n=3 voters and m=3 opinions:

\[
\begin{array}{ccccc}
 & m_0 & m_1 & m_2 \\
\hline
u_0 & 1 & 0 & 0 \\
u_1 & 1 & 1 & 0 \\
u_2 & 0 & 1 & 1 \\
\hline
\end{array}
\]
\noindent
 The task is to select a subset of $k=2$ opinions. Let $\varepsilon=0.7$ be the $\varepsilon$-neighbors threshold (in our example, this means that any two opinions which share at least one common voter are seen as $\varepsilon$-neighbors).\footnote{Specifically, computing distances gives (normalized Hamming) $d(m_0,m_1)=2/3\approx0.67$, $d(m_1,m_2)=1/3\approx0.33$, $d(m_0,m_2)=1.0$ (Jaccard produces the same neighbor graph here with distances $0.67$, $0.5$,$1.0$), so with $\varepsilon=0.7$ the $\varepsilon$-neighbors are $m_0-m_1$ and $m_1-m_2$ ($m_0$ and $m_2$ are not neighbors).} Let us work through the greedy selection algorithm for all of JR, BJR and DiverseBJR.

\begin{itemize}
    \item \textbf{JR:} 
    \begin{itemize}
        \item Stage 1: $m_0$ and $m_1$ tie on coverage, as both cover 2 unsatisfied voters (\{$u_0$,$u_1$\} and \{$u_1$,$u_2$\}, respectively). JR may arbitrarily select $m_0$ or $m_1$, after which JR will be satisfied: there is at least one user from every cohesive group of at least $n/k$ users who approves at least one selected item. 
        \item Stage 2: selects either $m_1$ or $m_2$ depending on the scoring rule $f$ given as input. This stage has no impact on the JR-guarantee which was already satisfied in Stage 1. 
        \item Final slate: $\{m_0,m_2\}$ or $\{m_0,m_1\}$ or $\{m_1,m_2\}$. We assume $\{m_0,m_1\}$ as the final slate for the remainder of this example.
    \end{itemize}
\end{itemize}

\noindent
For both BJR and DiverseBJR, BJR-budgets are set as [2, 1] (see Algorithm \ref{alg:diversebjr-simple}), meaning that the procedure will contain two rounds, where rounds 1 and 2 will aim at selecting opinions that satisfy 2 users and 1 user (respectively).   

\begin{itemize}
    \item \textbf{Standard BJR:} 
    \begin{itemize}
        \item Round 1: $m_0$ and $m_1$ tie on coverage, as both cover 2 unsatisfied voters (\{$u_0$,$u_1$\} and \{$u_1$,$u_2$\}, respectively). BJR may arbitrarily select $m_0$ or $m_1$, leaving only one user left unsatisfied.
        \item Round 2: if Round 1 selected $m_0$, Round 2 can select either $m_1$ or $m_2$ depending on the chosen tie-break rule (both cover the remaining unsatisfied voter $u_2$); if Round 1 selected $m_1$, Round 2 must select $m_0$ (to cover $u_0$). No user remains unsatisfied after Round 2.
        \item Final slate: $\{m_0,m_2\}$ or $\{m_0,m_1\}$. We assume $\{m_0,m_1\}$ as the final slate for the remainder of this example.
    \end{itemize}
    \item \textbf{DiverseBJR:}
    \begin{itemize}
        \item Round 1: selects $m_0$ after ($m_0$,$m_1$) tie-break ($m_0$ only has 1 $\varepsilon$-neighbor, whereas $m_1$ has 2), leaving only $u_2$ unsatisfied.
        \item Round 2: selects $m_2$ after ($m_1$,$m_2$) tie-break which is solved based on diversity ($m_2$ only has one neighbor, whereas $m_1$ has 2), leaving no user left unsatisfied.
        \item Final slate: $\{m_0,m_2\}$
    \end{itemize}
\end{itemize}
\noindent
Out of all three, DiverseBJR is the only one that guarantees the selection of the slate $\{m_0,m_2\}$, which covers the same voters but is \textbf{more diverse} than the JR/BJR slate candidates $\{m_0,m_1\}$ and $\{m_1,m_2\}$, as it contains less redundancy in voters among the selected opinions. While seemingly trivial, this distinction can have very concrete impacts when considering our task of online deliberation: by selecting opinions which differ the most from already selected opinions (subject to BJR constraints), the final subset of $k$ opinions typically covers a larger fraction of the opinion-space, and avoids including redundant opinions. These intuitions were validated in our experiments where DiverseBJR displayed less redundancy and better coverage than BJR (see for instance Figure \ref{fig:q1-all}: center-left and bottom)). 

\section{Detailed algorithms}

\subsection{DiverseBJR approximation algorithm}

\begin{algorithm}[htbp]
\caption{\textsc{DiverseBJR}. The algorithm follows a GreedyCC-style two stage procedure, similar to those used to operationalize the JR \cite{revel2025representative} and BJR \cite{fish2024generative} axioms. First, BJR-budgets are computed to ensure that each selected opinion is approved by  $\lfloor{n/k}\rfloor$ or $\lceil{n/k}\rceil$ users. In the main loop, Stage 1 greedily selects opinions to satisfy the current BJR budget by maximizing coverage, using the diversity criterion only for tie-breaking. When no remaining opinion can satisfy the current budget, the algorithm proceeds to Stage 2, where it fills any remaining slots with a given scoring rule $f$. At each round, the algorithm attempts to promote diversity by marking near-duplicate opinions as ineligible, but only when a separate greedy feasibility test certifies that the remaining BJR budgets can still be satisfied under the greedy policy. In this way, the algorithm preserves balanced justified representation under the greedy policy while encouraging diversity in the final selected set of opinions.}

\label{alg:diversebjr-simple}
\begin{algorithmic}[1]
\Require Approval matrix $A\in\{0,1\}^{n\times m}$, set size $k$, scoring rule $f$, diversity threshold $\varepsilon$
\Ensure A set $W$ of $k$ opinions
\State Define $[n]\!=\!\{1,\dots,n\}$ (the set of voters) and $[m]\!=\!\{1,\dots,m\}$ (the set of opinions)
\State Compute $\varepsilon$-neighbors for each opinion (distances are computed based on A)
\State Initialize $W\gets\emptyset$, $V\gets [n]$, \textit{ineligible-$\varepsilon$-neighbors}$\gets\emptyset$
\State Compute BJR-budgets \(r_1,\dots,r_k\) by distributing the remainder: \(r_j=\lfloor n/k\rfloor+1\) for \(j\le n\bmod k\), otherwise \(r_j=\lfloor n/k\rfloor\)
\For{$t=1$ to $k$}
    \algrenewcommand\algorithmiccomment[1]{\hfill$\triangleright$ #1}
    \State $S_t \gets \{\, i \in [m]\setminus W : |\{u\in V : A_{u,i}=1\}| \ge r_t \,\}$ (unselected opinions approved by at least $r_t$ voters in $V$)
    \State $E_t \gets ([m]\setminus W)\setminus \textit{ineligible-}\varepsilon\textit{-neighbors}$ (eligible unselected opinions - not currently excluded by $\varepsilon$-neighbor rule)
    \If{$S_t\neq\emptyset$}
        \State Pick $p\in S_t$ with maximum approval, break ties by (i) fewer $\varepsilon$-neighbors, then (ii) more "unique" approvers \\
    \hspace{1.5\parindent}(voters who approve $p$ and no $\varepsilon$-neighbor) \textbf{(Stage 1)}
    \Else
        \If{$E_t \neq \emptyset$}
            \State Pick $p \in \arg\max_{i\in E_t} f(i)$ (filling remaining slots with the scoring rule: \textbf{Stage 2})
        \Else
            \State Pick $p \in \arg\max_{i\in [m]\setminus W} f(i)$ (fallback: relax ineligibility)
        \EndIf
    \EndIf
    \State Add $p$ to $W$ and remove up to $r_t$ approving voters from $V$
    \State Add $\varepsilon$-neighbors of $p$ to \textit{ineligible-$\varepsilon$-neighbors} iff future BJR-budgets remain satisfiable (assessed via a separate\\
    \hspace{.5\parindent} greedy simulator for BJR feasibility, see Algorithm \ref{alg:greedy-bjr-simulator})
\EndFor
\State \Return $W$
\end{algorithmic}
\end{algorithm}

\subsection{BJR Feasibility Checker (Greedy simulator)}

\begin{algorithm}[H]
\caption{\textsc{BJR Feasibility Checker}. The feasibility checker is a greedy simulator: given the remaining voters $V$, candidate opinions $I$, and future BJR demands $R=(r_{t+1},\ldots,r_k)$, it attempts to construct an explicit assignment by simulating the remaining rounds (processing larger demands first). In each round it selects an opinion with at least $r$ remaining approvers using the same tie-breaking and voter-removal rules as the main algorithm, removes $r$ such voters, and continues. If all demands are met, the simulator returns \textsf{True} with a constructive certificate; otherwise it returns \textsf{False}, which is inconclusive.}

\label{alg:greedy-bjr-simulator}
\begin{algorithmic}
\Require Approval matrix $A\in\{0,1\}^{n\times m}$, unrepresented voter set $V\subseteq[n]$, candidate opinions $I\subseteq[m]$ (not in $W$ or hypothetically excluded), remaining BJR demands $R=(r_{t+1},\dots,r_k)$, number of trials $T\ge1$, \textsf{removal\_policy} and \textsf{tie\_break} (same as used by the main algorithm)
\Ensure \textsf{True} and an assignment map $\mathcal{A}$ if a feasible greedy schedule is found; otherwise \textsf{False}
\State Sort $R$ in nonincreasing order (process largest demands first)
\For{$trial \gets 1$ to $T$} 
    \State $V' \gets V$, $I' \gets I$, $\mathcal{A}\gets\varnothing$ \Comment{clear assignment map}
    \If{$trial>1$} 
    optionally randomize tie-breaking (to produce alternative greedy trajectories)
    \EndIf
    \State \textbf{flag\_fail} $\gets$ \textsf{False}
    \For{each $r$ in $R$}
        \State $C \gets \{\, i\in I' : |\{v\in V' : A_{v,i}=1\}| \ge r \,\}$ \Comment{candidates with $\ge r$ approvers in $V'$}
        \If{$C=\varnothing$}
            \State \textbf{flag\_fail} $\gets$ \textsf{True}; \textbf{break} \Comment{trial failed}
        \EndIf
        \State Choose $i^\star\in C$ according to \textsf{tie\_break} (use same heuristic as main algorithm)
        \State Select $S \subseteq \{v\in V':A_{v,i^\star}=1\}$ with $|S|=r$ according to \textsf{removal\_policy} (e.g. lowest residual degree)
        \State $\mathcal{A}[i^\star]\gets S$
        \State $V' \gets V' \setminus S$, \quad $I' \gets I' \setminus \{i^\star\}$
    \EndFor
    \If{\textbf{not} \textbf{flag\_fail}} \Comment{all demands filled in this trial}
        \State \Return \textsf{True}, $\mathcal{A}$ \Comment{certificate: greedy schedule exists}
    \EndIf
\EndFor
\State \Return \textsf{False} \Comment{inconclusive: greedy simulator failed to construct a schedule}
\end{algorithmic}
\end{algorithm}

\section{Results on additional questions}\label{appendix:full-results}
The data is split in two question-types: controversial and consensual (defined in Section \ref{sec:data}). Due to space constraints, results on only one controversial question (Q1) were shown in the main body of the paper. Results on all remaining questions are shown below. Note that when analyzing and discussing results in the main body of the paper, results on all questions (including ones presented here) were considered. 

\subsection{Controversial questions}
Results on the second controversial question in the data (Q2) are presented in Figure \ref{fig:q2-all} below. 

\begin{figure}[H]%
\centering
    \begin{subfigure}[c]{.49\textwidth}
      \centering
      \includeinkscape[height=5cm]{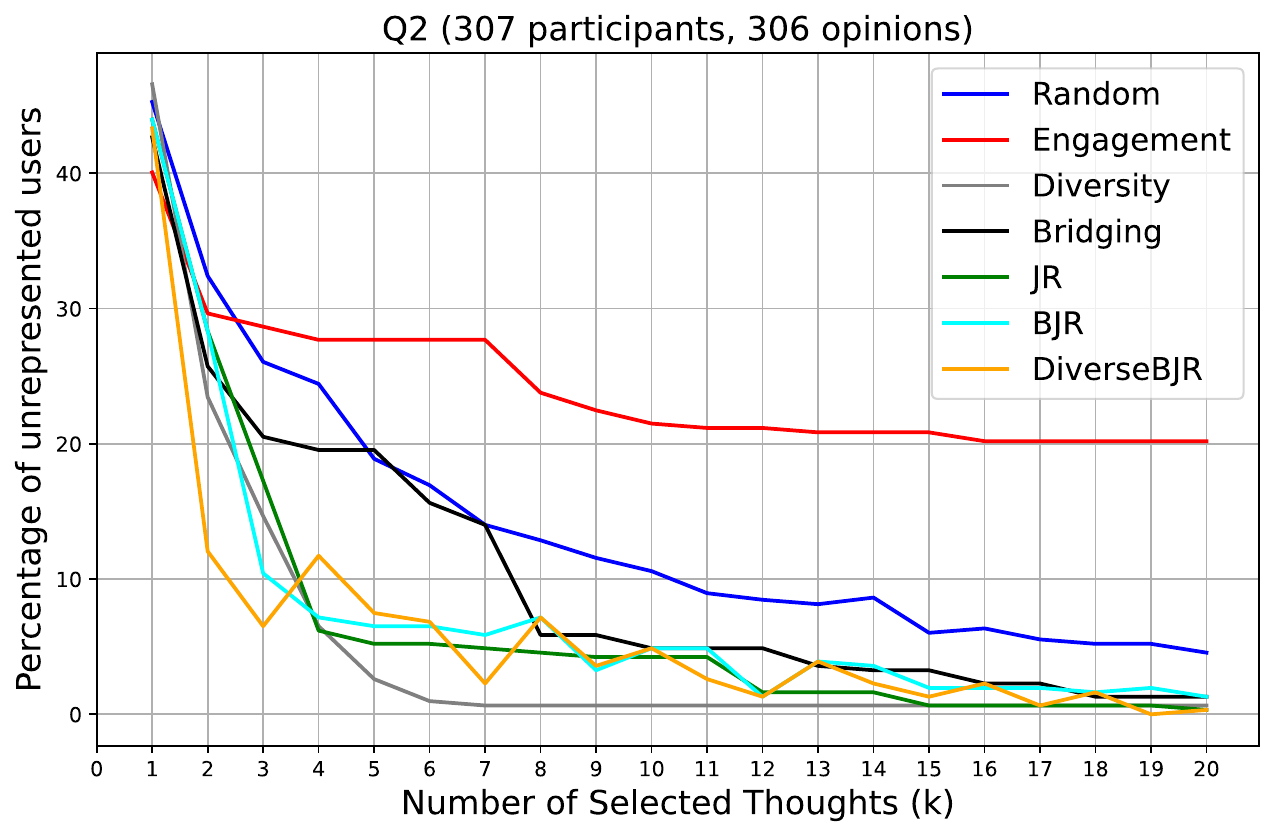}
    \end{subfigure}\hfill
  \begin{subfigure}[c]{.49\textwidth}
    \centering
    \includeinkscape[height=5cm]{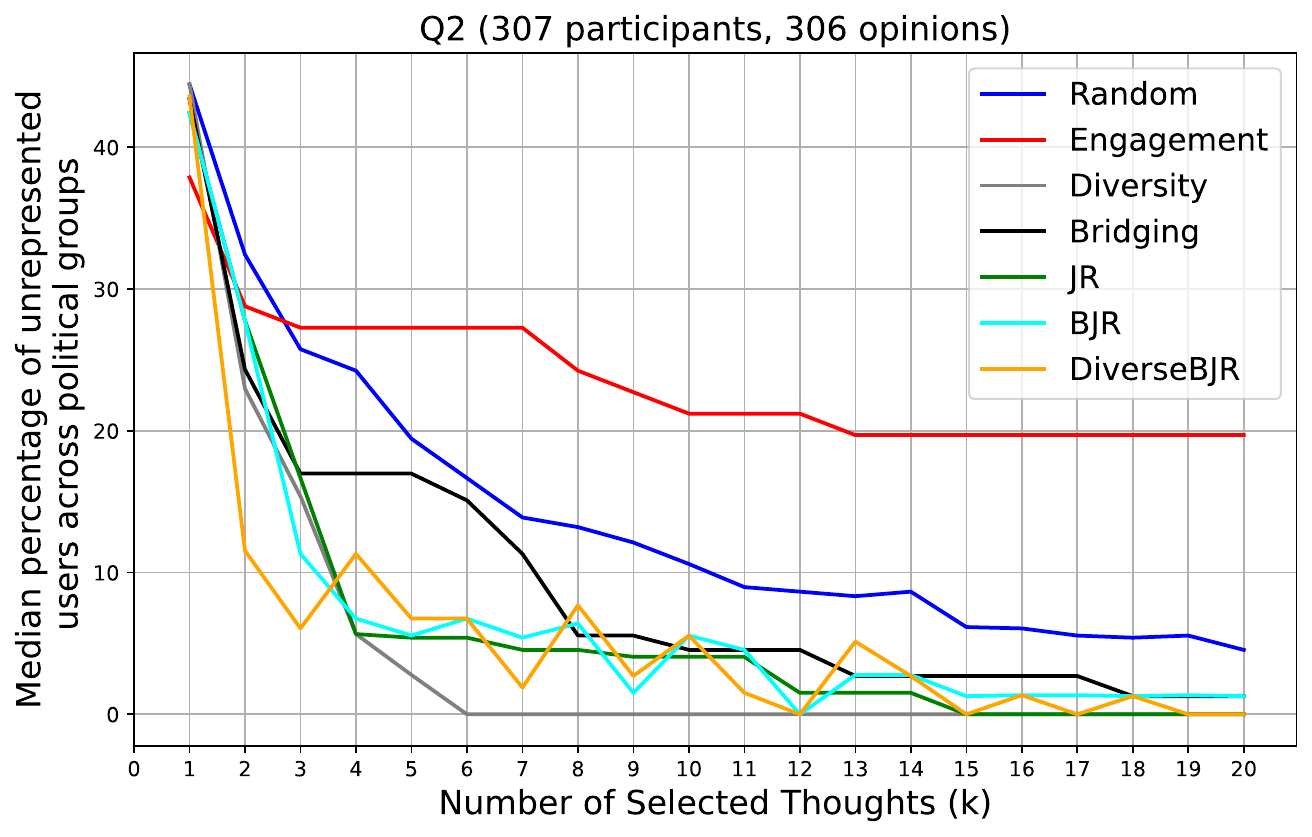}%
  \end{subfigure}\\
    \begin{subfigure}[c]{.49\textwidth}
      \centering
      \includeinkscape[height=5cm]{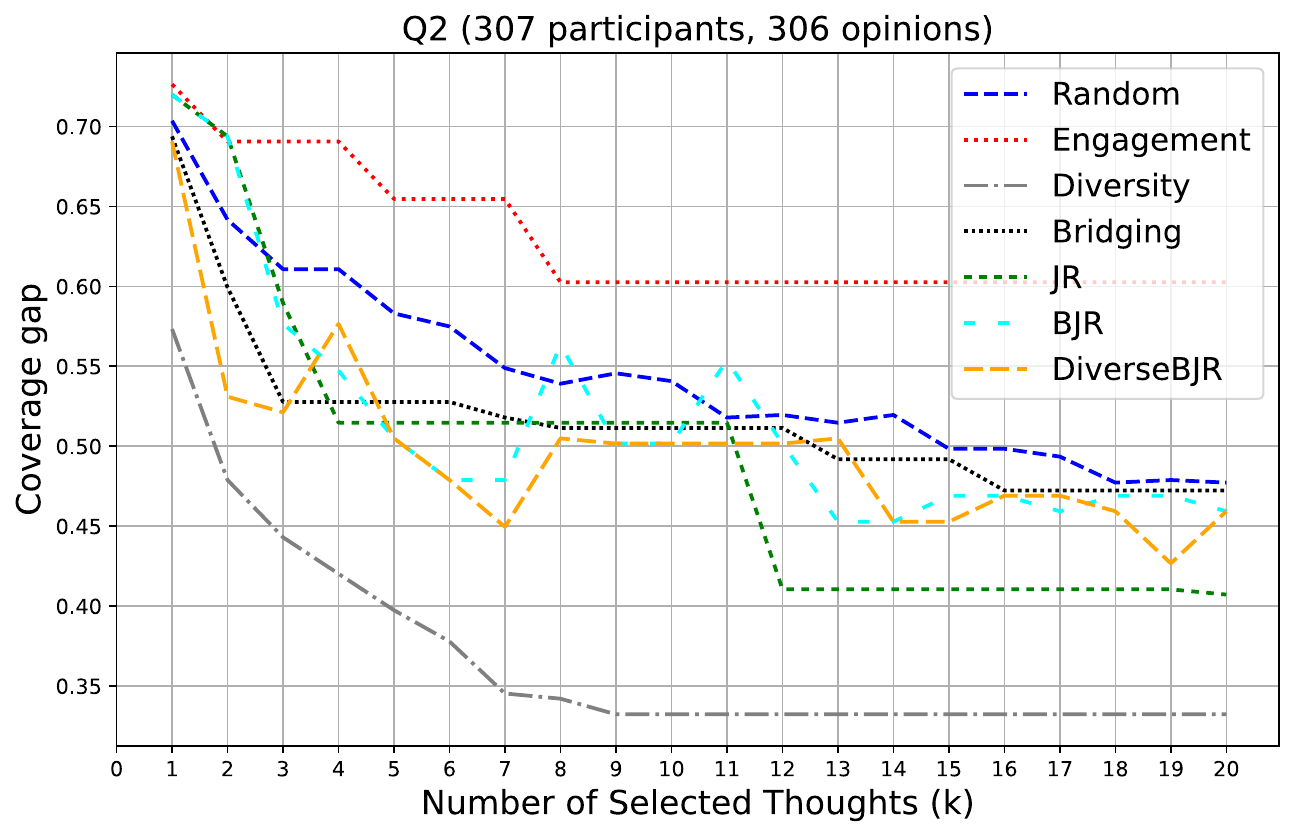}
    \end{subfigure}\hfill
  \begin{subfigure}[c]{.49\textwidth}
    \centering
    \includeinkscape[height=5.5cm]{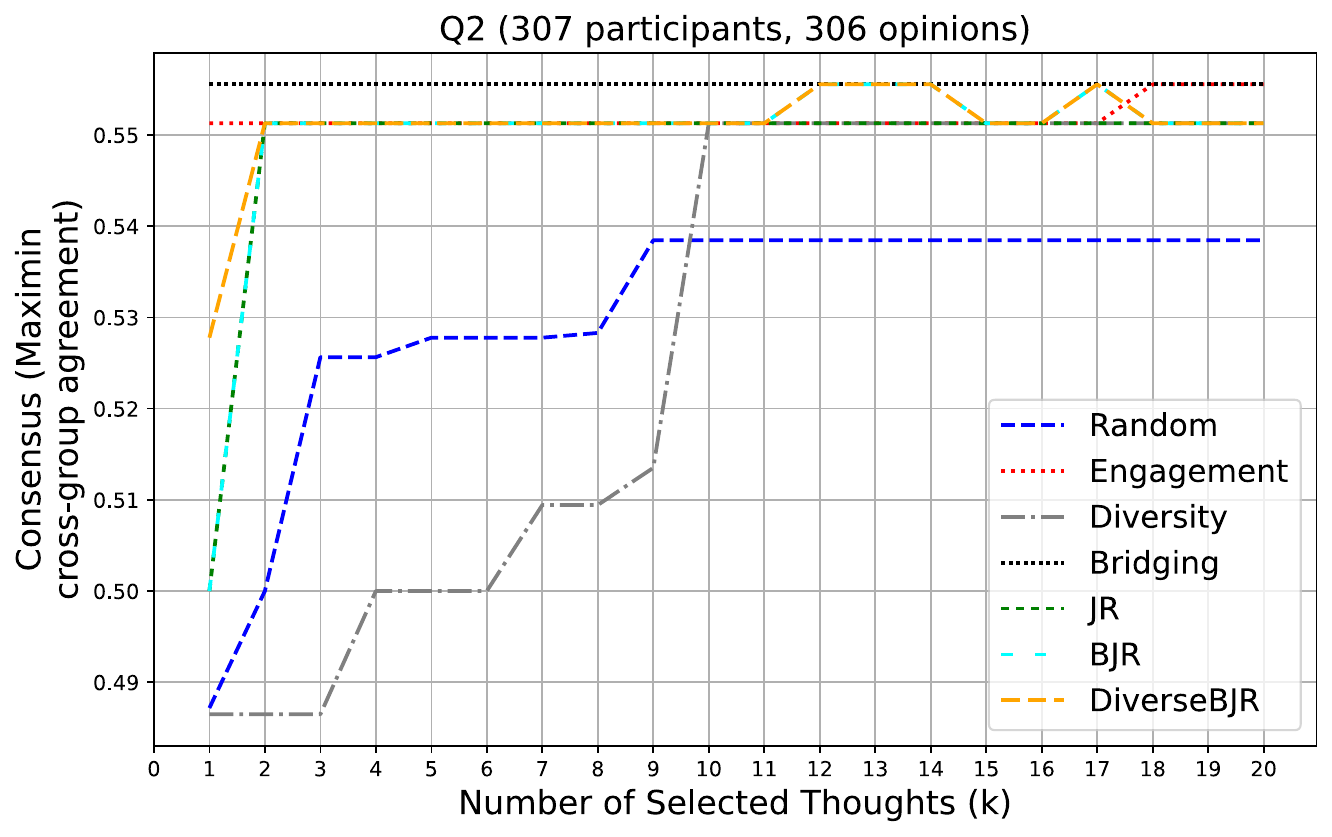}%
  \end{subfigure}
  \begin{subfigure}[c]{.49\textwidth}
    \centering
    \includeinkscape[height=5cm]{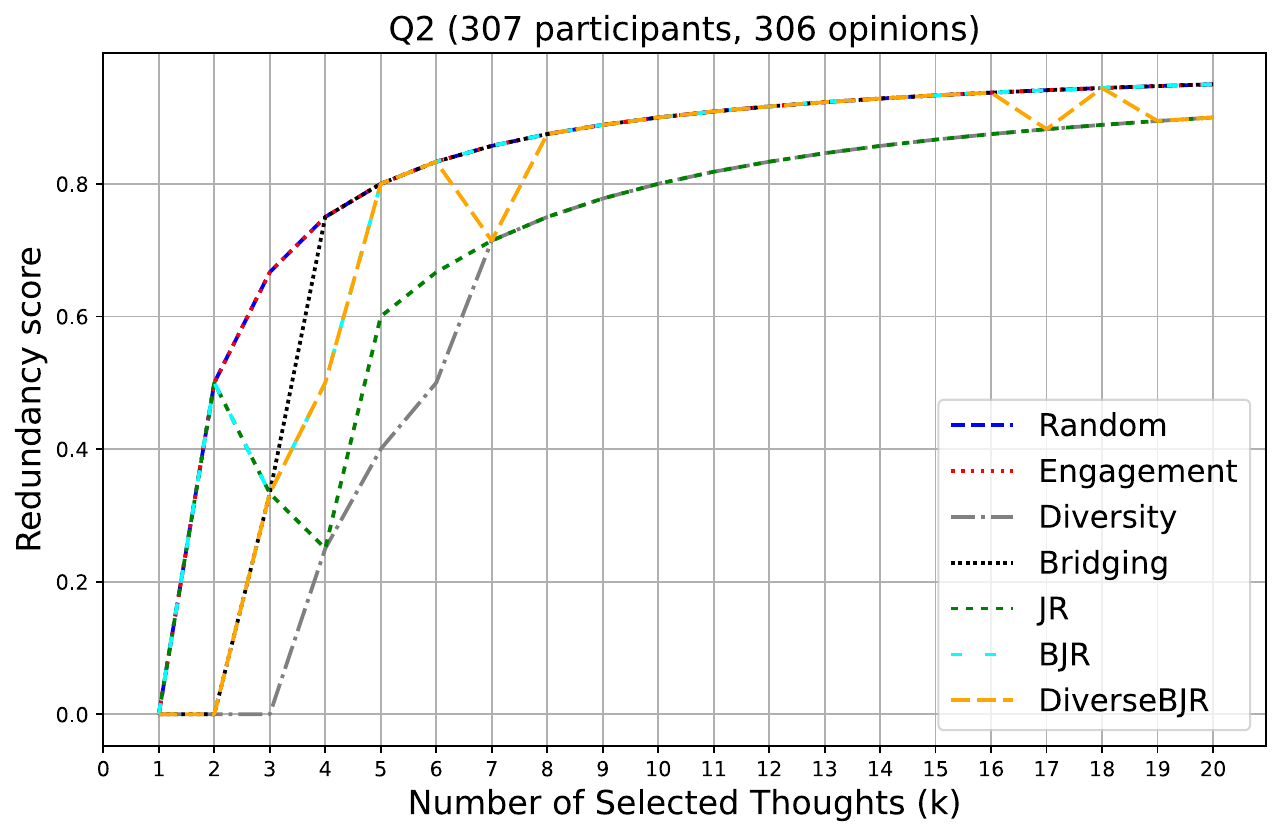}%
  \end{subfigure}\\
  \caption{Q2: "What are features or characteristics that make a protest appropriate?" (307 participants, 306 opinions)%
      \label{fig:q2-all}}
\end{figure}

\vspace{5em}

\subsection{Consensual questions}
Results on consensual questions in the data (Q3 to Q6) are presented in the figures below.  
\begin{figure}[H]%
\centering
    \begin{subfigure}[c]{.49\textwidth}
      \centering
      \includeinkscape[height=5cm]{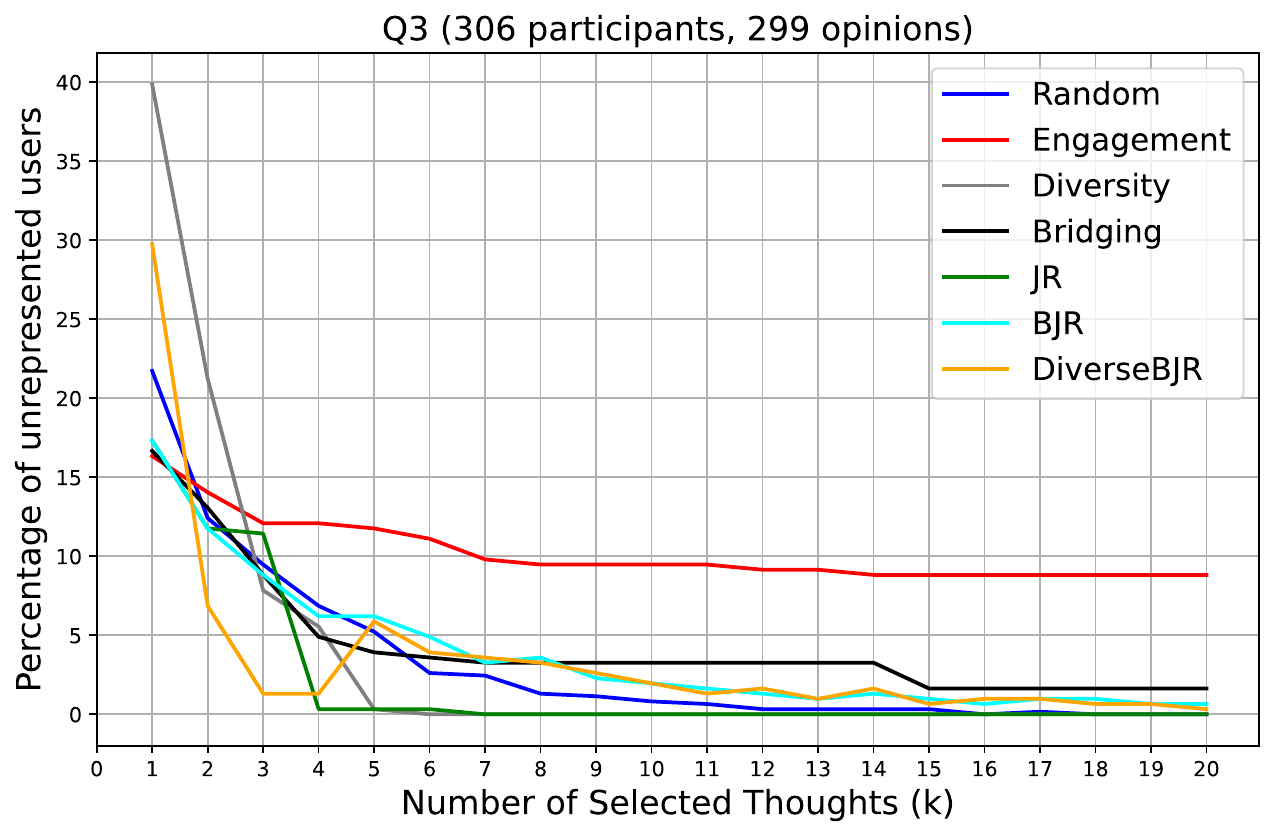}
    \end{subfigure}\hfill
  \begin{subfigure}[c]{.49\textwidth}
    \centering
    \includeinkscape[height=5cm]{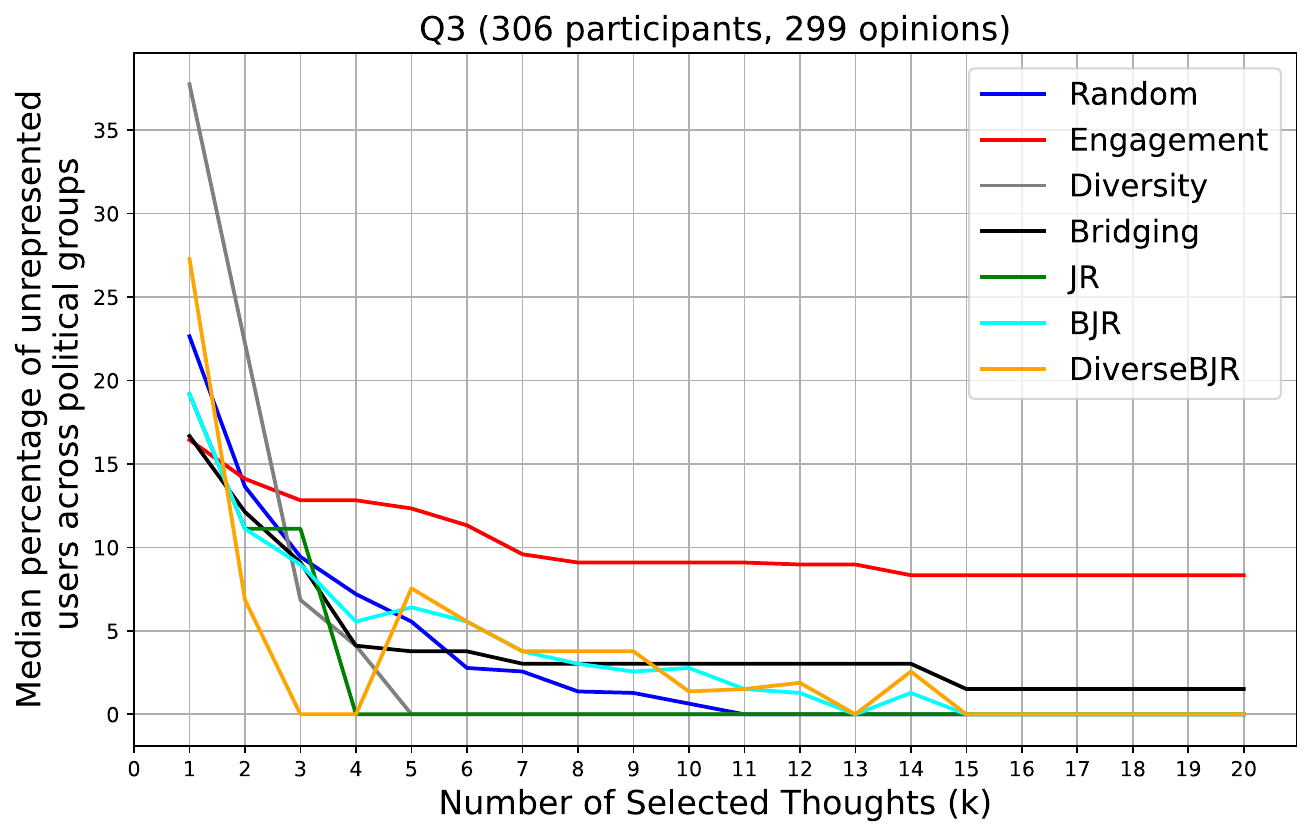}%
  \end{subfigure}\\
    \begin{subfigure}[c]{.49\textwidth}
      \centering
      \includeinkscape[height=5cm]{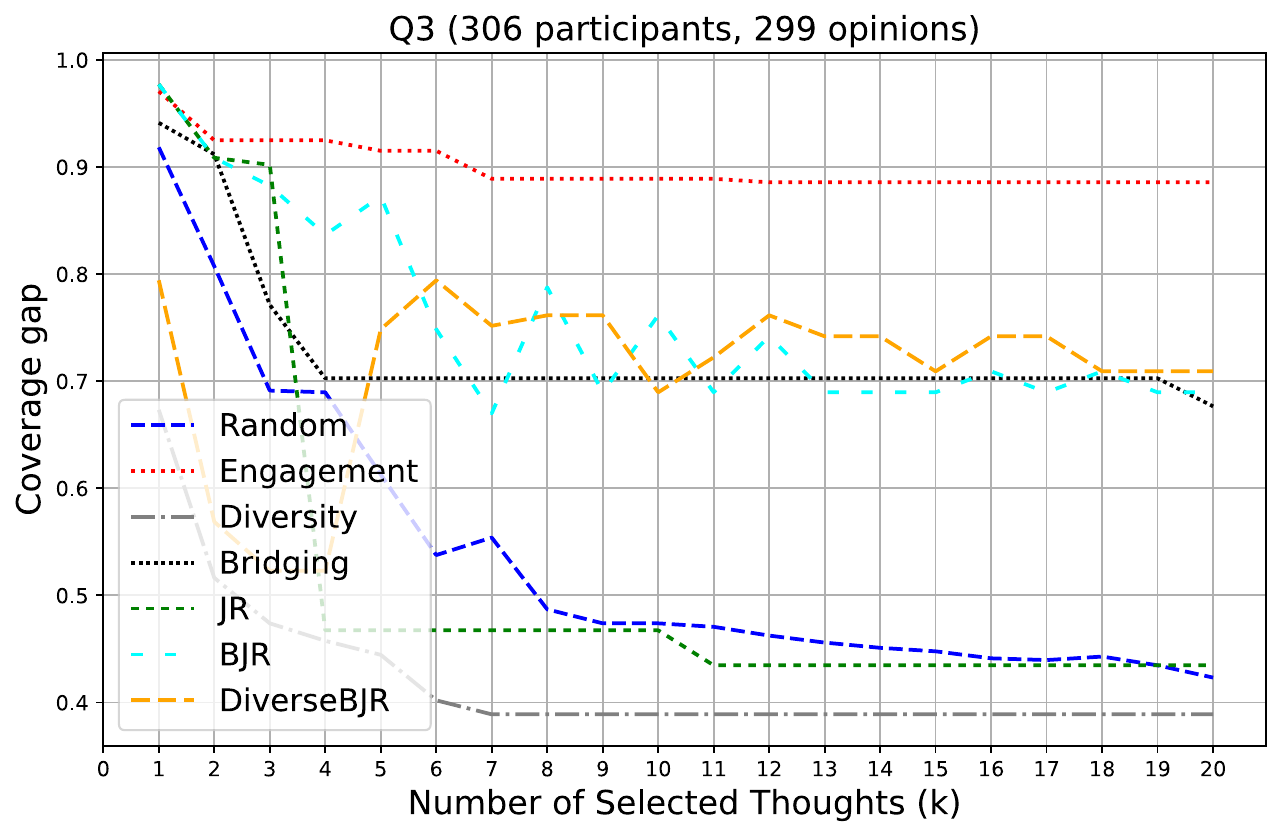}
    \end{subfigure}\hfill
  \begin{subfigure}[c]{.49\textwidth}
    \centering
    \includeinkscape[height=5.5cm]{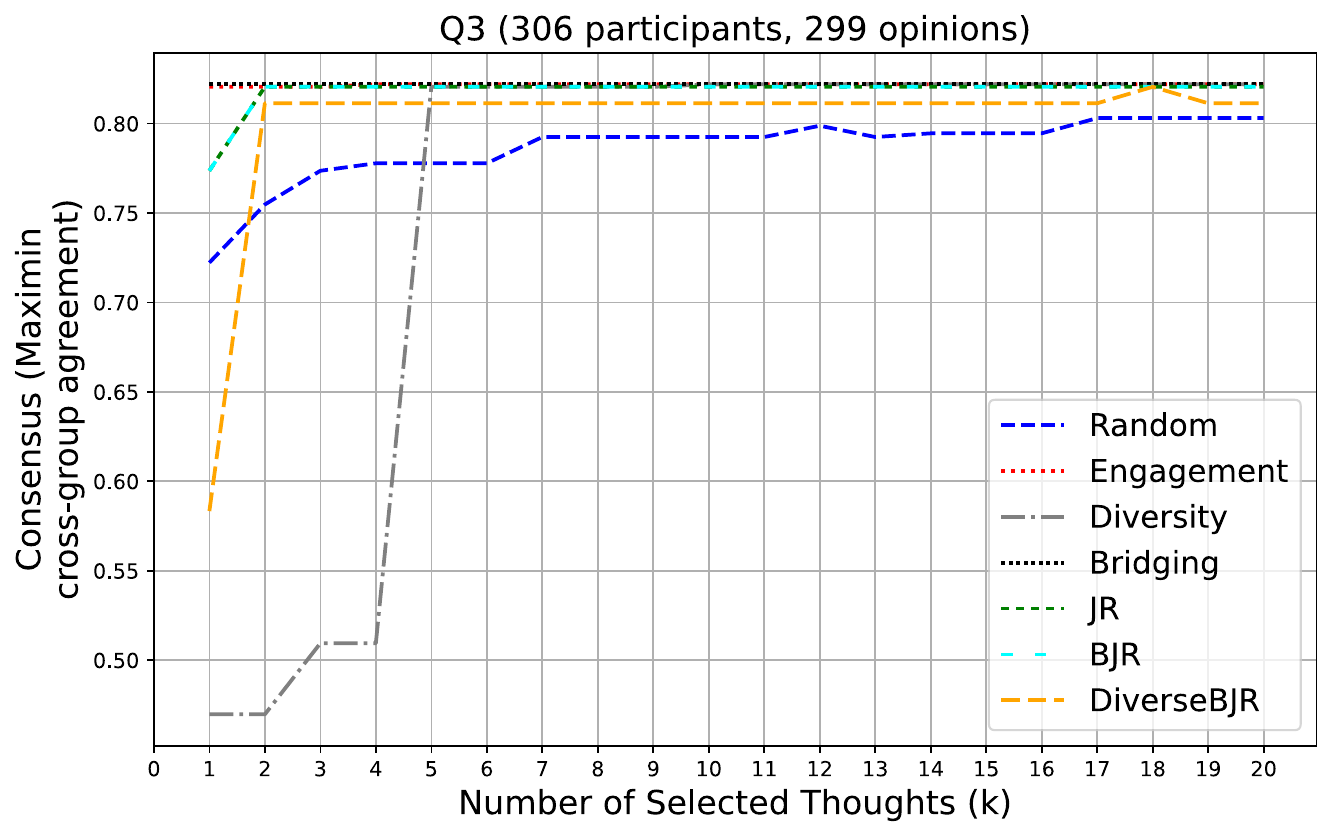}%
  \end{subfigure}
  \begin{subfigure}[c]{.49\textwidth}
    \centering
    \includeinkscape[height=5cm]{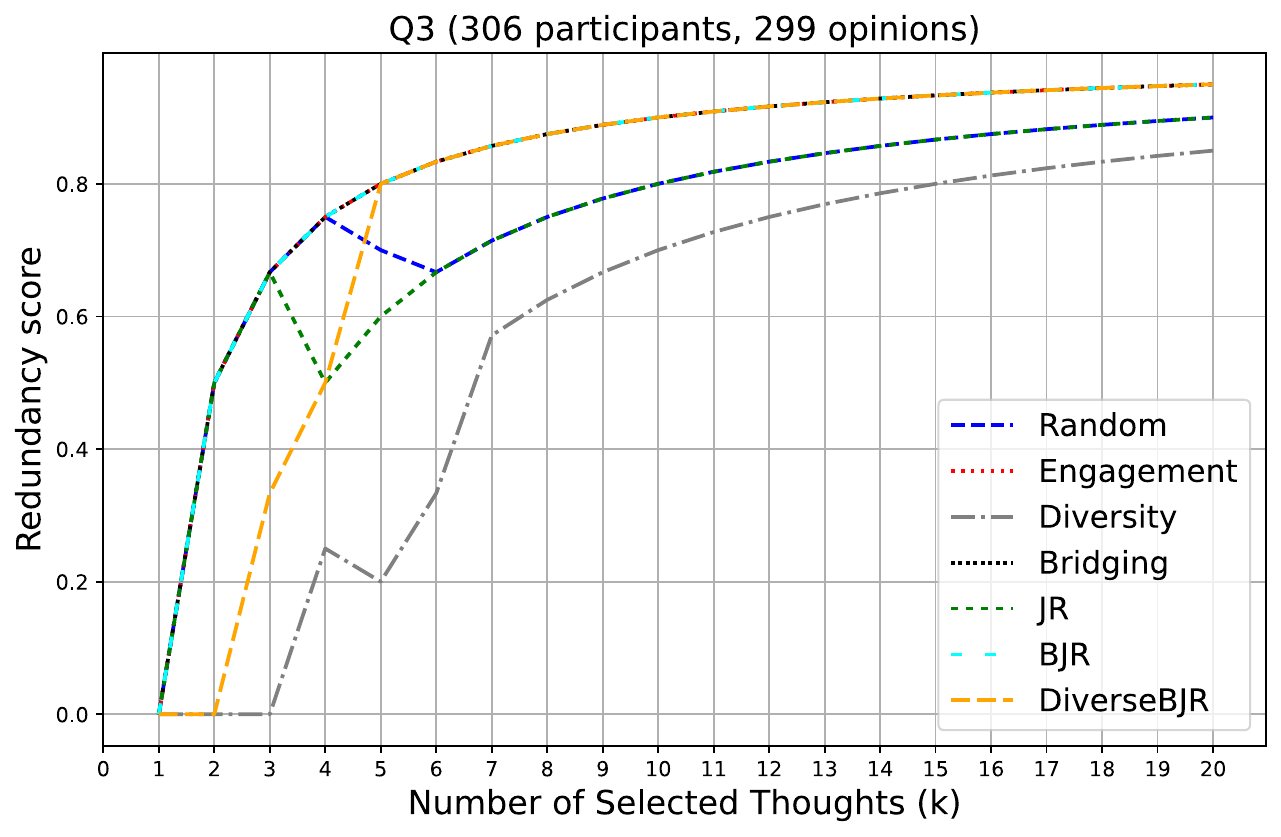}%
  \end{subfigure}\\
  \caption{Q3: "What characteristics or actions, in your view, deem a protest inappropriate?" (306 participants, 299 opinions)%
      \label{fig:q3-all}}
\end{figure}
 
\begin{figure}[H]%
\centering
    \begin{subfigure}[c]{.49\textwidth}
      \centering
      \includeinkscape[height=5cm]{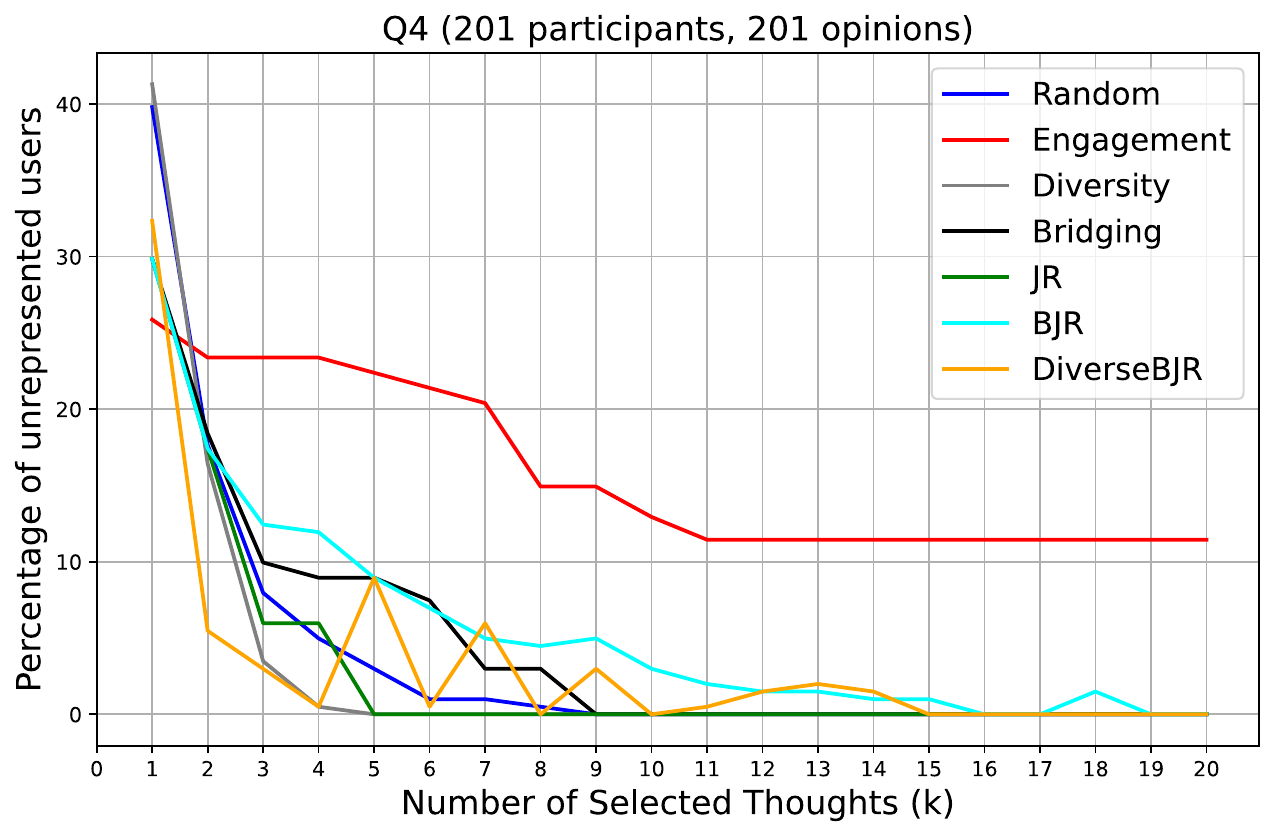}
    \end{subfigure}\hfill
  \begin{subfigure}[c]{.49\textwidth}
    \centering
    \includeinkscape[height=5cm]{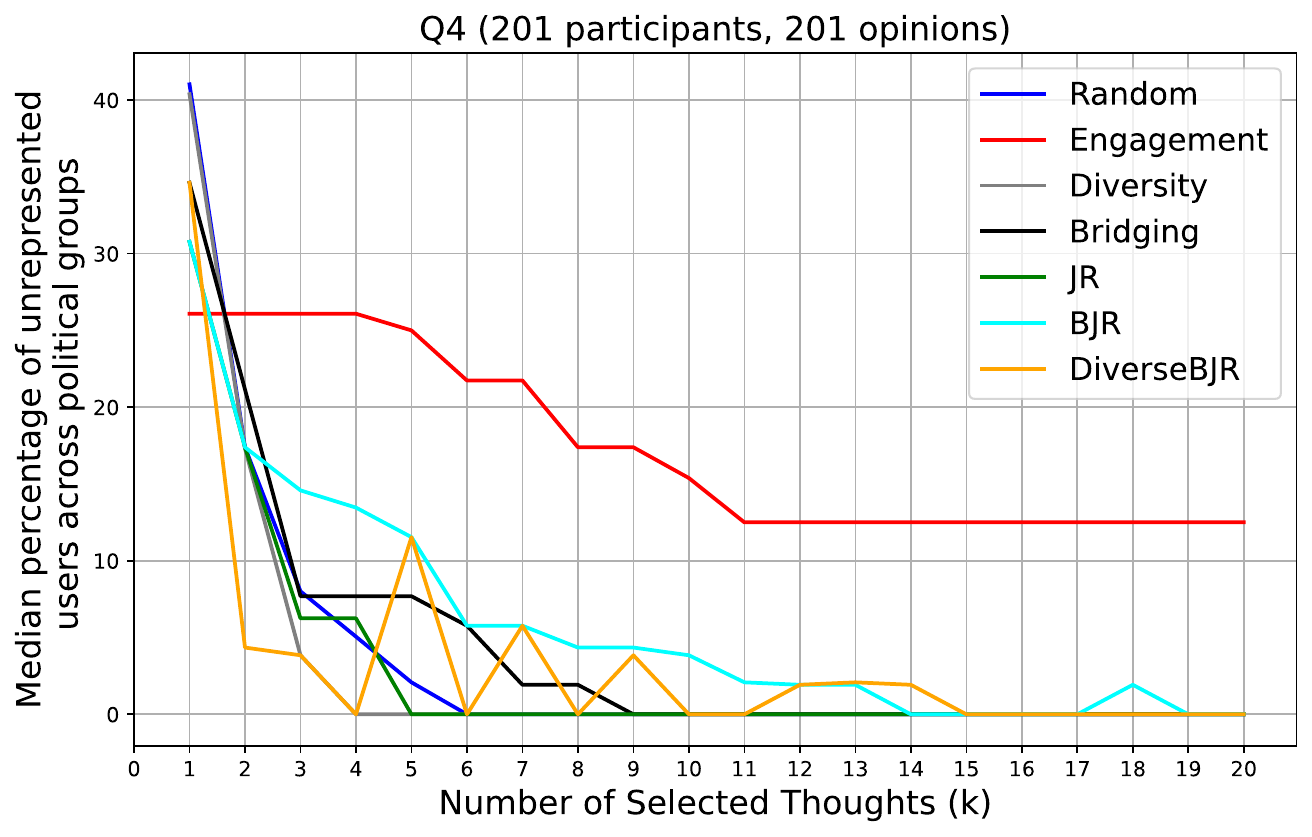}%
  \end{subfigure}\\
    \begin{subfigure}[c]{.49\textwidth}
      \centering
      \includeinkscape[height=5cm]{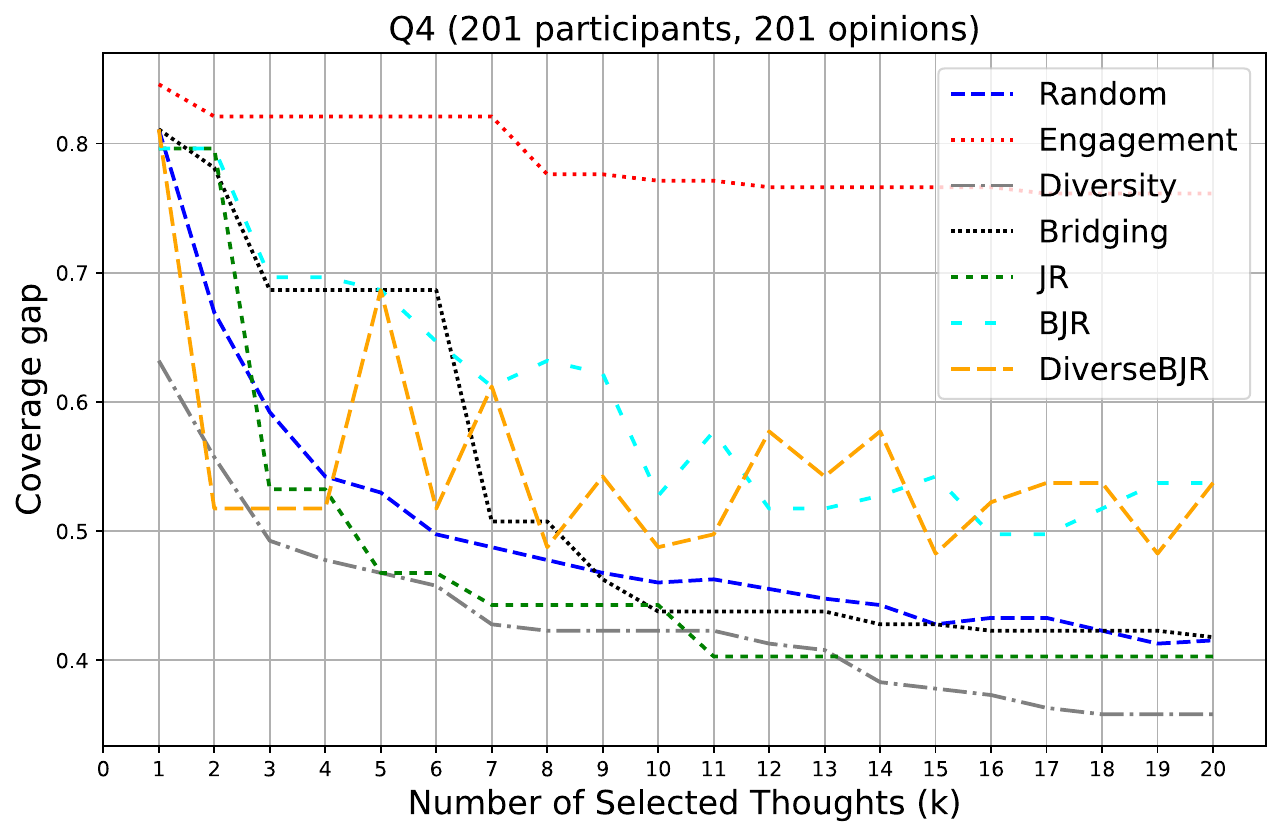}
    \end{subfigure}\hfill
  \begin{subfigure}[c]{.49\textwidth}
    \centering
    \includeinkscape[height=5.5cm]{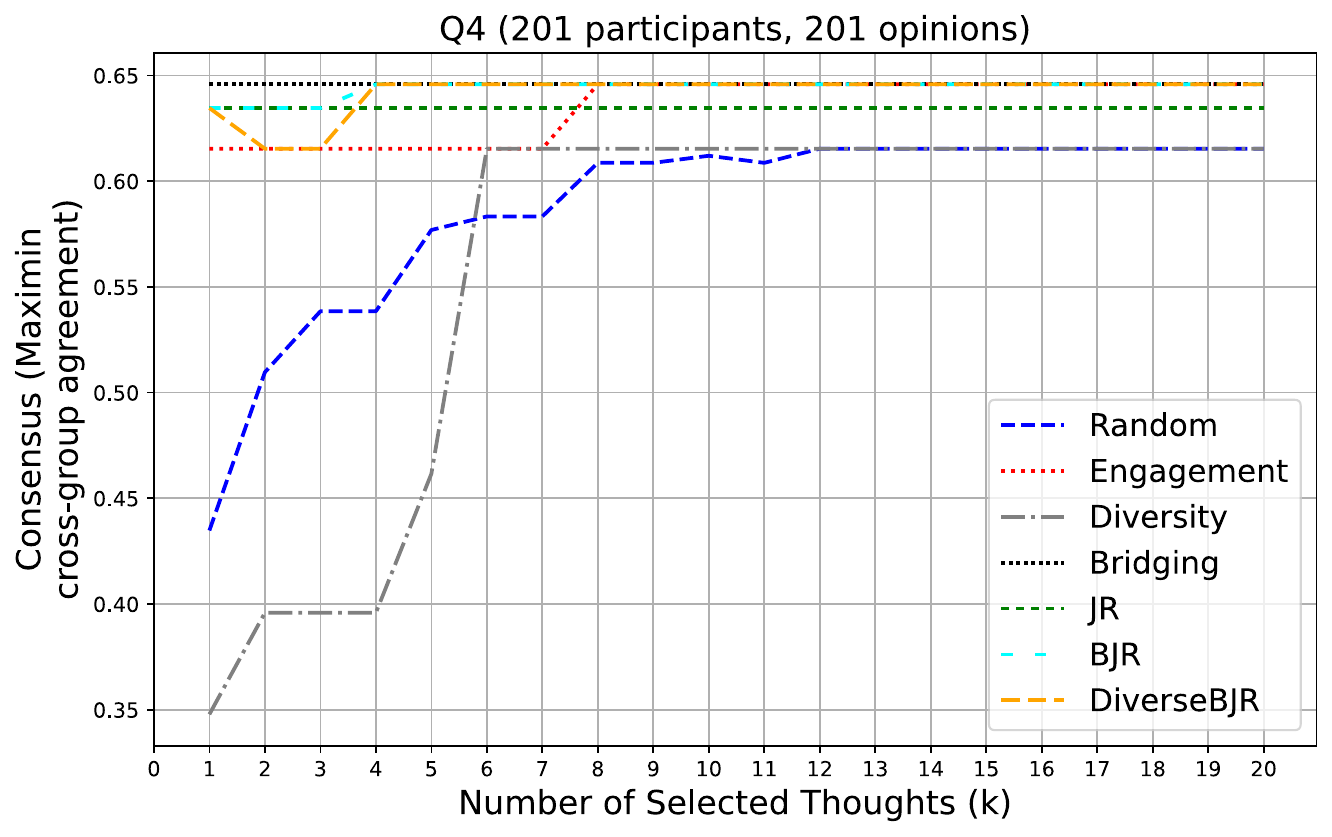}%
  \end{subfigure}
  \begin{subfigure}[c]{.49\textwidth}
    \centering
    \includeinkscape[height=5cm]{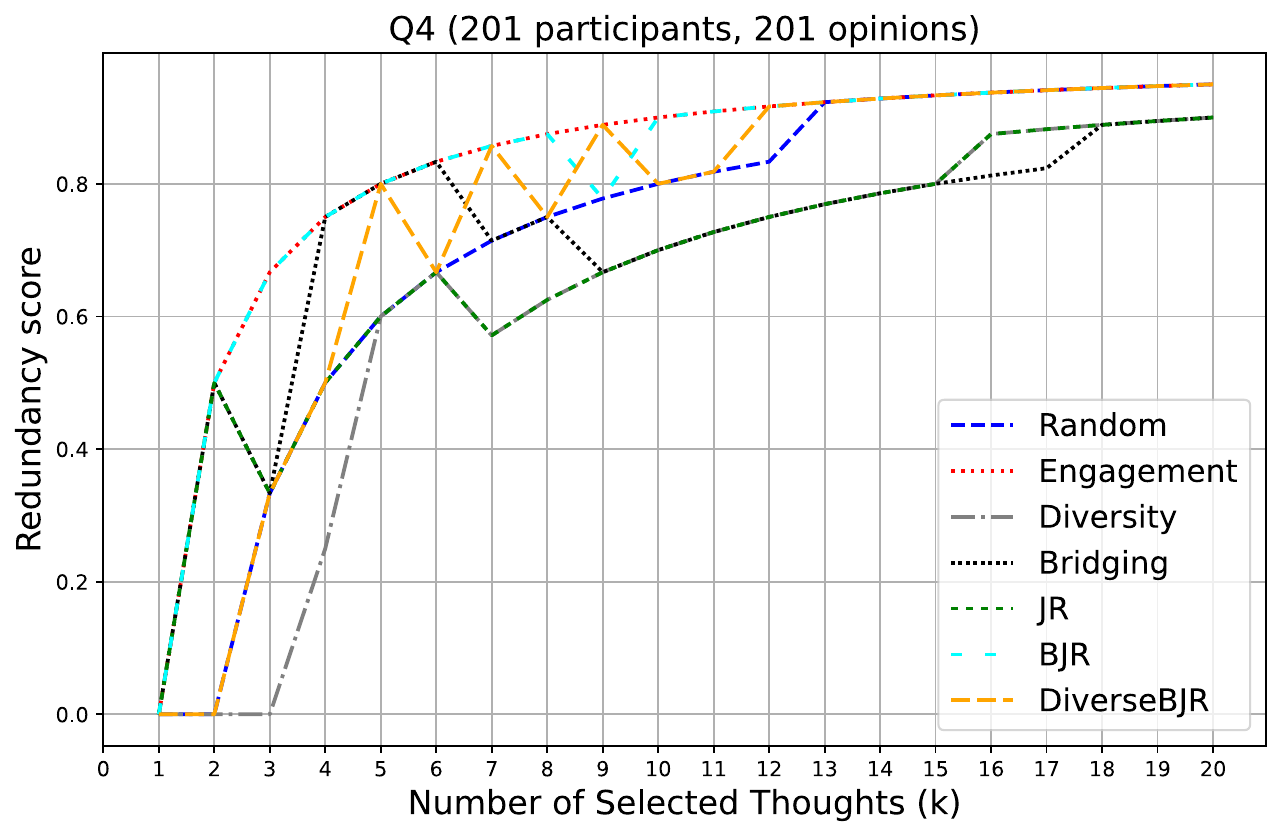}%
  \end{subfigure}\\
  \caption{Q4: "What are some of the reasons you have not participated in or attended a protest?" (201 participants, 201 opinions)%
      \label{fig:q4-all}}
\end{figure}

\begin{figure}[H]%
\centering
    \begin{subfigure}[c]{.49\textwidth}
      \centering
      \includeinkscape[height=5cm]{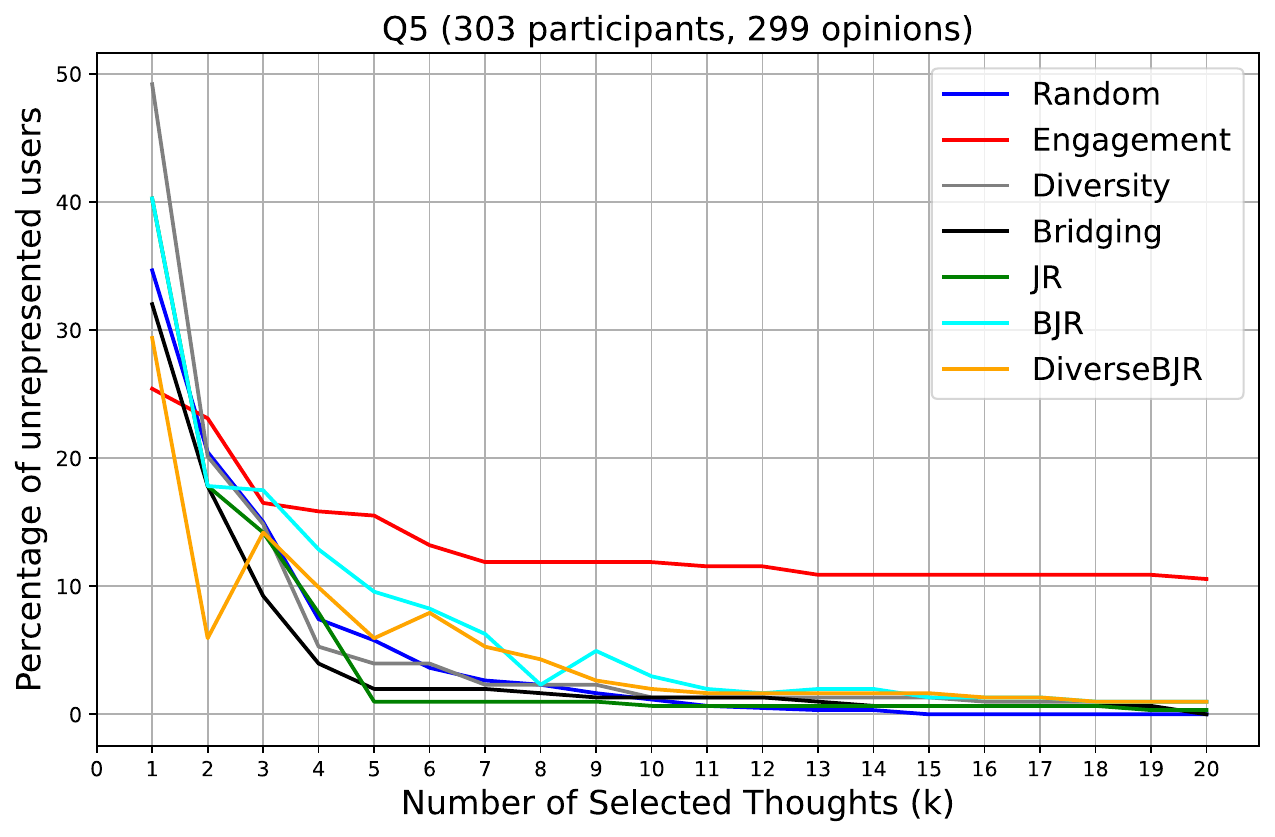}
    \end{subfigure}\hfill
  \begin{subfigure}[c]{.49\textwidth}
    \centering
    \includeinkscape[height=5cm]{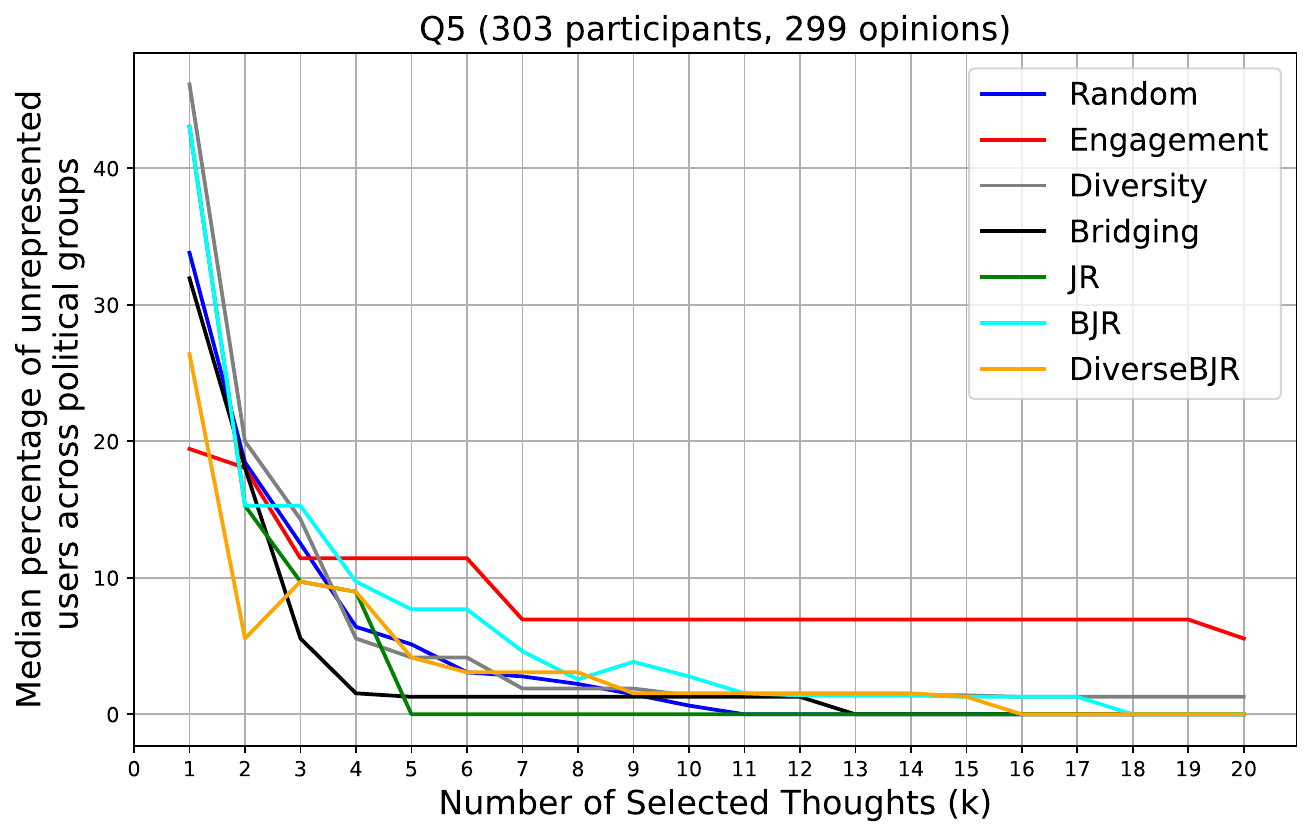}%
  \end{subfigure}\\
    \begin{subfigure}[c]{.49\textwidth}
      \centering
      \includeinkscape[height=5cm]{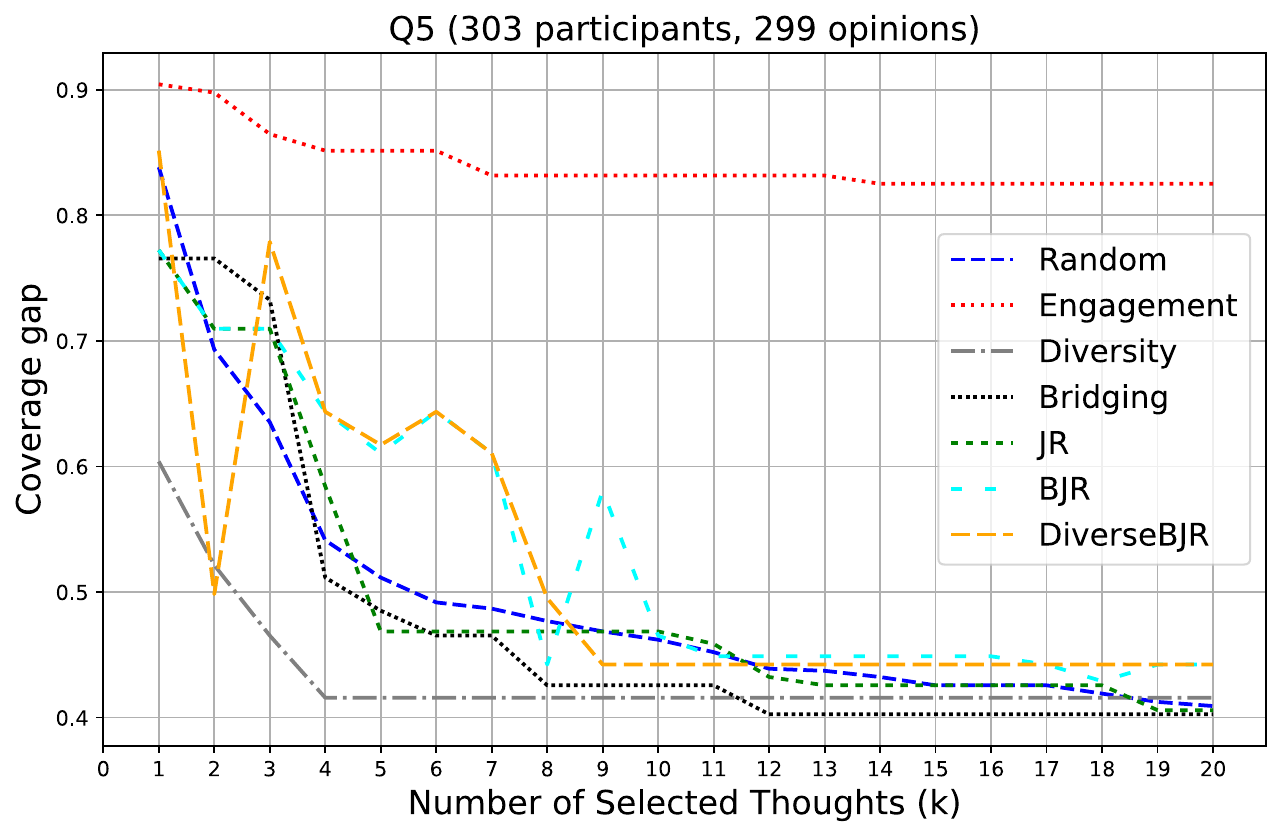}
    \end{subfigure}\hfill
  \begin{subfigure}[c]{.49\textwidth}
    \centering
    \includeinkscape[height=5.5cm]{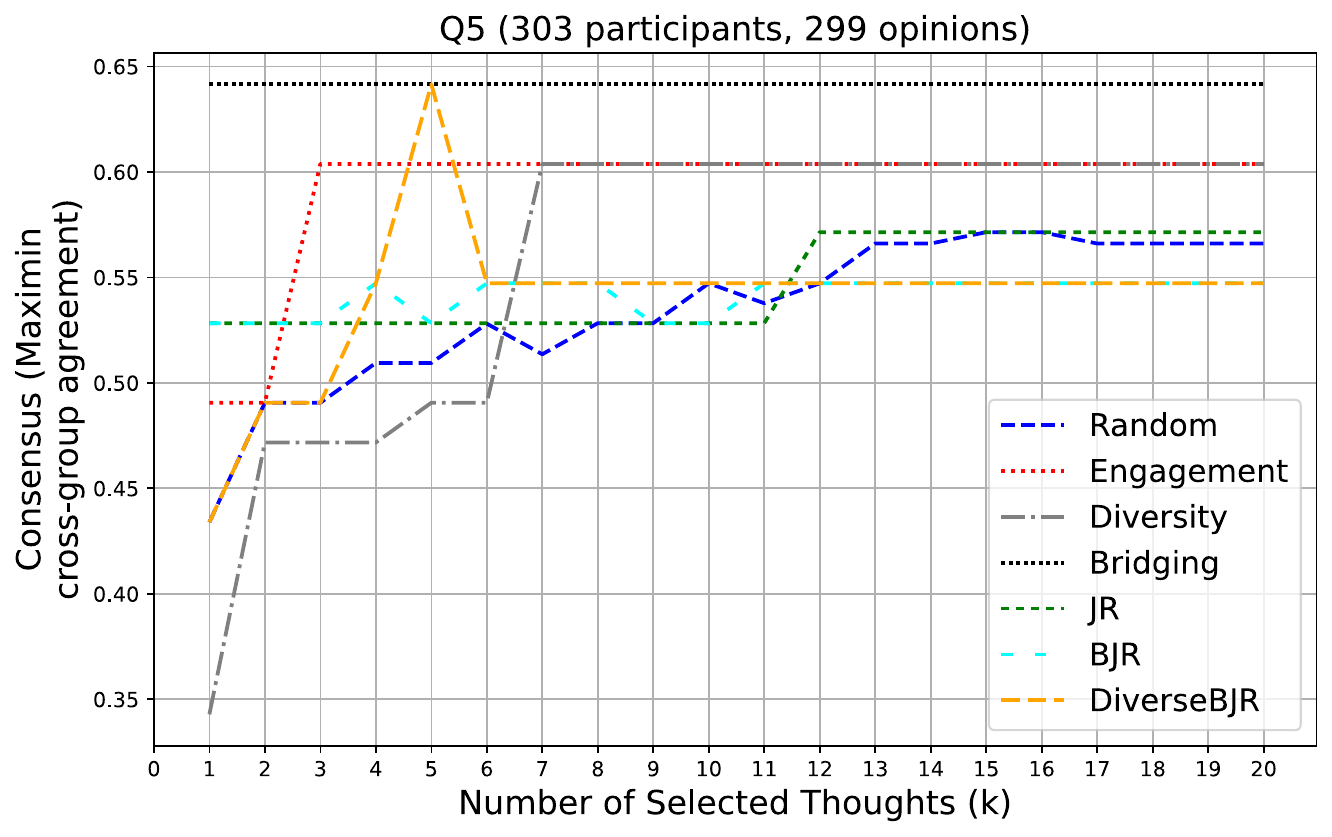}%
  \end{subfigure}
  \begin{subfigure}[c]{.49\textwidth}
    \centering
    \includeinkscape[height=5cm]{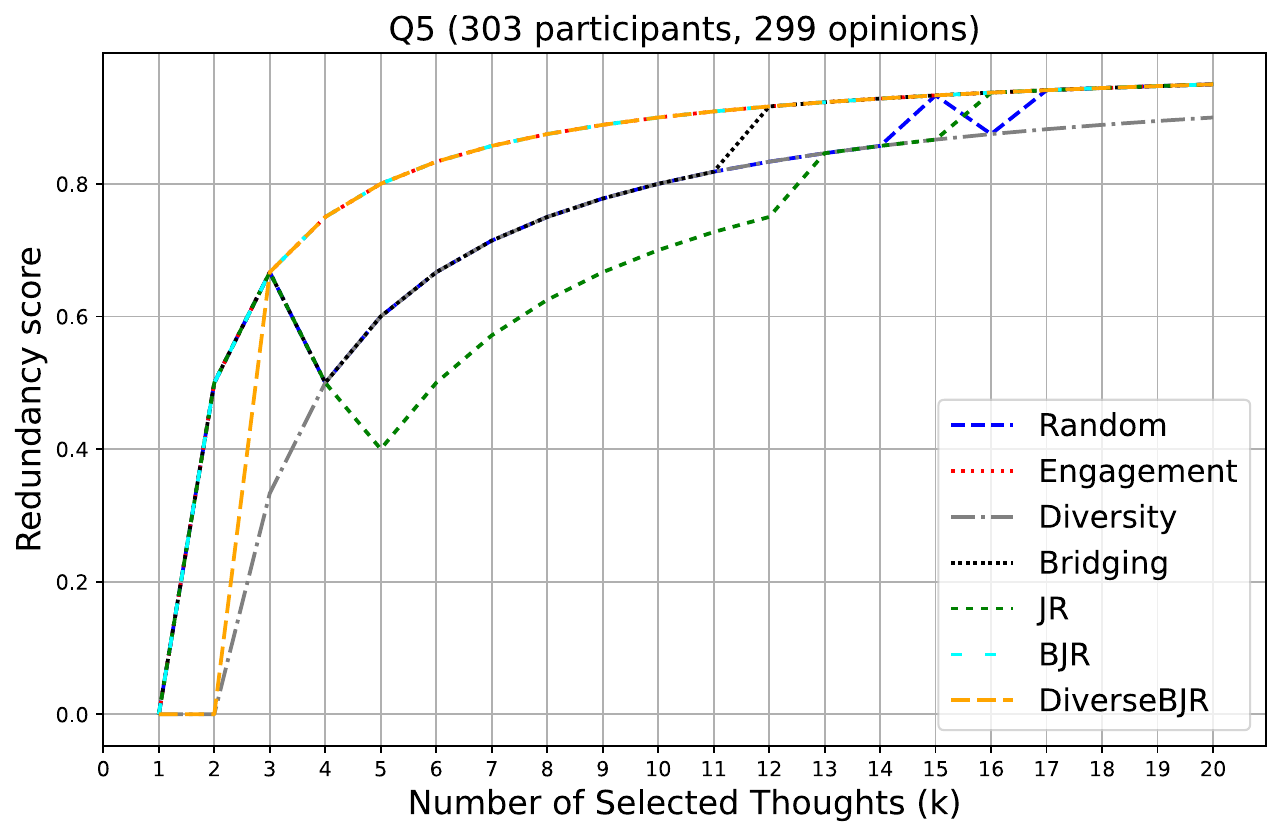}%
  \end{subfigure}\\
  \caption{Q5: "What measures (if any) could be taken to restrict or limit inappropriate protests?" (303 participants, 299 opinions)%
      \label{fig:q5-all}}
\end{figure}

\begin{figure}[H]%
\centering
    \begin{subfigure}[c]{.49\textwidth}
      \centering
      \includeinkscape[height=5cm]{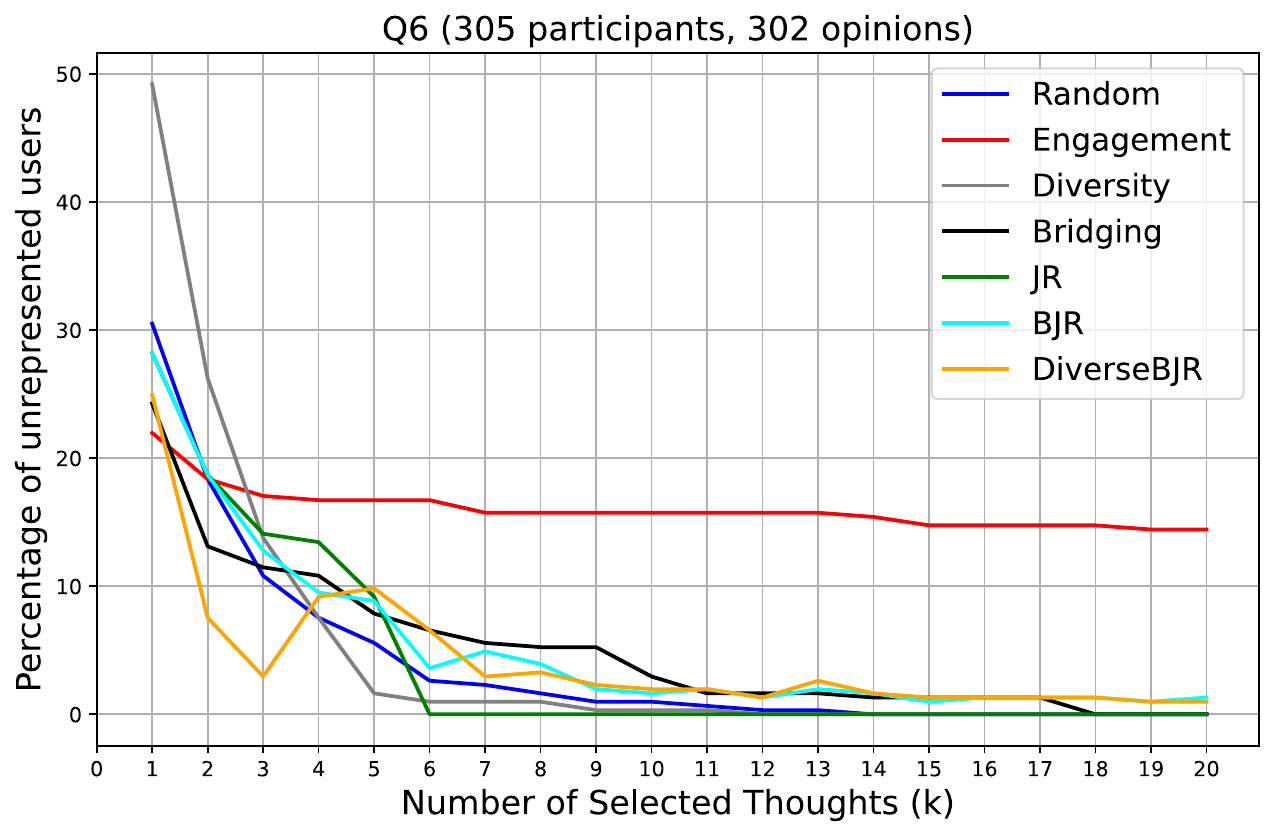}
    \end{subfigure}\hfill
  \begin{subfigure}[c]{.49\textwidth}
    \centering
    \includeinkscape[height=5cm]{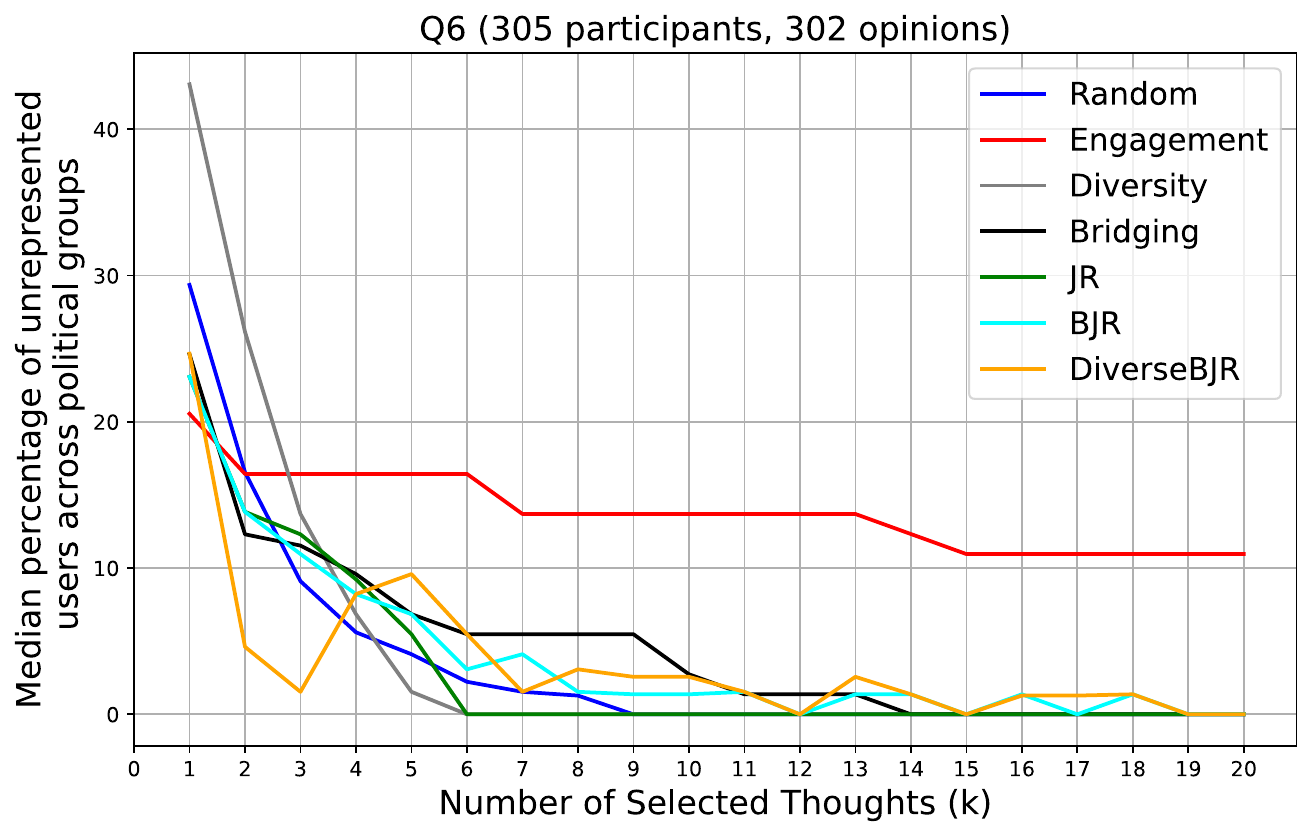}%
  \end{subfigure}\\
    \begin{subfigure}[c]{.49\textwidth}
      \centering
      \includeinkscape[height=5cm]{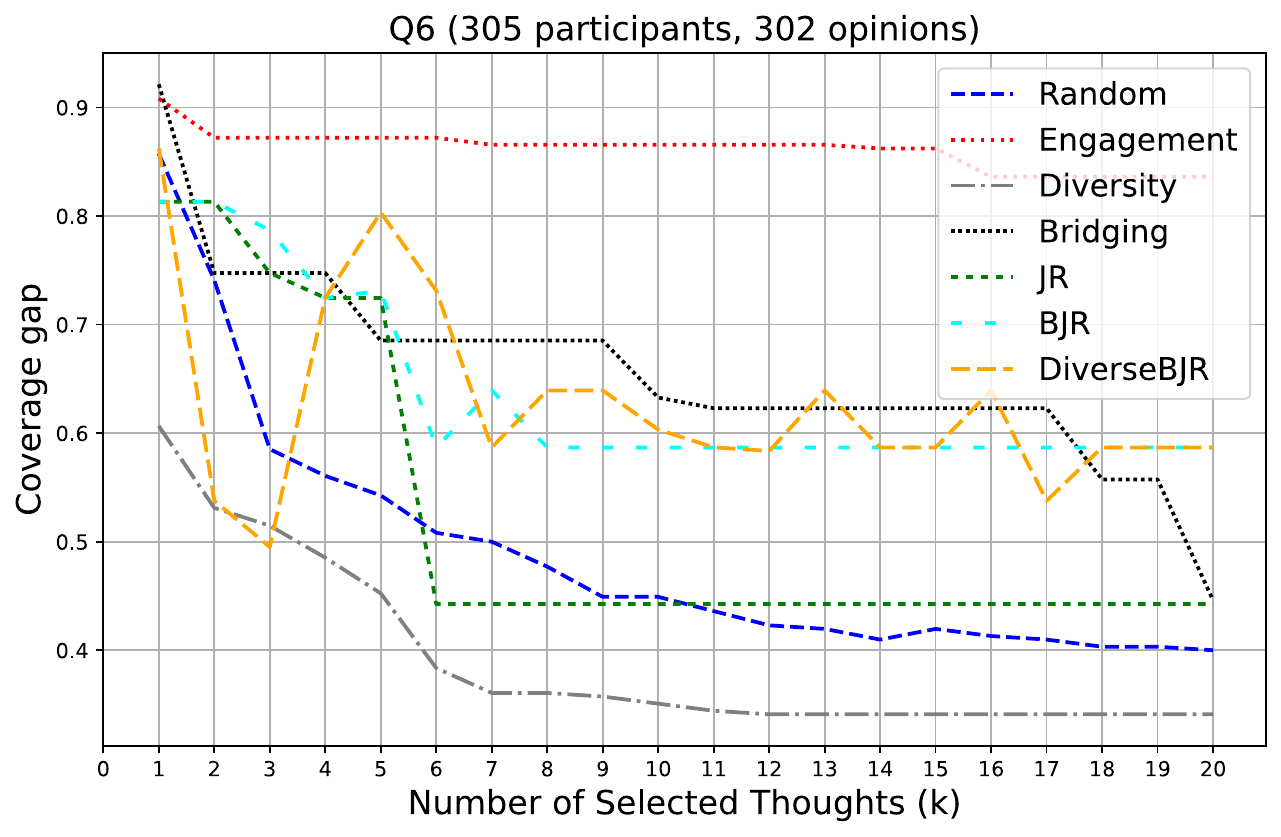}
    \end{subfigure}\hfill
  \begin{subfigure}[c]{.49\textwidth}
    \centering
    \includeinkscape[height=5.5cm]{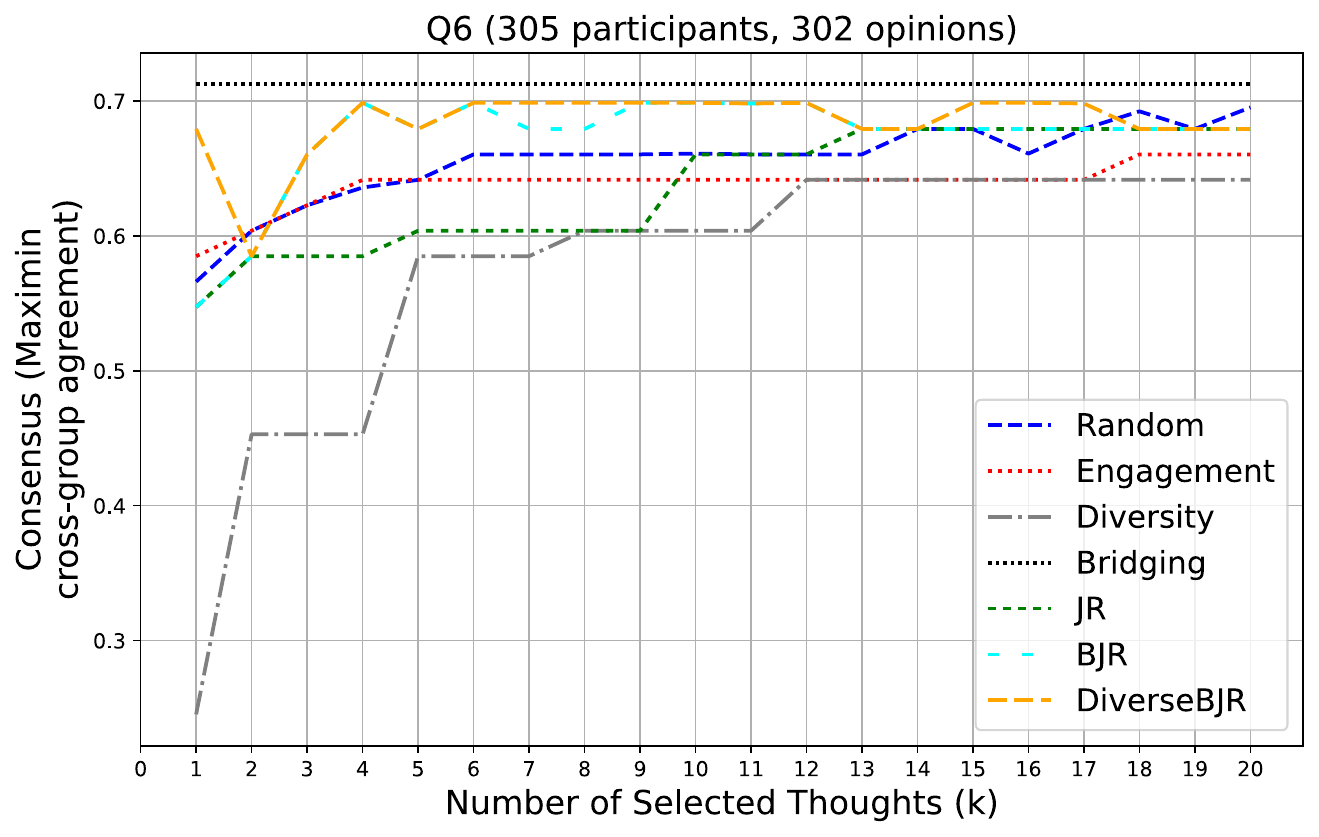}%
  \end{subfigure}
  \begin{subfigure}[c]{.49\textwidth}
    \centering
    \includeinkscape[height=5cm]{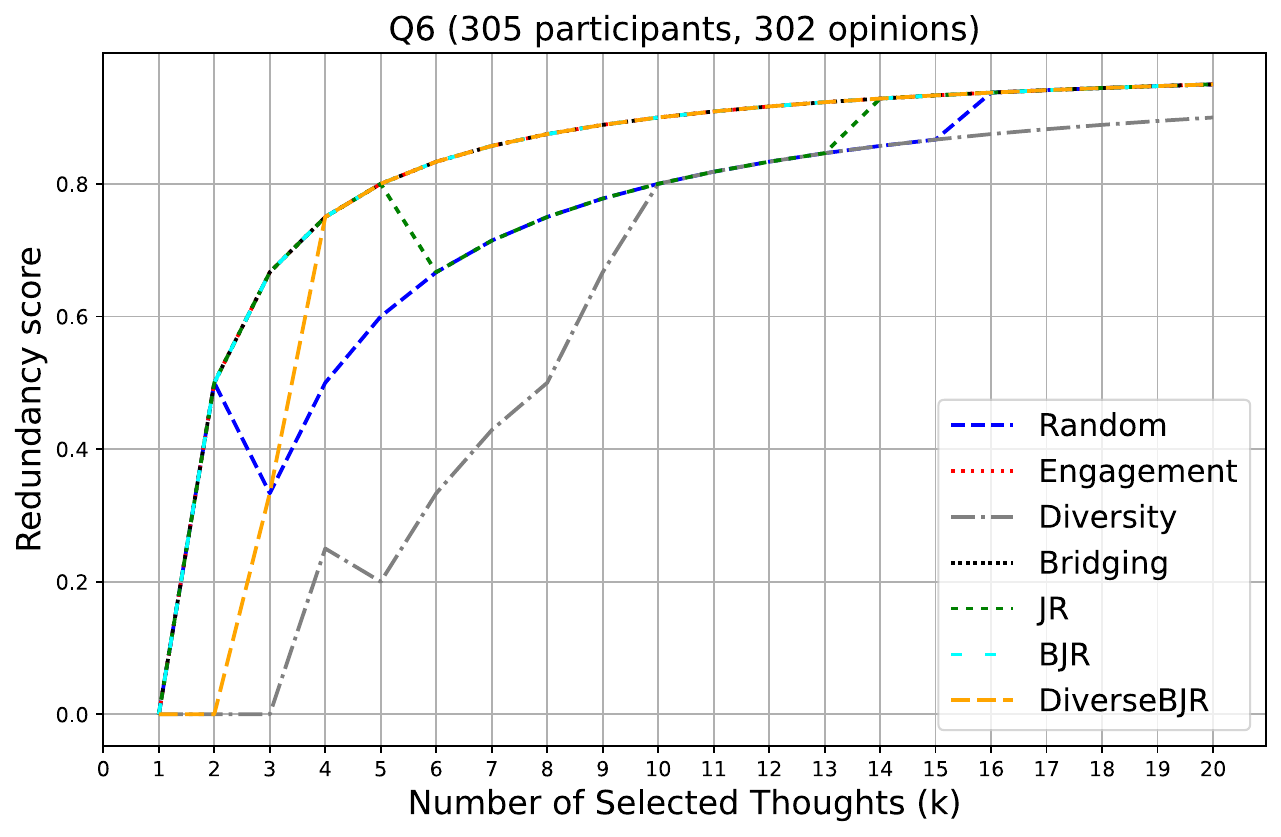}%
  \end{subfigure}\\
  \caption{Q6: "What measures (if any) could be taken to ensure appropriate protests are protected?" (305 participants, 302 opinions)%
      \label{fig:q6-all}}
\end{figure}

\section{Replicating results using a Virtual Citizen Assembly dataset}
\label{appendix:validation-habermas}
\begin{figure}[H]
\centering
    \begin{subfigure}[c]{.50\columnwidth}
      \centering
      \includeinkscape[height=5.5cm]{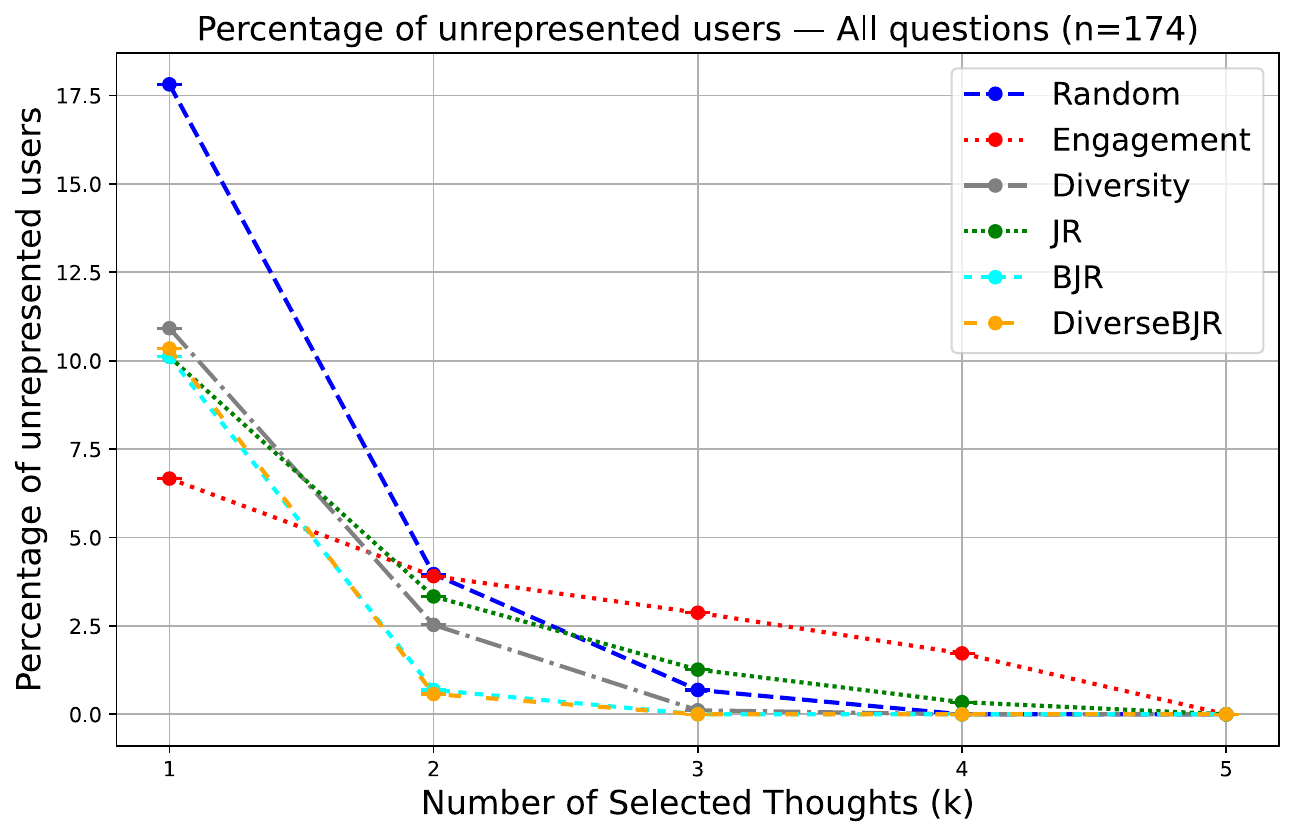}
    \end{subfigure}\hfill
  \begin{subfigure}[c]{.50\columnwidth}
    \centering
    \includeinkscape[height=5.5cm]{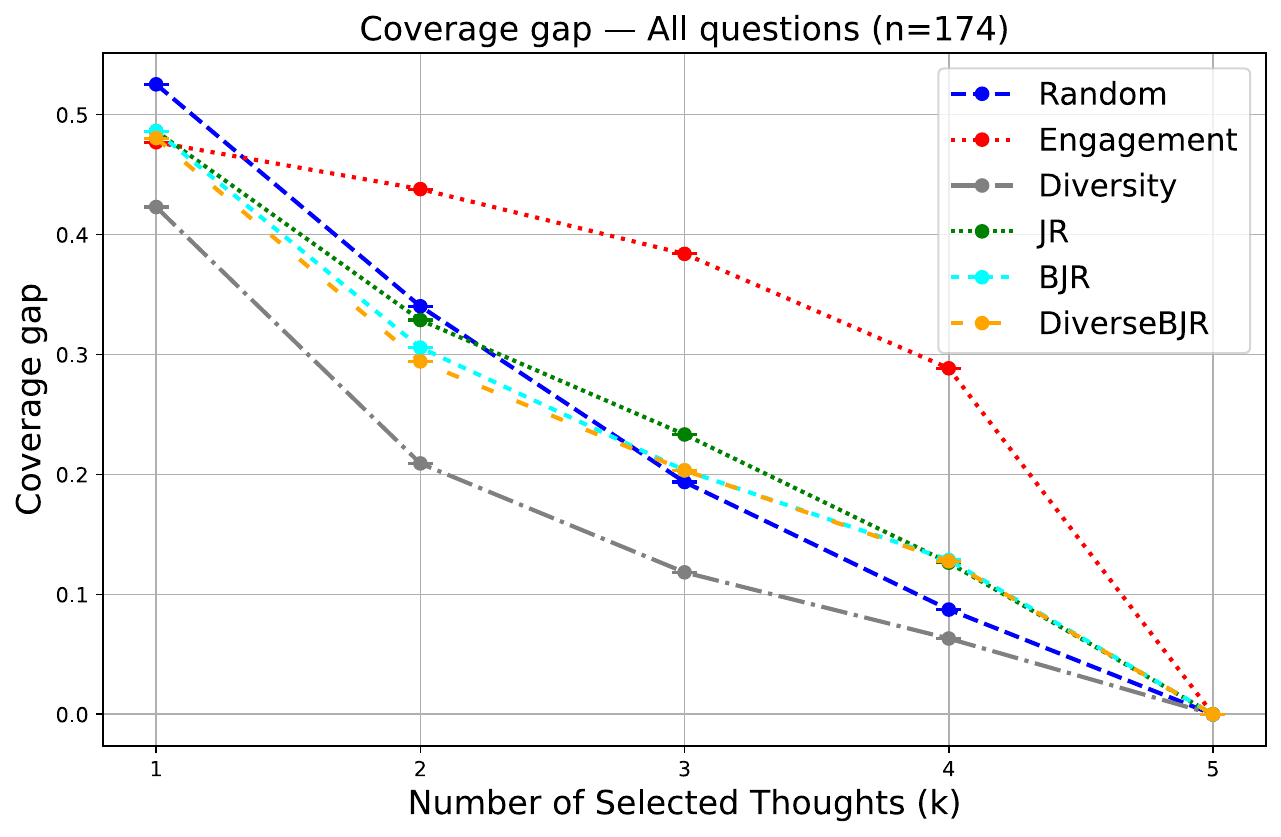}%
  \end{subfigure}\hfill
  \begin{subfigure}[c]{.50\columnwidth}
    \centering
    \includeinkscape[height=5.5cm]{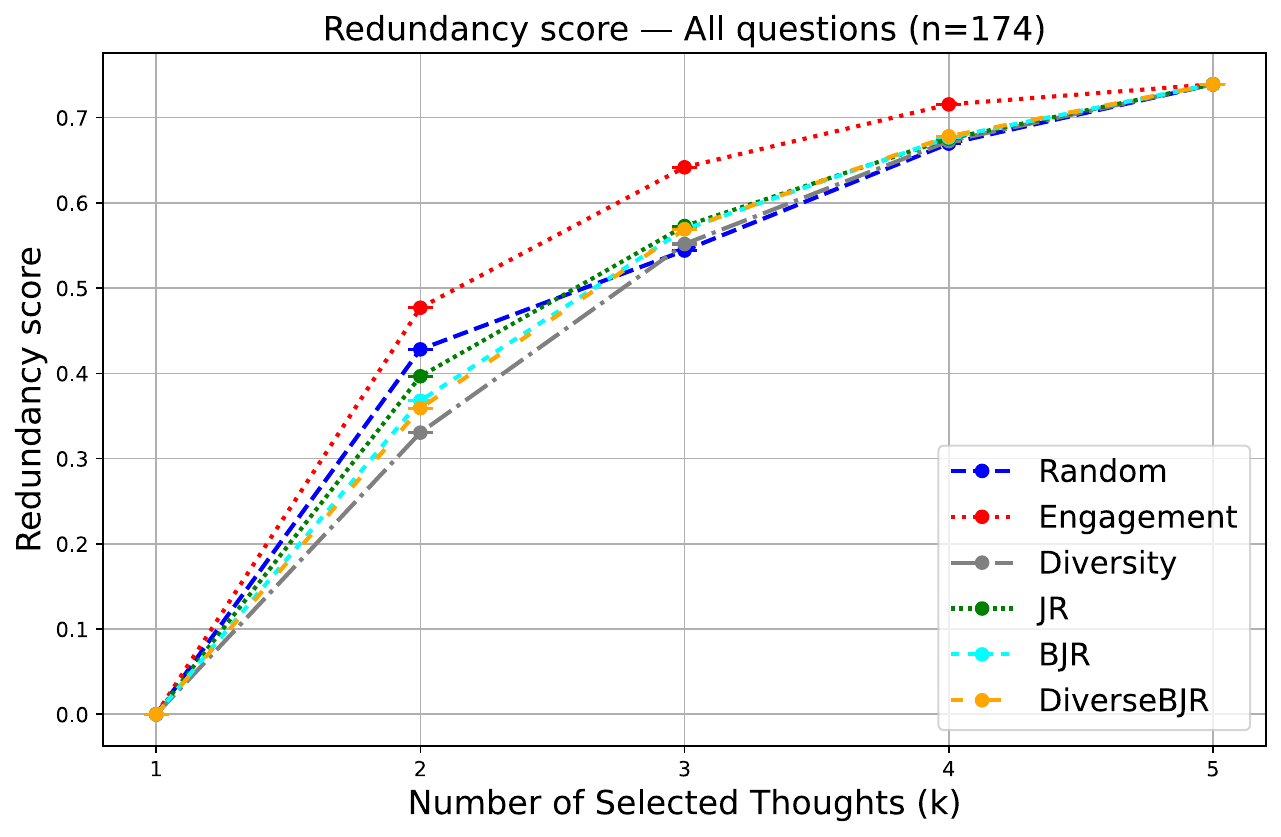}%
  \end{subfigure}\\
  \caption{\textbf{We find that our findings are replicated on a separate dataset originating from the Habermas Machine study} \cite{tessler2024ai}, involving a demographically representative sample of the UK public. \textbf{In this dataset, the approval votes are complete}, the resulting approval matrix is thus originally dense and does not require any inference (dotted lines in the plot are used for better distinction of overlapping zones).
  \label{fig:habermas_results}}
\end{figure}

In this section, we control for our findings not being dependent on the probabilistic inference run by Remesh to fill missing votes (see Section \ref{sec:rel-work} for a description of how Remesh's process works). To do so, we reproduce our experimental pipeline on a distinct dataset, namely the Virtual Citizen Assembly data from the Habermas Machine study \cite{tessler2024ai}. The study used a sample of $\approx$200 participants that are demographically representative of the UK population, and a set of 9 questions related to UK public policy, e.g., "Should we reduce voting age to 16?". Participants were divided in groups of 4-5 participants. Each group received a question. Participants individually expressed their opinion on the question in free-text format, and then voted on all of the remaining group's opinions. Votes were collected using the following 7-point Likert scale: \{"STRONGLY\_DISAGREE", "DISAGREE", "SOMEWHAT\_DISAGREE", "NEUTRAL", "SOMEWHAT\_AGREE", "AGREE", "STRONGLY\_AGREE"\}. We convert data to binary votes to match our problem setting, where we map \{"STRONGLY\_DISAGREE", "DISAGREE", "SOMEWHAT\_DISAGREE", "NEUTRAL"\} to a value of zero, and \{"SOMEWHAT\_AGREE", "AGREE", "STRONGLY\_AGREE"\} to a value of one. We preprocess data to only keep groups with exactly 5 participants for homogeneous evaluation. This results in a total of 174 approval matrices of shape 5x5. In practice, a set of 5 opinions does not need a representative subset, let alone an algorithm to do so. However, this data offers a valuable validation set on which algorithms can be evaluated. We reproduce all metrics and all baselines except the bridging baseline and the consensus metric (both of which require political leaning attributes not present in the publicly available version of the data). We also discard the Zero-shot ChatGPT baseline for the same reasons as with the Remesh data: it failed to reliably respect the requested cardinality $k$ (e.g., selecting 2 opinions when prompted to only select 1), which precludes fair comparison with other baselines (details in Appendix \ref{appendix:llm-baseline}). We compute median values to aggregate results on all 9 questions. Results for this Virtual Assembly validation set are shown in Fig. \ref{fig:habermas_results}.

\noindent
Our results replicate the findings from the main experiments in Section \ref{sec:results}: JR/BJR/DiverseBJR offer large and quick early gains in representation (upper-left panel - Fig \ref{fig:habermas_results}), and BJR/DiverseBJR offer a better coverage gap (upper-right panel - Fig \ref{fig:habermas_results}) and lower redundancy scores (lower-panel - Fig \ref{fig:habermas_results}) than Engagement. A single unexpected result is observed with the Engagement on representation for $k=1$ (upper-left panel - Fig \ref{fig:habermas_results}), where Engagement exhibits better performance than remaining baselines. However, Engagement quickly lags behind all others starting from $k=2$.

\section{Computational Complexity Analysis}
Results for time and memory complexity are presented in Table \ref{tab:baseline-complexity}. For the Diversity and DiverseBJR baselines which require pairwise distance comparisons, a vectorized matrix of distances is pre-computed once before the greedy loop (leading to the $\mathcal{O}(m^2 n)$ time complexity). A naïve implementation where distances are computed on the fly would have led to a time complexity of $\mathcal{O}(km^2 n)$. For DiverseBJR, the time complexity of $\mathcal{O}(Tk^2m n) + \mathcal{O}(m^2 n) + \mathcal{O}(kmn)$ includes the greedy BJR feasibility checker which, when invoked, costs $\mathcal{O}(Tkm n)$ per invocation and $\mathcal{O}(Tk^2m n)$ in aggregate (each of $T$ trials simulates up to $k$ future rounds). Note that in practice, in our experiments, $T$ is a small constant (e.g., 5-10) and that the simulator which checks BJR is only triggered if at the end of a round neighbors exist and future demands remain.
\begin{table}[htbp]
\centering
\resizebox{.8\columnwidth}{!}{
\begin{tabular}{l|l|l}
\toprule
Baseline & Time Complexity & Memory Complexity \\%& Notes \\
\midrule
Random 
& $\mathcal{O}(k)$ 
& $\mathcal{O}(1)$ 
\\%& Uniform sampling; negligible overhead \\

Engagement 
& $\mathcal{O}(nm + m \log m)$ 
& $\mathcal{O}(m)$ 
\\%& Counting dominates; selection can be linear-time \\

Bridging
& $\mathcal{O}(nm + \gamma m)$ 
& $\mathcal{O}(m)$ 
\\%& Group-wise checks affect constants only \\

Diversity (Greedy Maximin) 
& $\mathcal{O}(m^2 n) + \mathcal{O}(k mn)$ 
& $\mathcal{O}(m^2)$ 
\\%& Distance precomputation dominates; naive implementations are much slower \\

JR / BJR (Greedy CC) 
& $\mathcal{O}(k m n)$ 
& $\mathcal{O}(m + n)$ 
\\%& Polynomial-time greedy approximation \\

DiverseBJR 
& $\mathcal{O}(Tk^2m n) + \mathcal{O}(m^2 n) + \mathcal{O}(k mn)$ 
& $\mathcal{O}(m^2)$ \\

Zero-shot ChatGPT 
& $\mathcal{O}(k)\times$ LLM inference 
& Model-dependent \\
\bottomrule
\end{tabular}}
\caption{Theoretical complexity of baseline methods. 
$n$ = number of users, $m$ = total number of opinions, $k$ = number of selected opinions, $\gamma$ = number of groups, $T$ = number of BJR feasibility trials.}
\label{tab:baseline-complexity}
\end{table}

\section{Zero-shot LLM baseline}\label{appendix:llm-baseline}
\textbf{Implementation:} We used ChatGPT (\textit{"gpt-4.1-mini"}) in a zero-shot setting as our LLM-baseline. We used a temperature of 0 and a nucleus sampling (top-p) of 1 for more deterministic results and enhanced reproducibility. %
We experimented with various prompting strategies but none mitigated the behavior where the model returned less than $k$ indices. 

\begin{figure}[H]
\centering
    \begin{subfigure}[c]{.49\textwidth}
      \centering
      \includeinkscape[height=5cm]{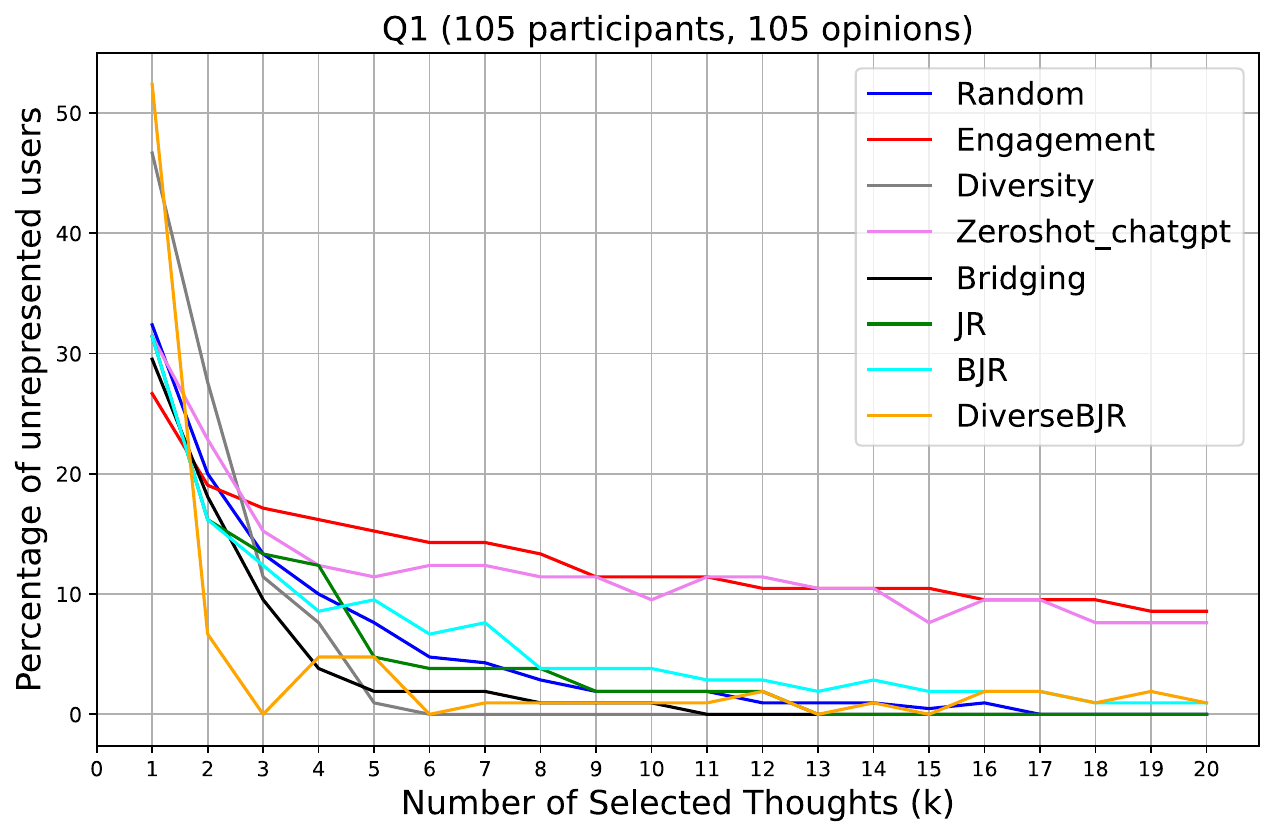}
    \end{subfigure}\hfill
  \begin{subfigure}[c]{.49\textwidth}
    \centering
    \includeinkscape[height=5cm]{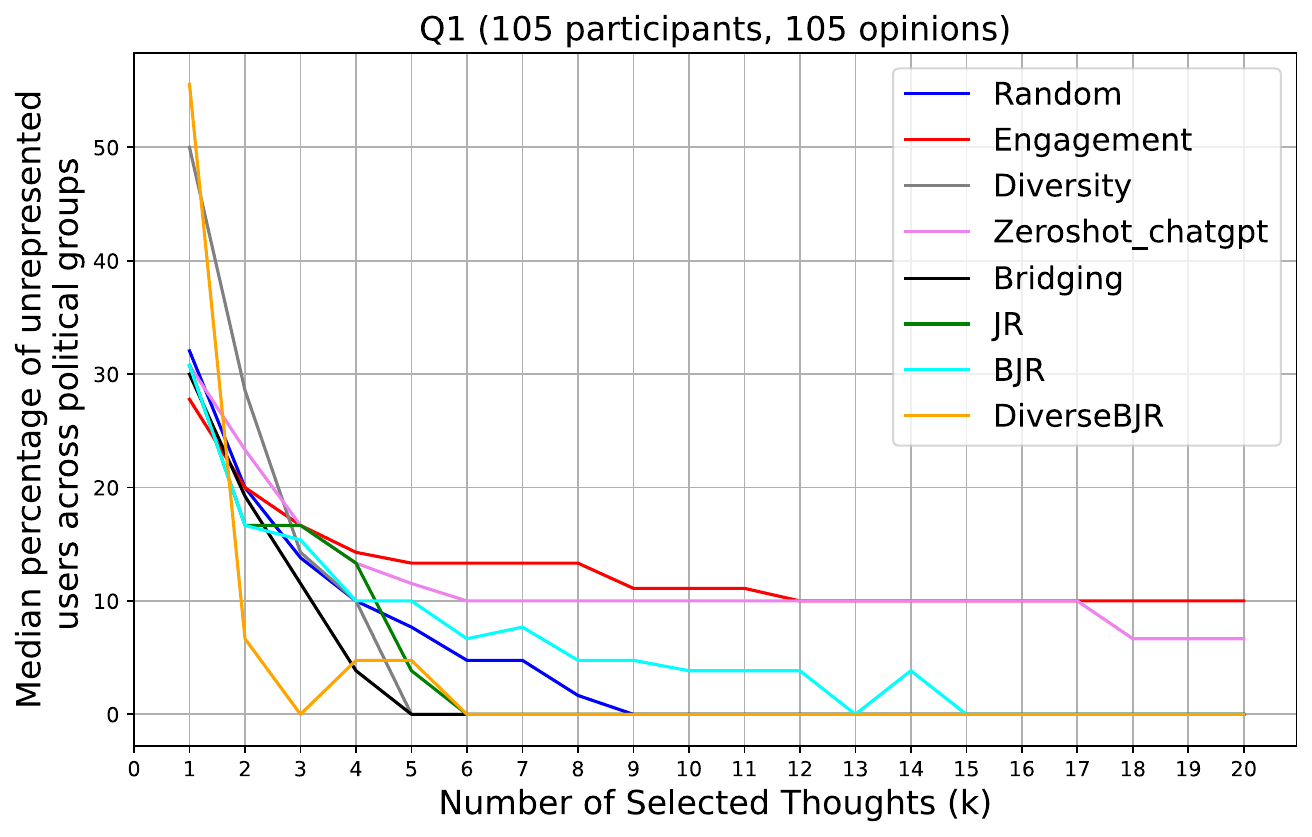}%
  \end{subfigure}\\
    \begin{subfigure}[c]{.49\textwidth}
      \centering
      \includeinkscape[height=5cm]{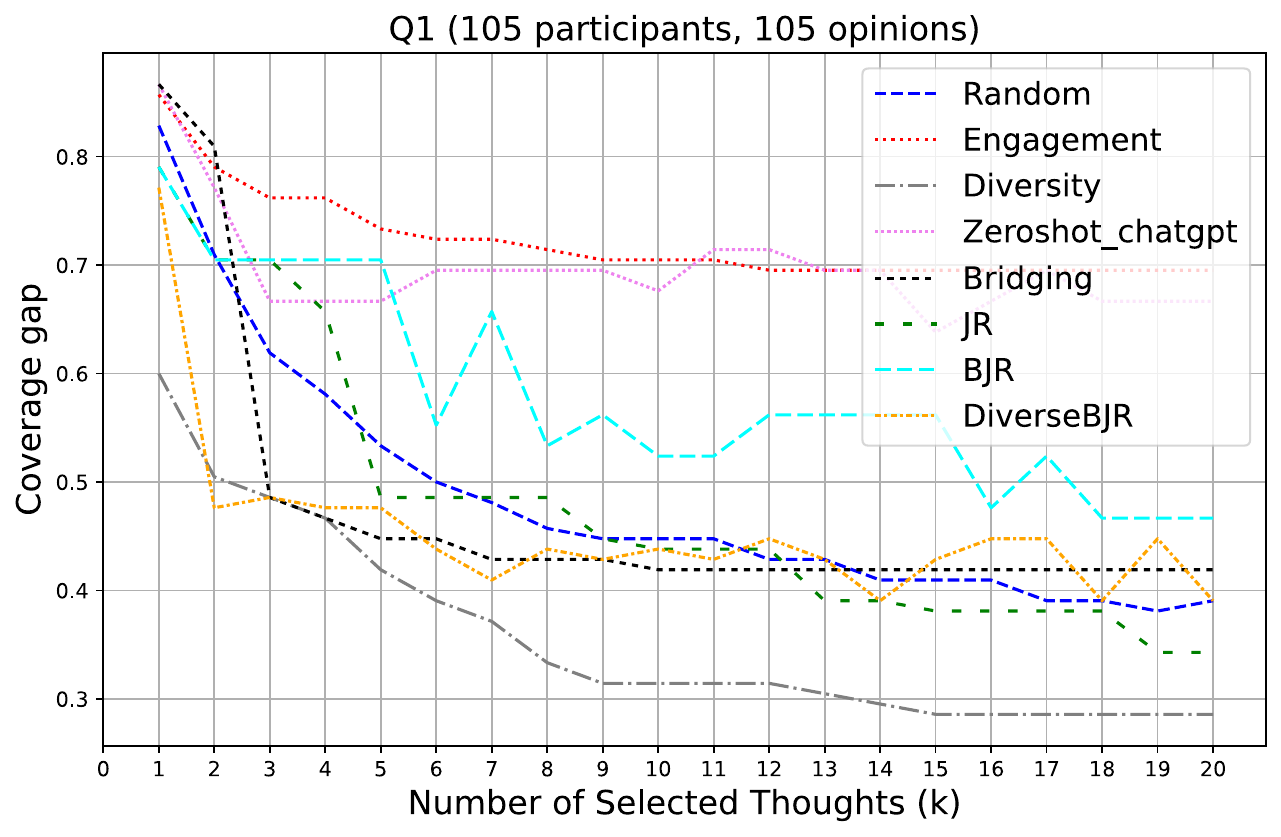}
    \end{subfigure}\hfill
  \begin{subfigure}[c]{.49\textwidth}
    \centering 
    \includeinkscape[height=5.5cm]{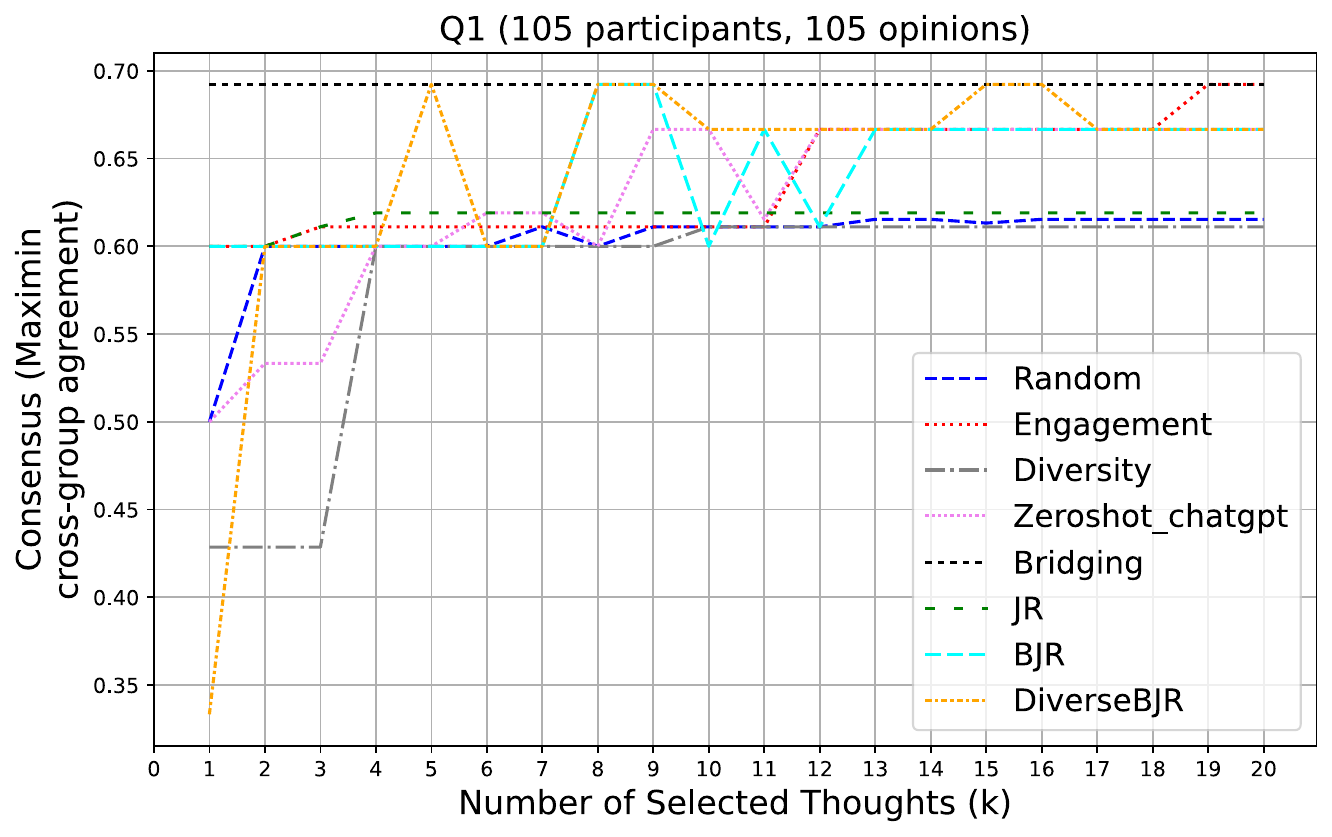}%
  \end{subfigure}
  \begin{subfigure}[c]{.49\textwidth}
    \centering
    \includeinkscape[height=5cm]{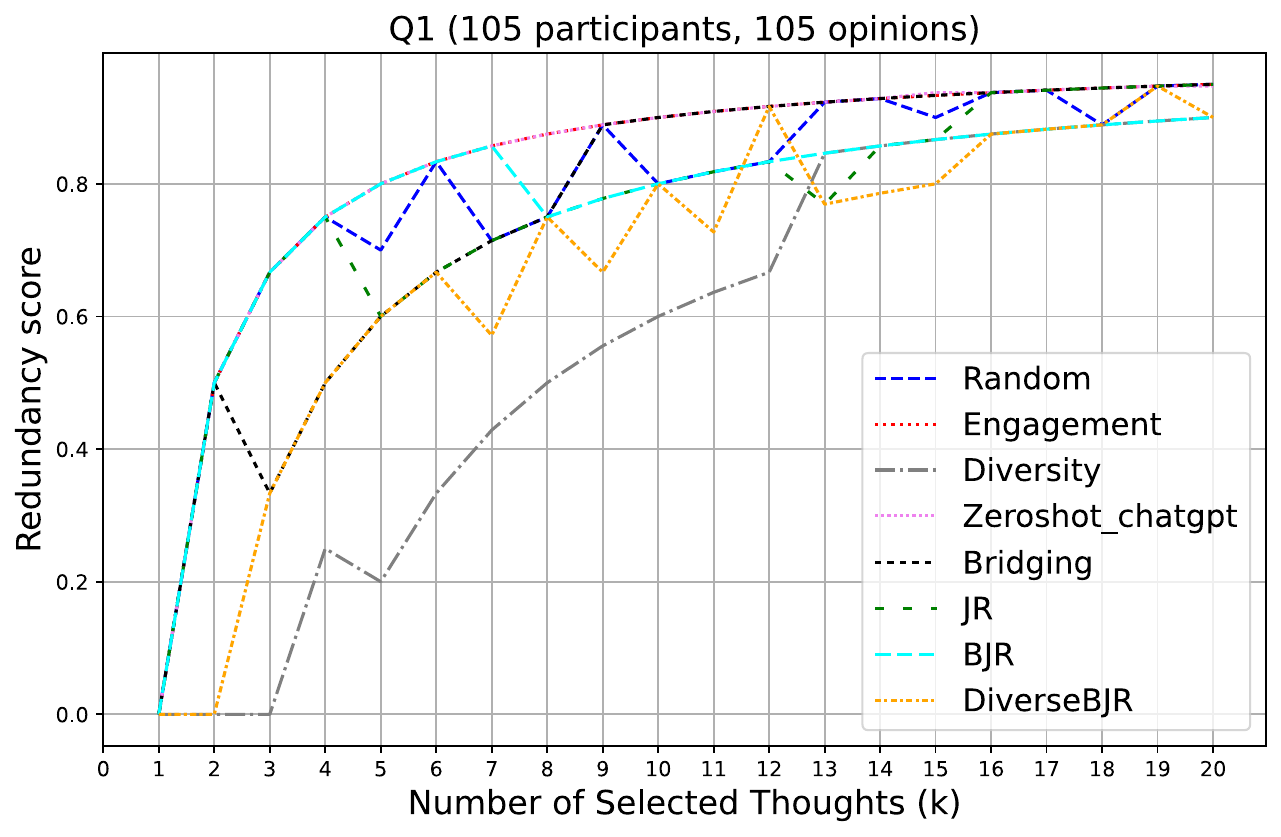}%
  \end{subfigure}\\
  \caption{Results of the zero-shot ChatGPT baseline, which we omit from the main results because it occasionally returns subsets with less than k opinions, see Appendix \ref{appendix:llm-baseline}%
      \label{fig:chatgpt-q1-all}}
\end{figure}

\textbf{Results:} The model occasionally returned less indices (i.e., selected less opinions) than the requested $k$, especially for high $k$ values ($k\approx10-20$). Because representation, coverage, and redundancy are computed on the returned indices, variable subset sizes break comparability with other methods: fewer-than-k outputs can systematically understate redundancy and distort representation/coverage, yielding curves that reflect noise rather than selection quality. Since this behavior is not controllable post hoc without introducing additional assumptions (e.g., padding with a fallback rule) and would bias the comparison, we exclude the ChatGPT baseline entirely from the main results. We include one illustrative plot below for completeness (Fig. \ref{fig:chatgpt-q1-all}). The ChatGPT baseline performed worse than most other baselines for most $k$ values, but note that this plot \textbf{should be interpreted with caution} because it is biased by inconsistent slate cardinality.

\textbf{Prompt:}
\textit{
"Context: You are an assistant that converts a list of \{size of the opinions\} human free-form opinions on a single question into a set of \{k\} statements that are collectively representative of the whole set. Treat each input opinion both as a voter and as a candidate statement. Use semantic similarity to infer approvals, cluster opinions to detect groups. The question is: \{question\}, and the opinions are: \{opinions\}.
\noindent\\ 
Output: Your output should be a JSON object with the following shape:
\{\{"selected\_statements": [\{\{"idx": \textless idx statement 1\textgreater, "statement":\textless statement 1 \textgreater\}\}, ..., \{\{"idx":\textless idx statement K \textgreater, "statement":\textless statement K \textgreater\}\}], where each statement in selected\_statements is a verbatim copy of the selected statement, and each idx is the corresponding integer index of the selected statements in the opinions list (not in the selected statements list). Please don't write any single additional word, and don't modify any character (don't uppercase/lowercase, don't remove punctuation)."
}

\section{Sensitivity analysis of the diversity threshold parameter $\varepsilon$ }\label{appendix:epsilon-radius}
\begin{figure}[htbp]%
\centering
    \begin{subfigure}[c]{.49\textwidth}
      \centering
      \includeinkscape[height=5cm]{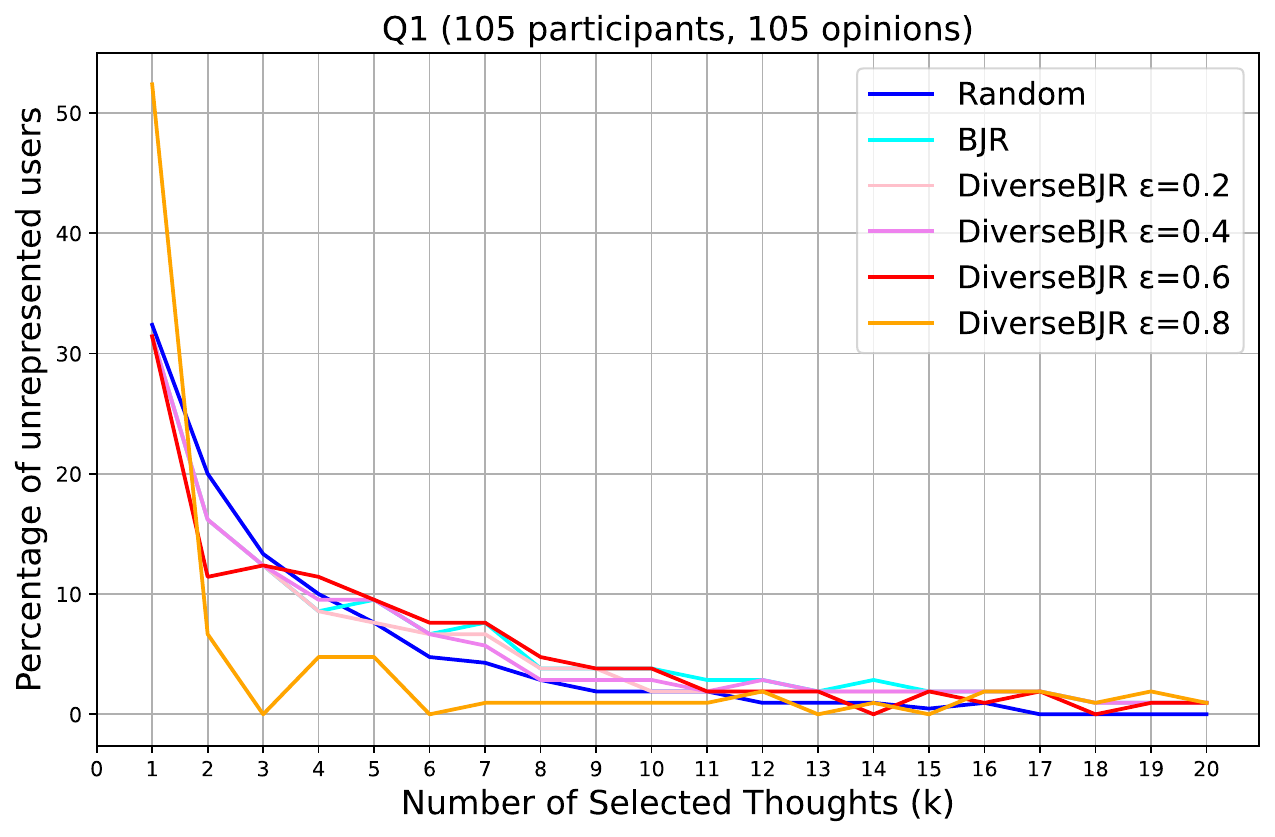}
    \end{subfigure}\hfill
  \begin{subfigure}[c]{.49\textwidth}
    \centering
    \includeinkscape[height=5cm]{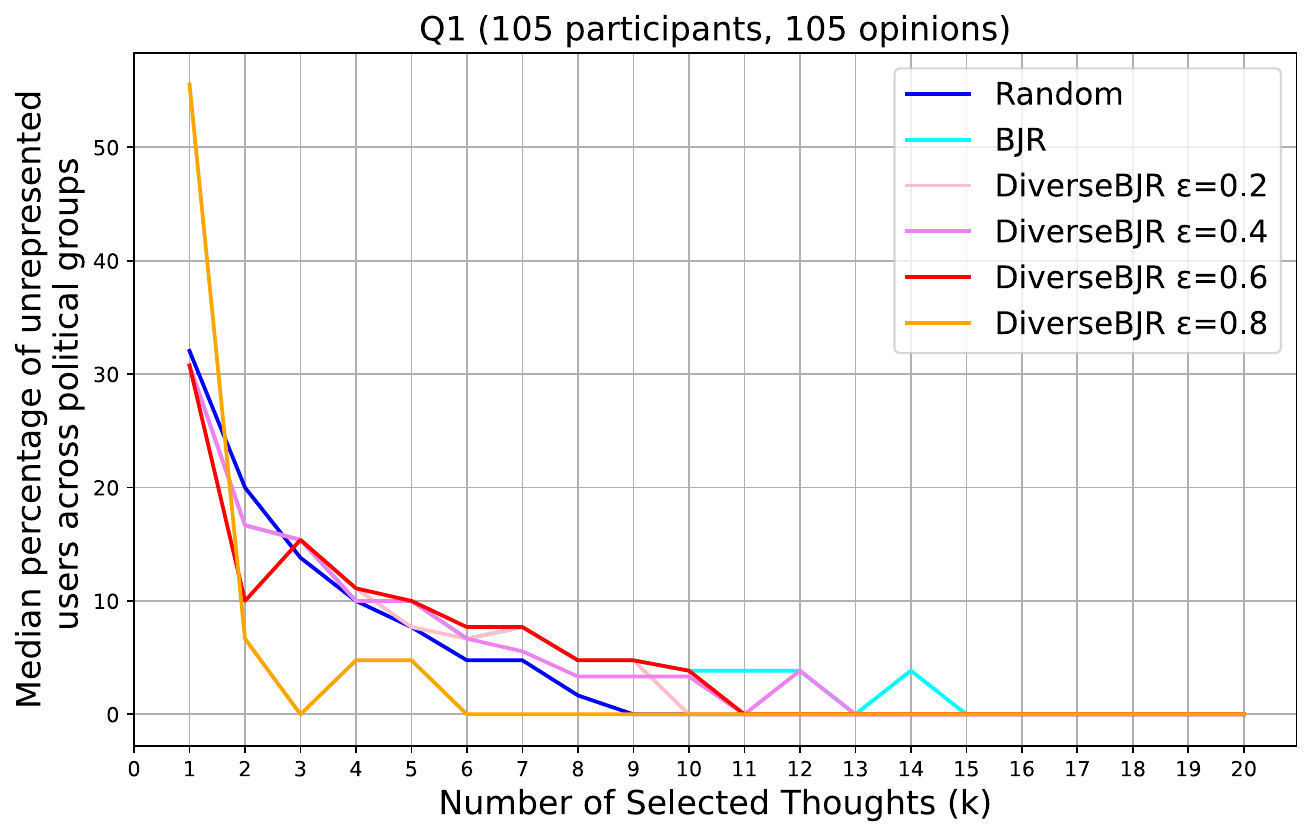}%
  \end{subfigure}\\
    \begin{subfigure}[c]{.49\textwidth}
      \centering
      \includeinkscape[height=5cm]{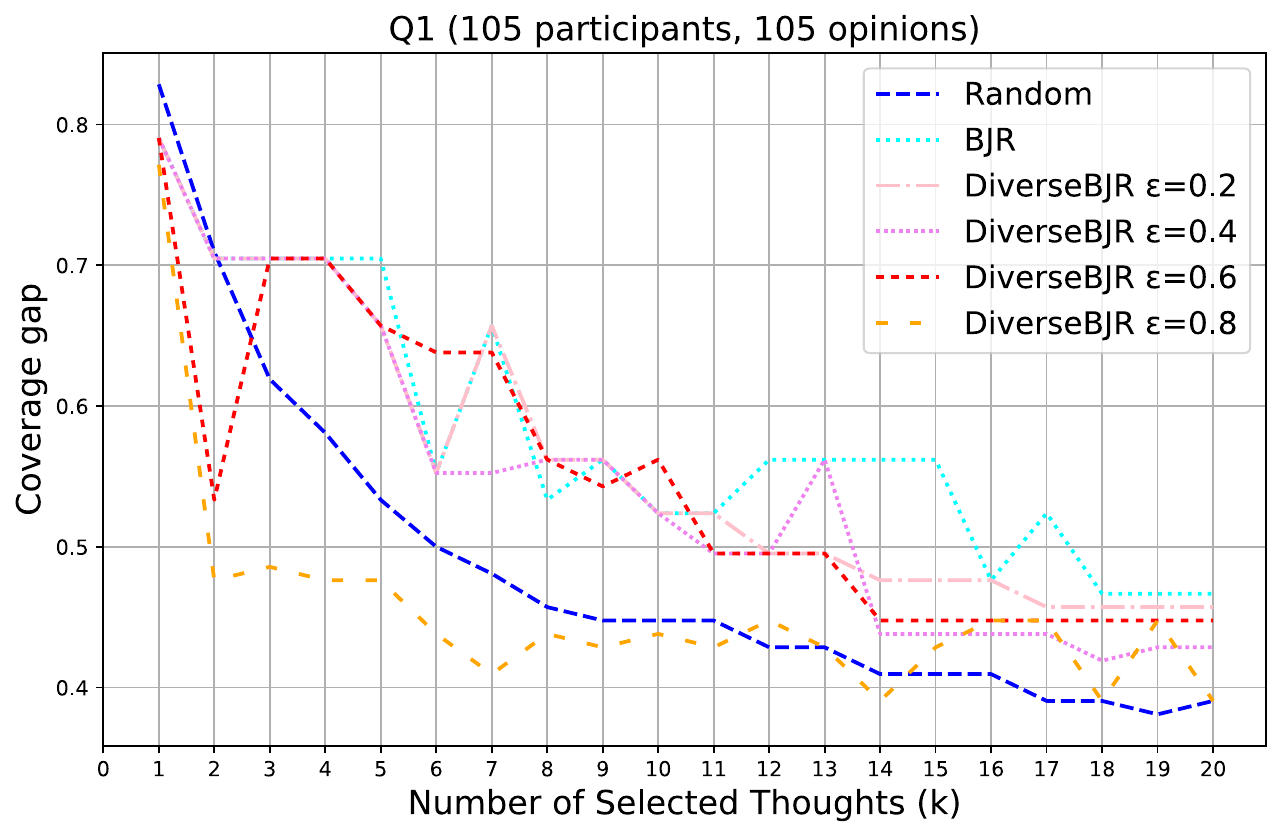}
    \end{subfigure}\hfill
  \begin{subfigure}[c]{.49\textwidth}
    \centering
    \includeinkscape[height=5.5cm]{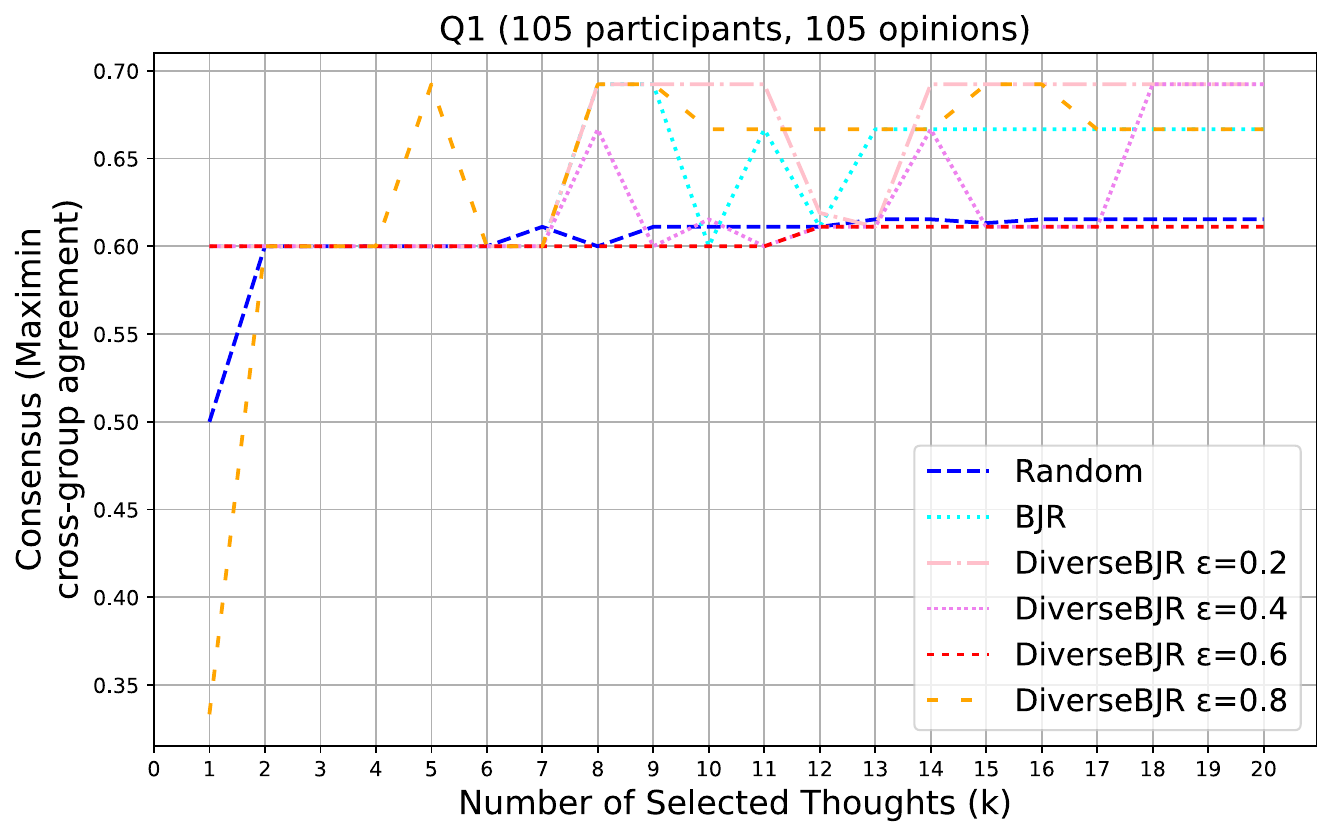}%
  \end{subfigure}
  \begin{subfigure}[c]{.49\textwidth}
    \centering
    \includeinkscape[height=5cm]{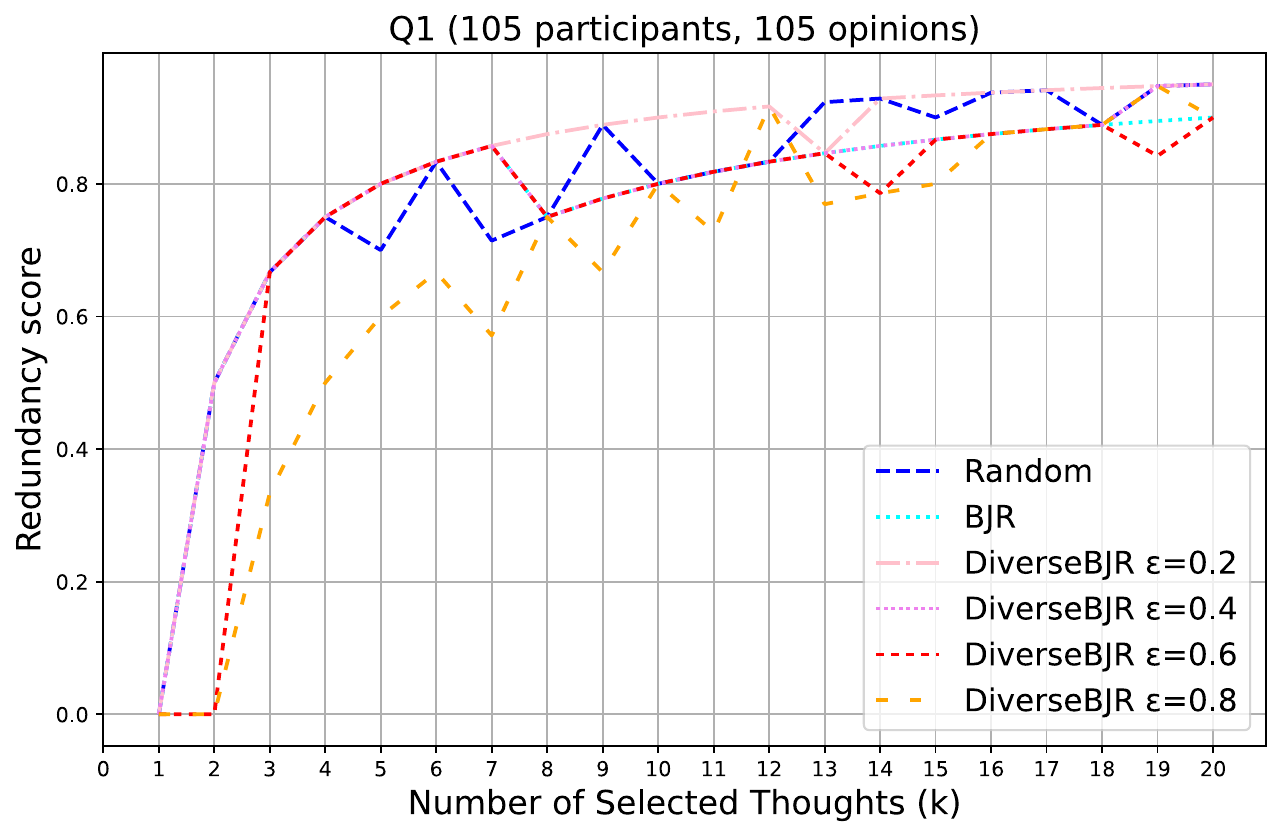}%
  \end{subfigure}\\
  \caption{Q1: "How has your personal experience with protests influenced your viewpoint on the right to assemble?" (105 participants, 105 opinions)%
      \label{fig:q1-all-sensitivity-analysis}}
\end{figure}

\begin{figure}[htbp]%
\centering
    \begin{subfigure}[c]{.49\textwidth}
      \centering
      \includeinkscape[height=5cm]{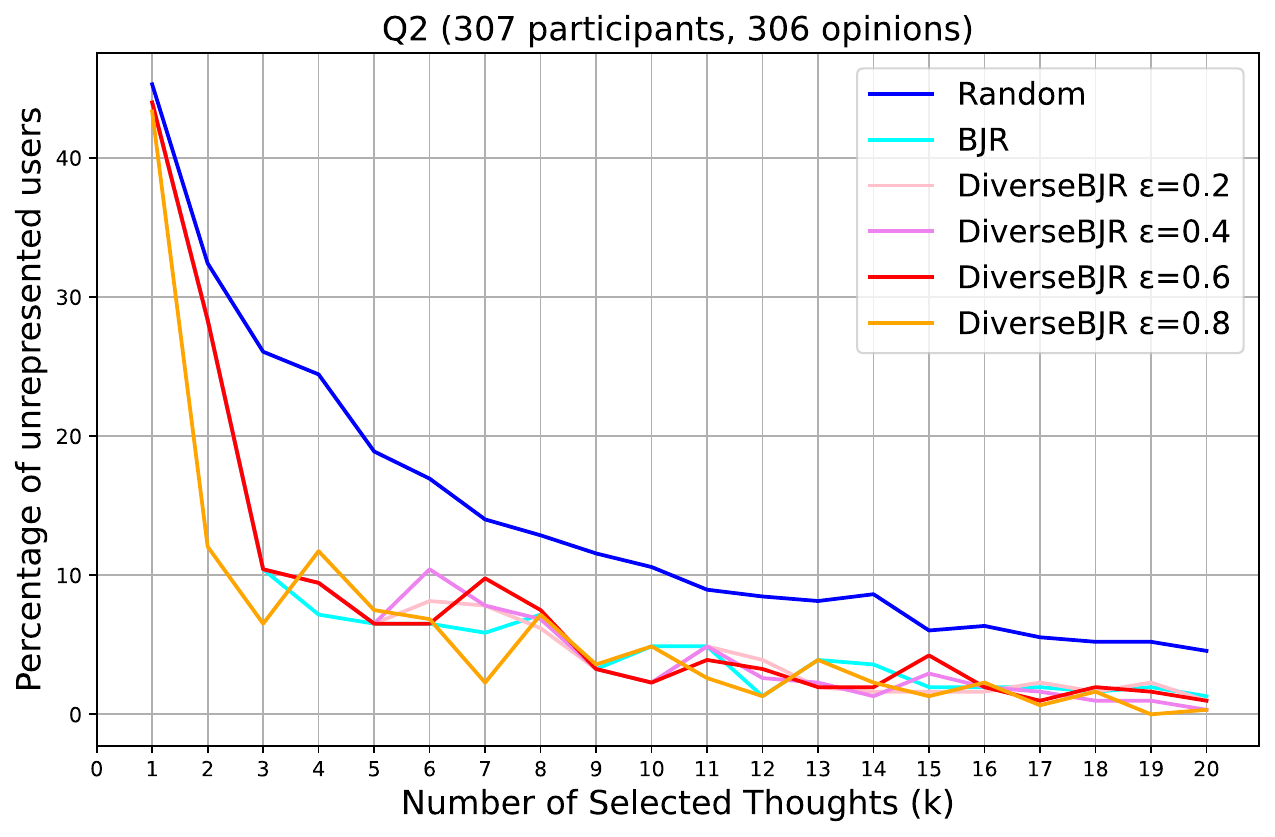}
    \end{subfigure}\hfill
  \begin{subfigure}[c]{.49\textwidth}
    \centering
    \includeinkscape[height=5cm]{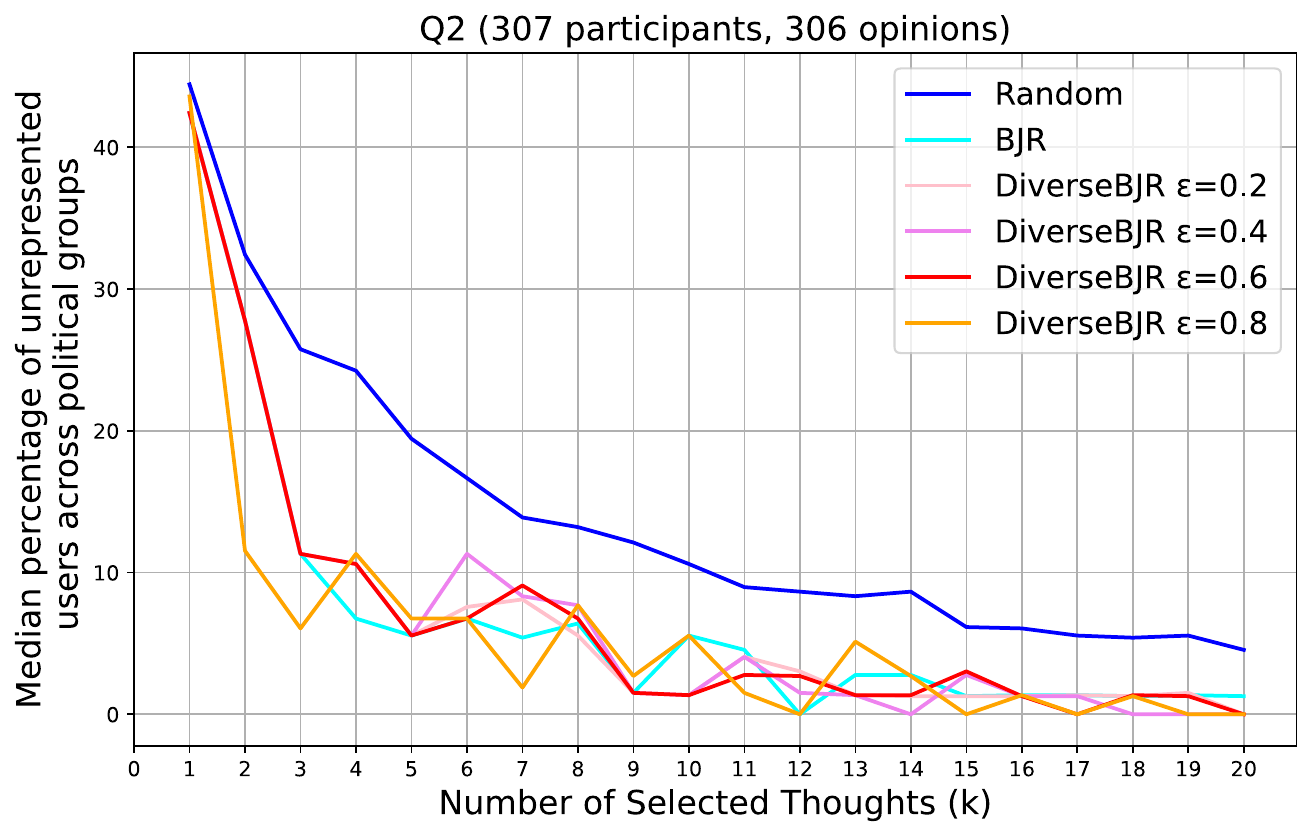}%
  \end{subfigure}\\
    \begin{subfigure}[c]{.49\textwidth}
      \centering
      \includeinkscape[height=5cm]{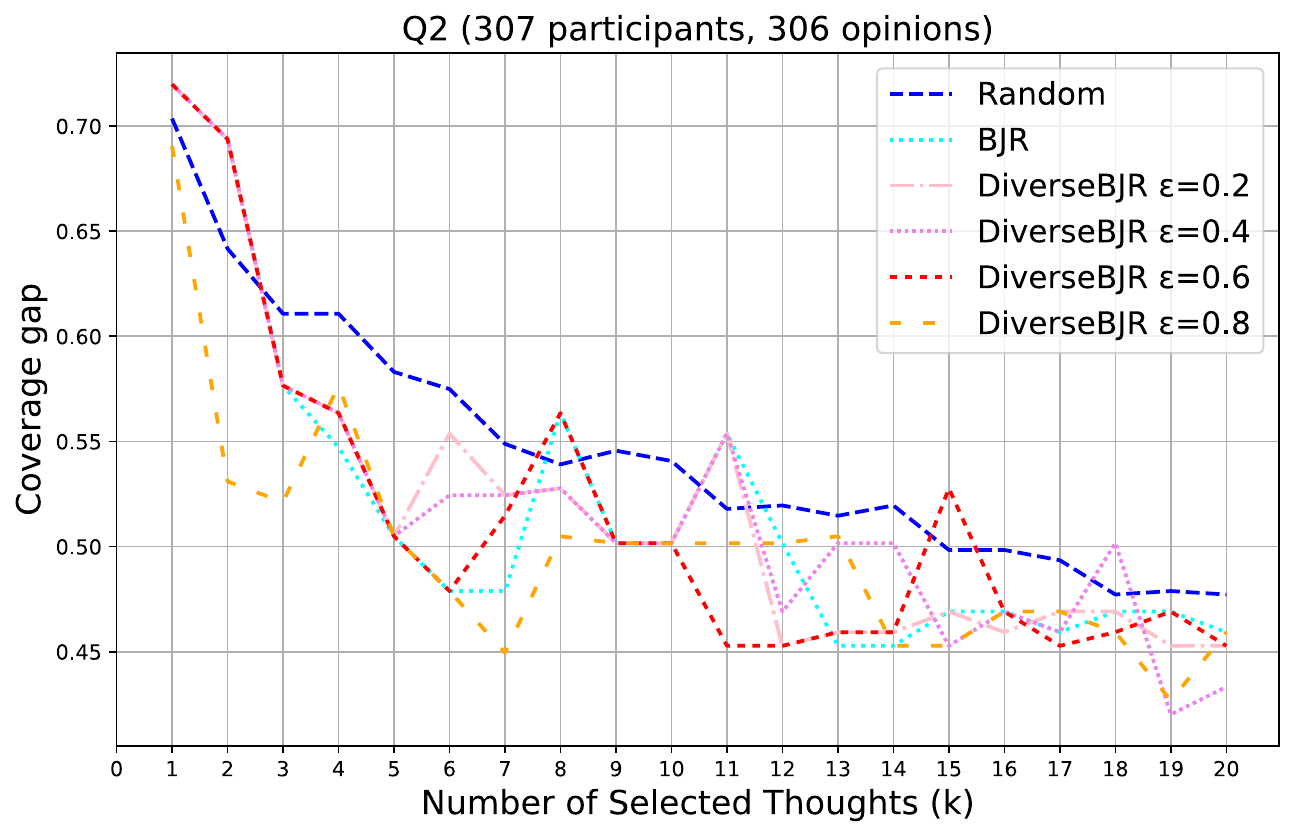}
    \end{subfigure}\hfill
  \begin{subfigure}[c]{.49\textwidth}
    \centering
    \includeinkscape[height=5.5cm]{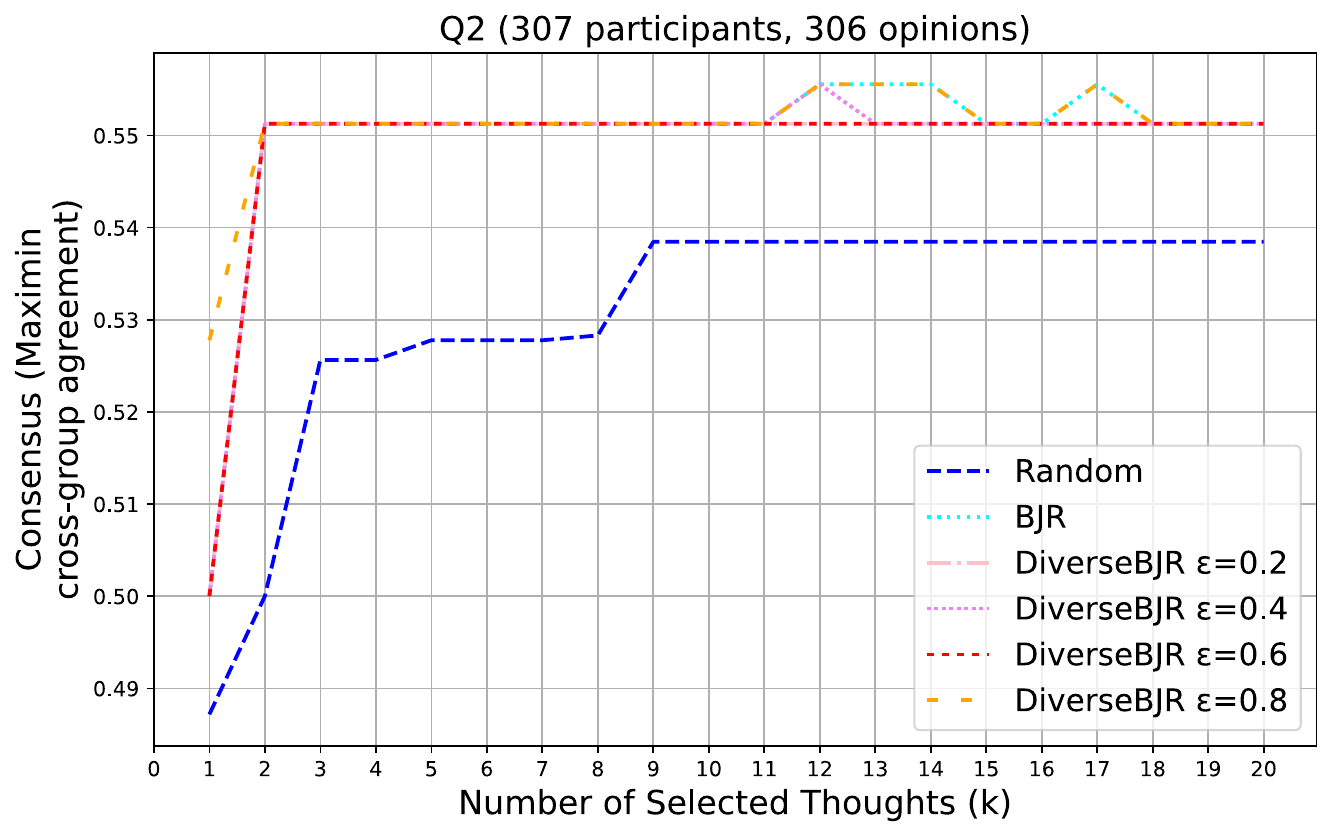}%
  \end{subfigure}
  \begin{subfigure}[c]{.49\textwidth}
    \centering
    \includeinkscape[height=5cm]{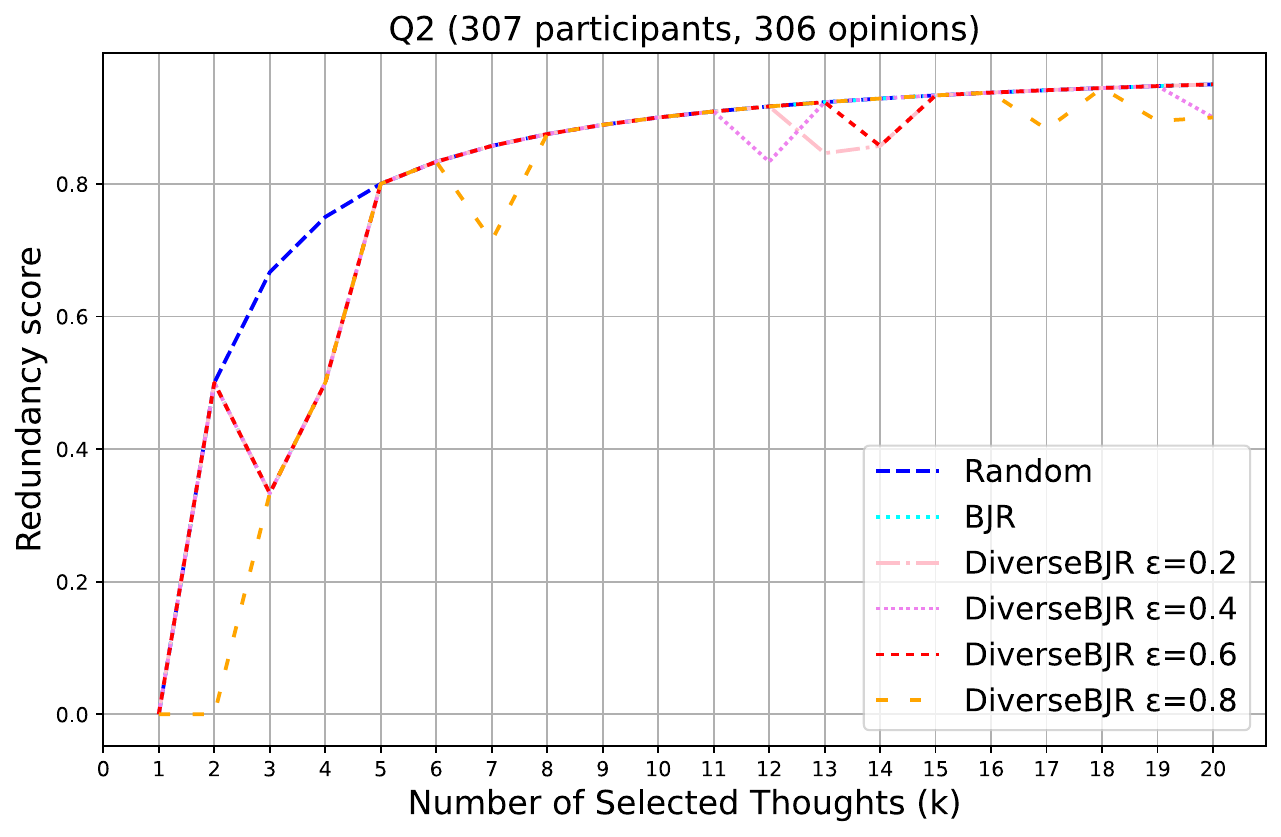}%
  \end{subfigure}\\
  \caption{Q2: "What are features or characteristics that make a protest appropriate?" (307 participants, 306 opinions)%
      \label{fig:q2-all-sensitivity-analysis}}
\end{figure}

\begin{figure}[H]%
\centering
    \begin{subfigure}[c]{.49\textwidth}
      \centering
      \includeinkscape[height=5cm]{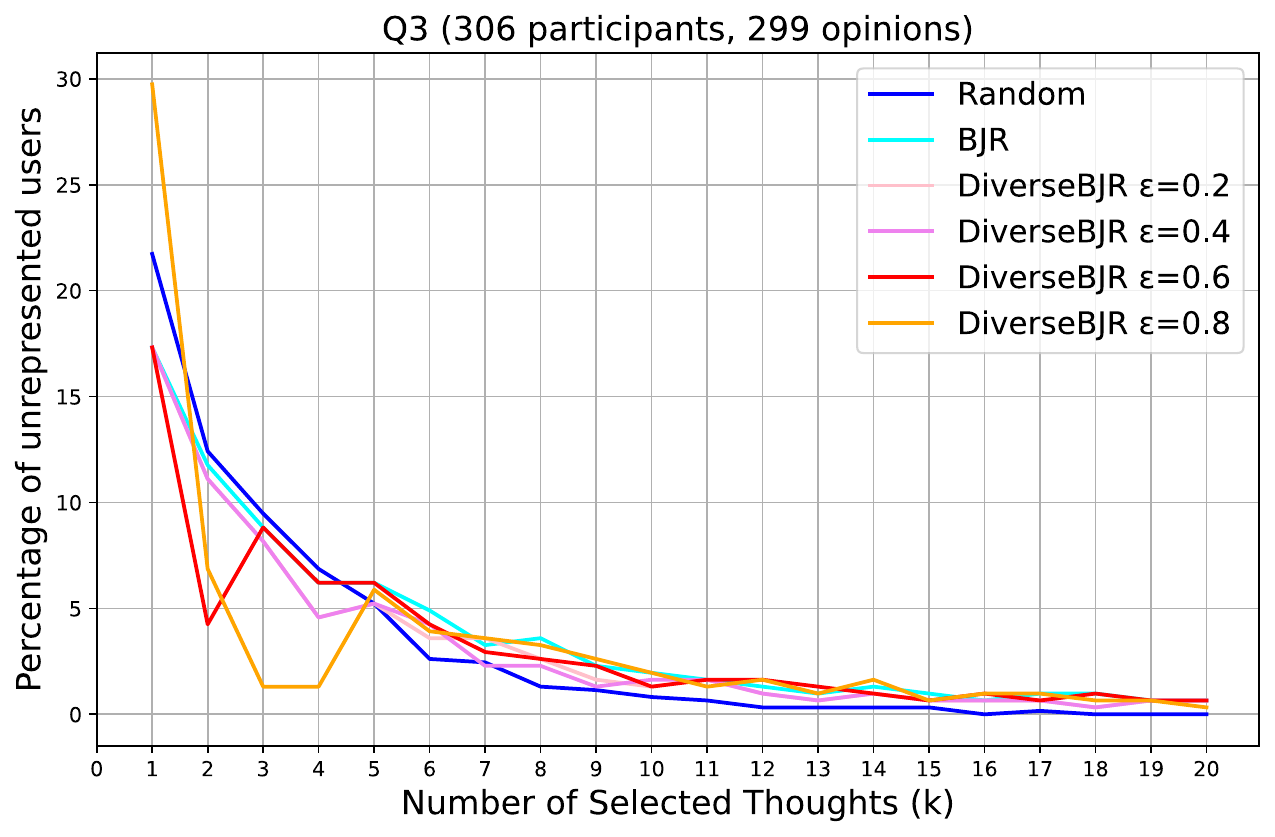}
    \end{subfigure}\hfill
  \begin{subfigure}[c]{.49\textwidth}
    \centering
    \includeinkscape[height=5cm]{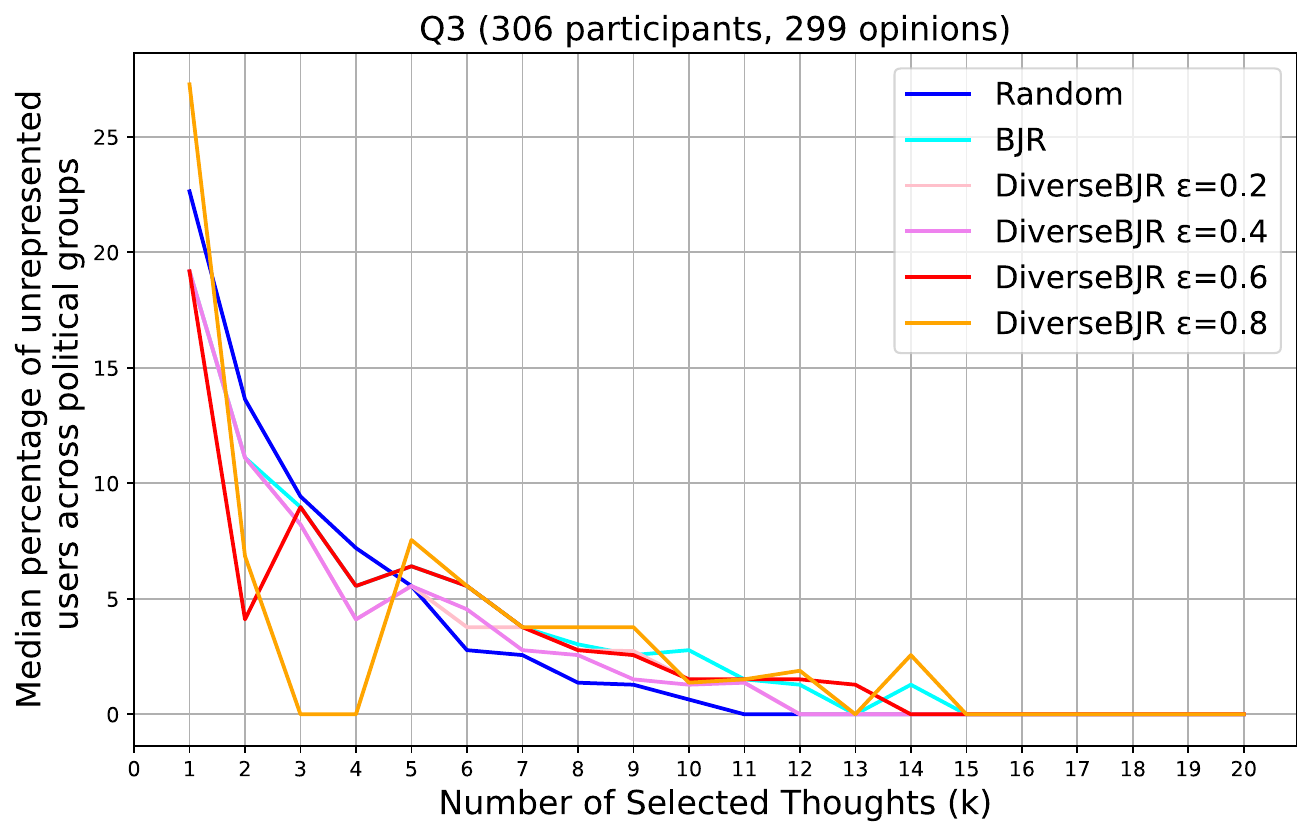}%
  \end{subfigure}\\
    \begin{subfigure}[c]{.49\textwidth}
      \centering
      \includeinkscape[height=5cm]{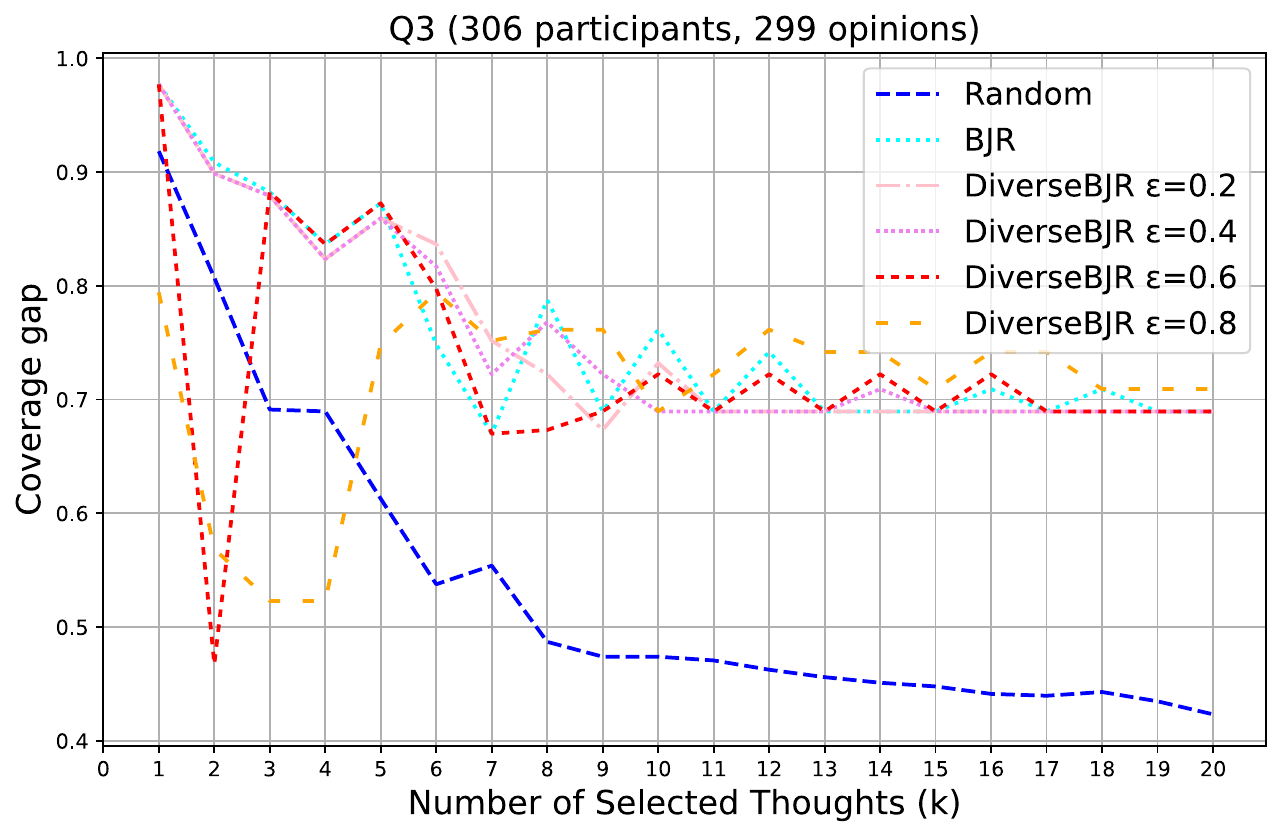}
    \end{subfigure}\hfill
  \begin{subfigure}[c]{.49\textwidth}
    \centering
    \includeinkscape[height=5.5cm]{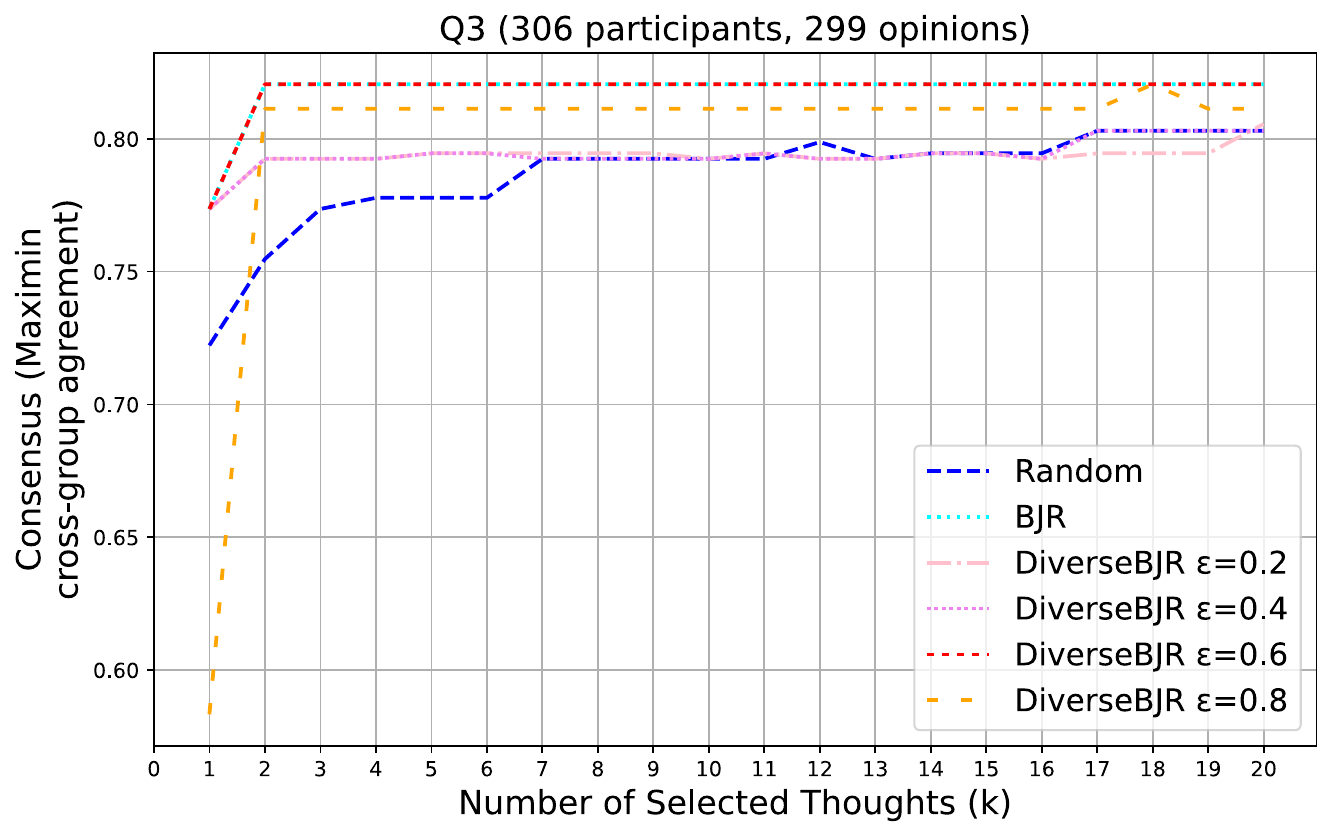}%
  \end{subfigure}
  \begin{subfigure}[c]{.49\textwidth}
    \centering
    \includeinkscape[height=5cm]{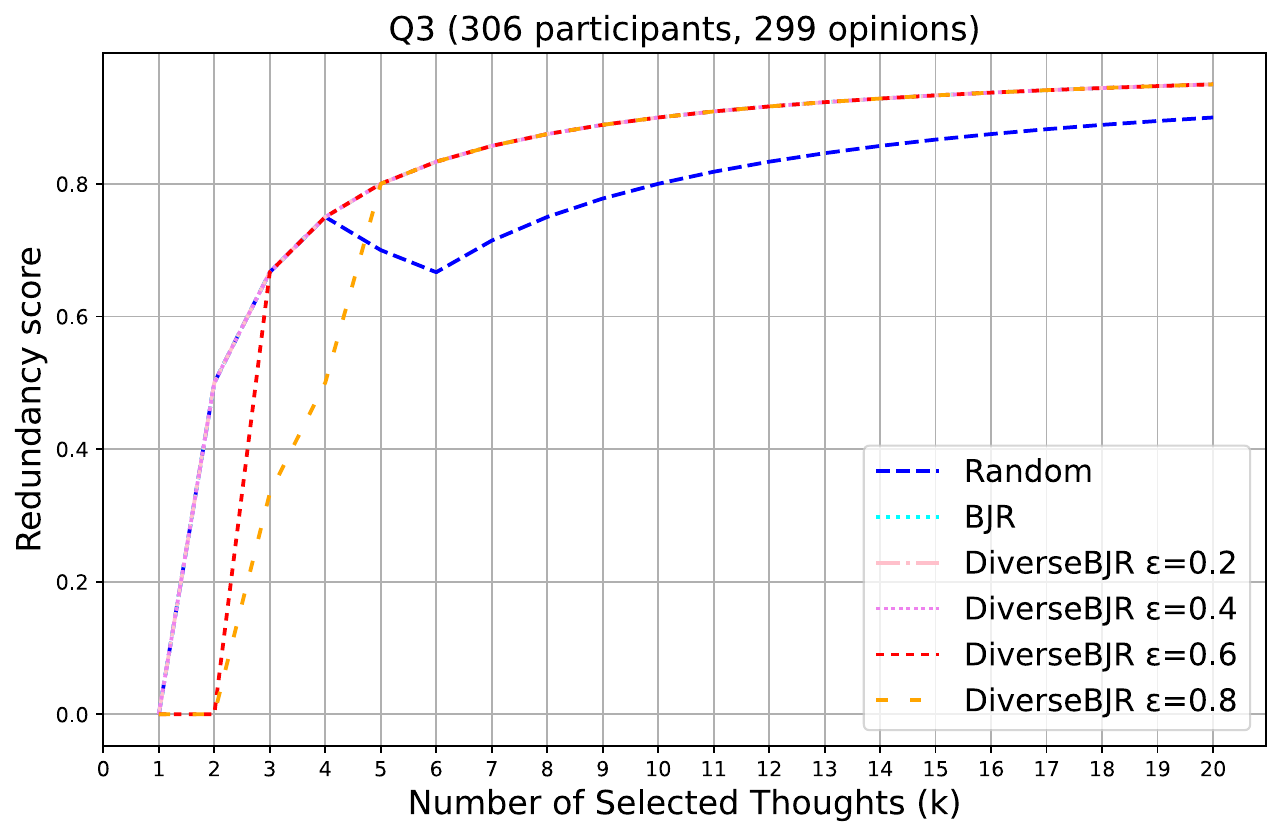}%
  \end{subfigure}\\
  \caption{Q3: "What characteristics or actions, in your view, deem a protest inappropriate?" (306 participants, 299 opinions)%
      \label{fig:q3-all-sensitivity-analysis}}
\end{figure}

\begin{figure}[H]%
\centering
    \begin{subfigure}[c]{.49\textwidth}
      \centering
      \includeinkscape[height=5cm]{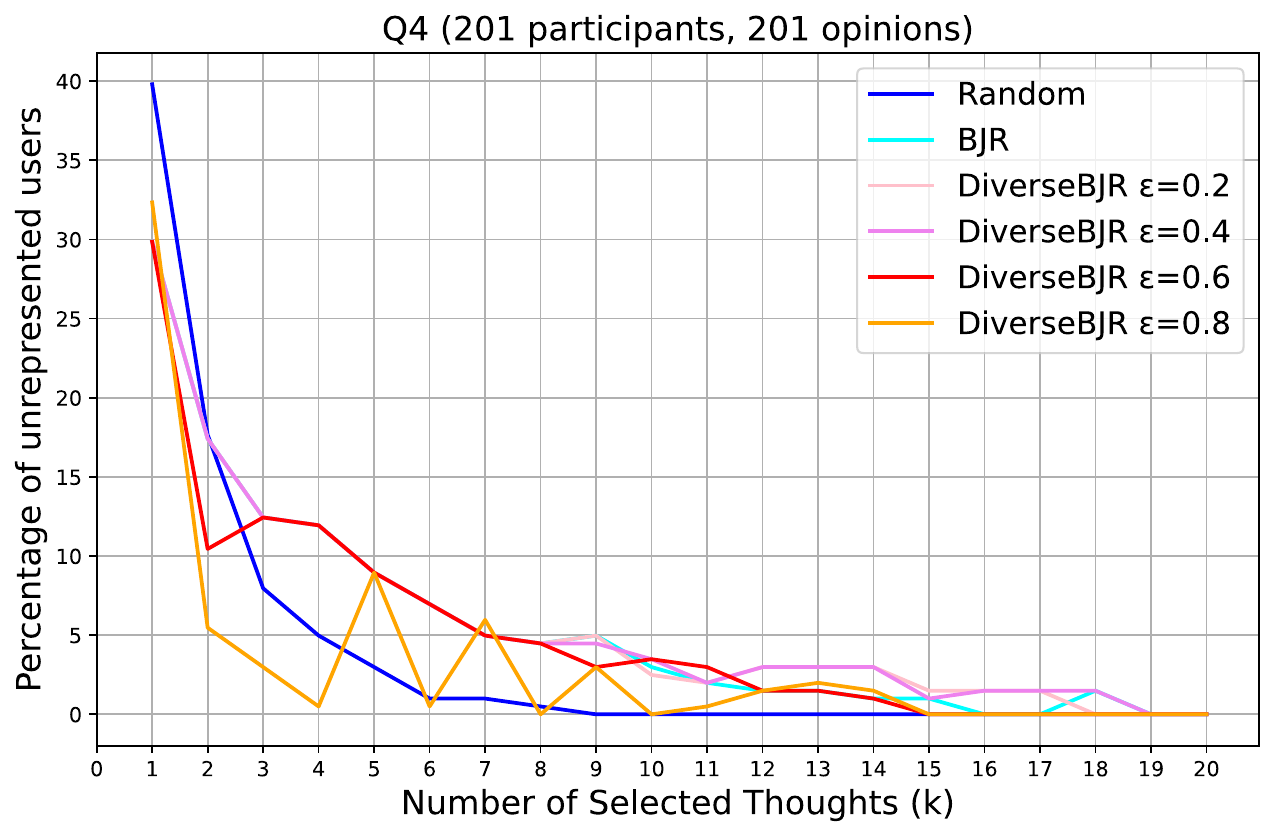}
    \end{subfigure}\hfill
  \begin{subfigure}[c]{.49\textwidth}
    \centering
    \includeinkscape[height=5cm]{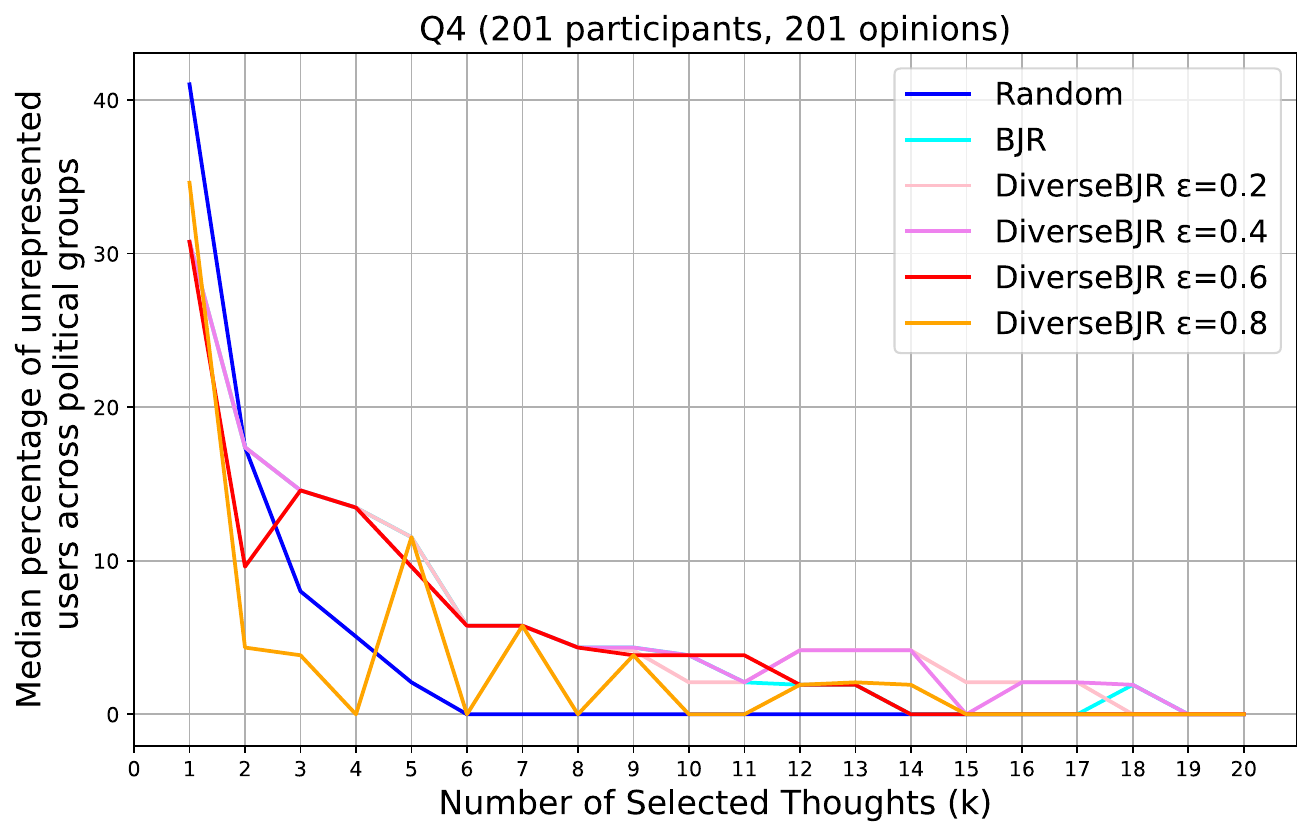}%
  \end{subfigure}\\
    \begin{subfigure}[c]{.49\textwidth}
      \centering
      \includeinkscape[height=5cm]{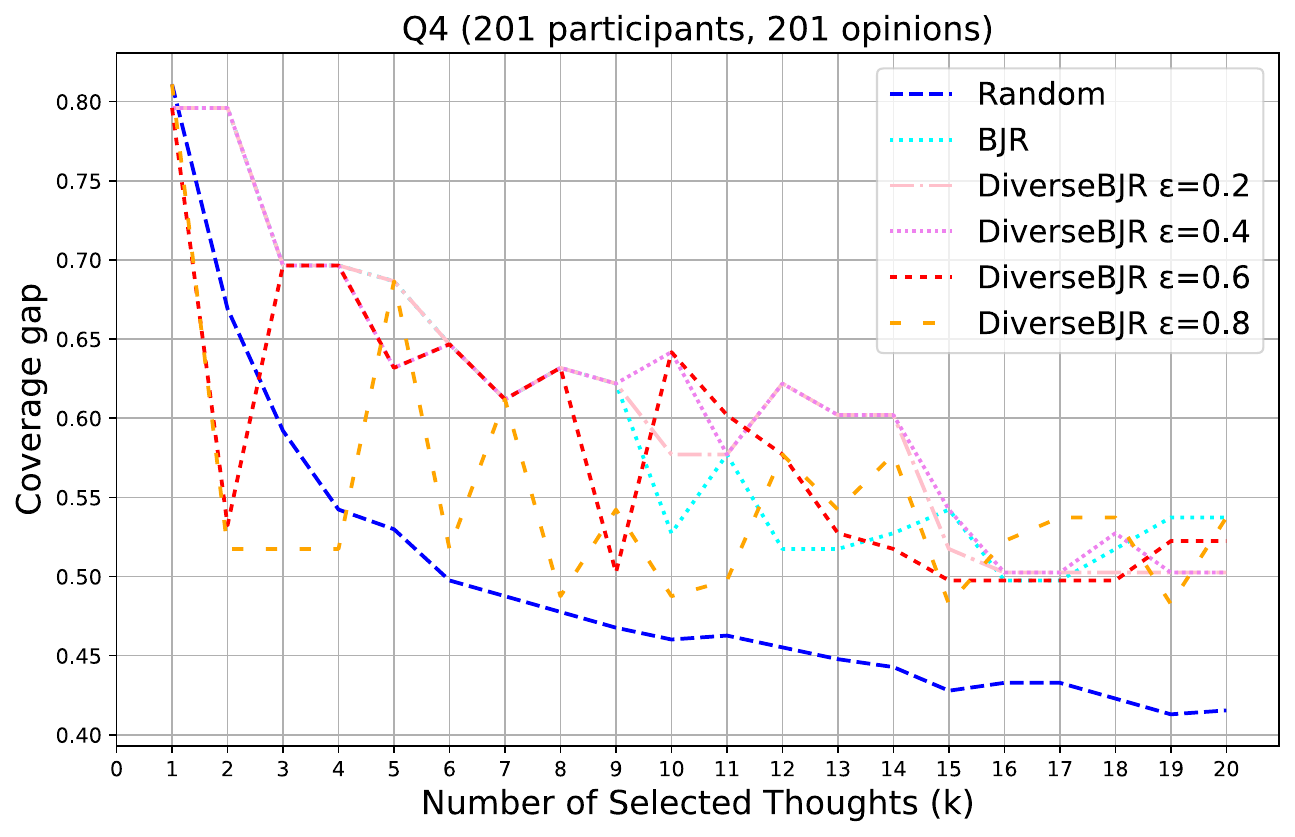}
    \end{subfigure}\hfill
  \begin{subfigure}[c]{.49\textwidth}
    \centering
    \includeinkscape[height=5.5cm]{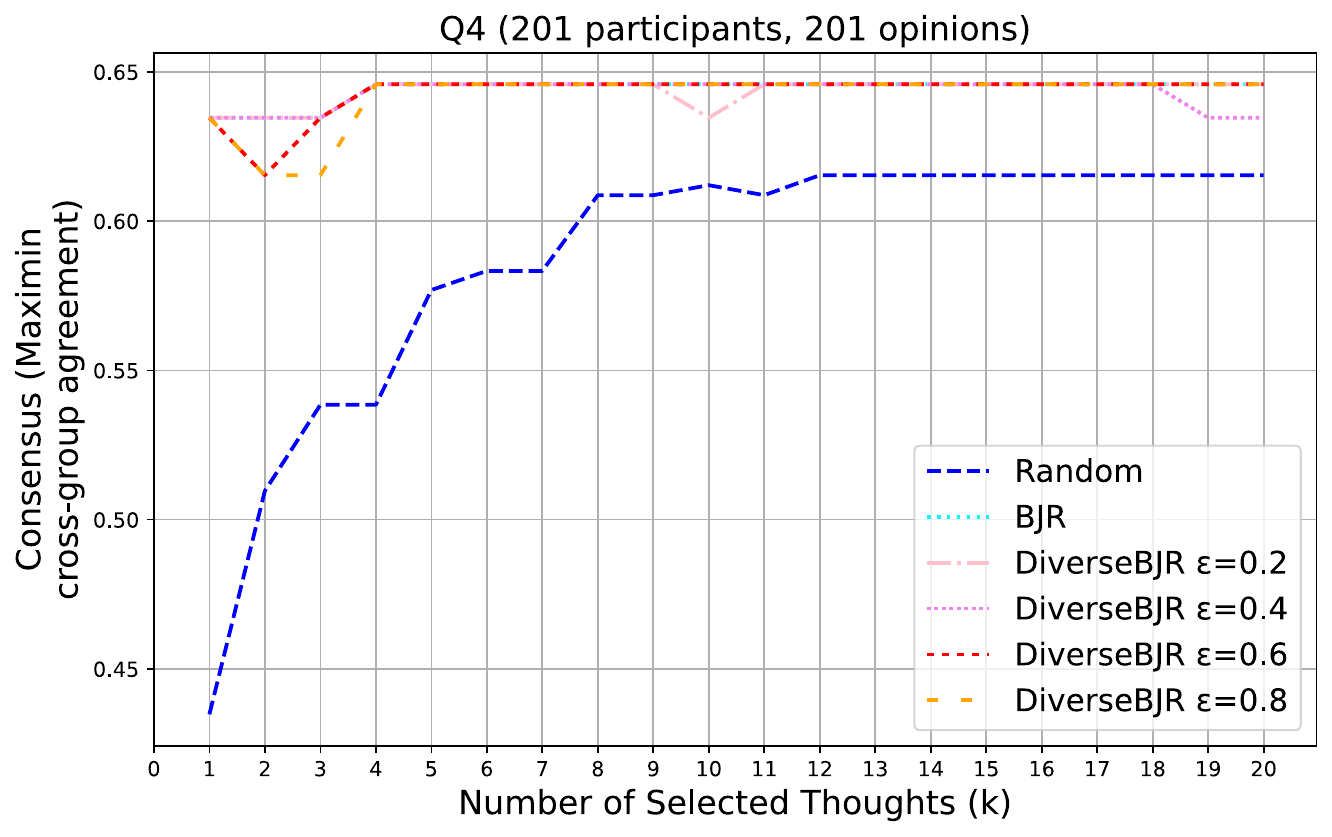}%
  \end{subfigure}
  \begin{subfigure}[c]{.49\textwidth}
    \centering
    \includeinkscape[height=5cm]{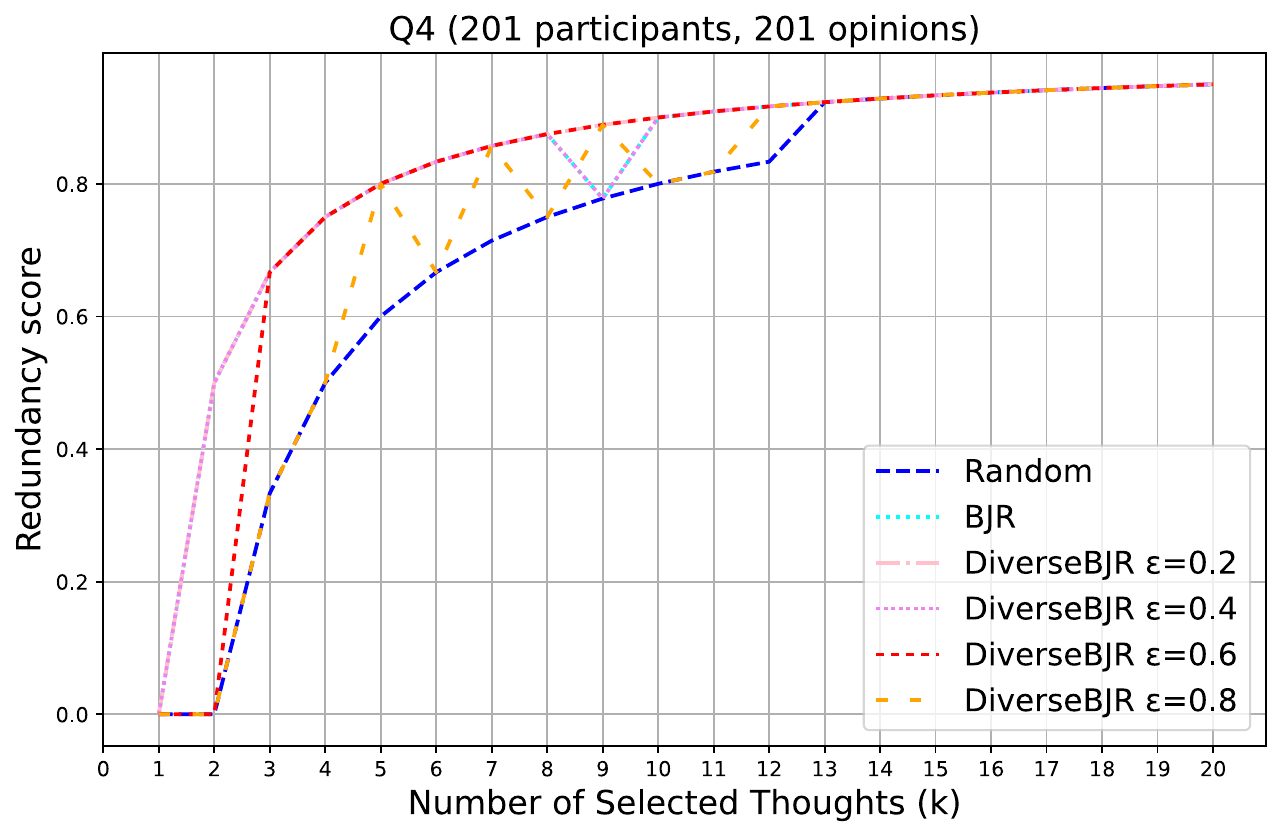}%
  \end{subfigure}\\
  \caption{Q4: "What are some of the reasons you have not participated in or attended a protest?" (201 participants, 201 opinions)%
      \label{fig:q4-all-sensitivity-analysis}}
\end{figure}

\begin{figure}[H]%
\centering
    \begin{subfigure}[c]{.49\textwidth}
      \centering
      \includeinkscape[height=5cm]{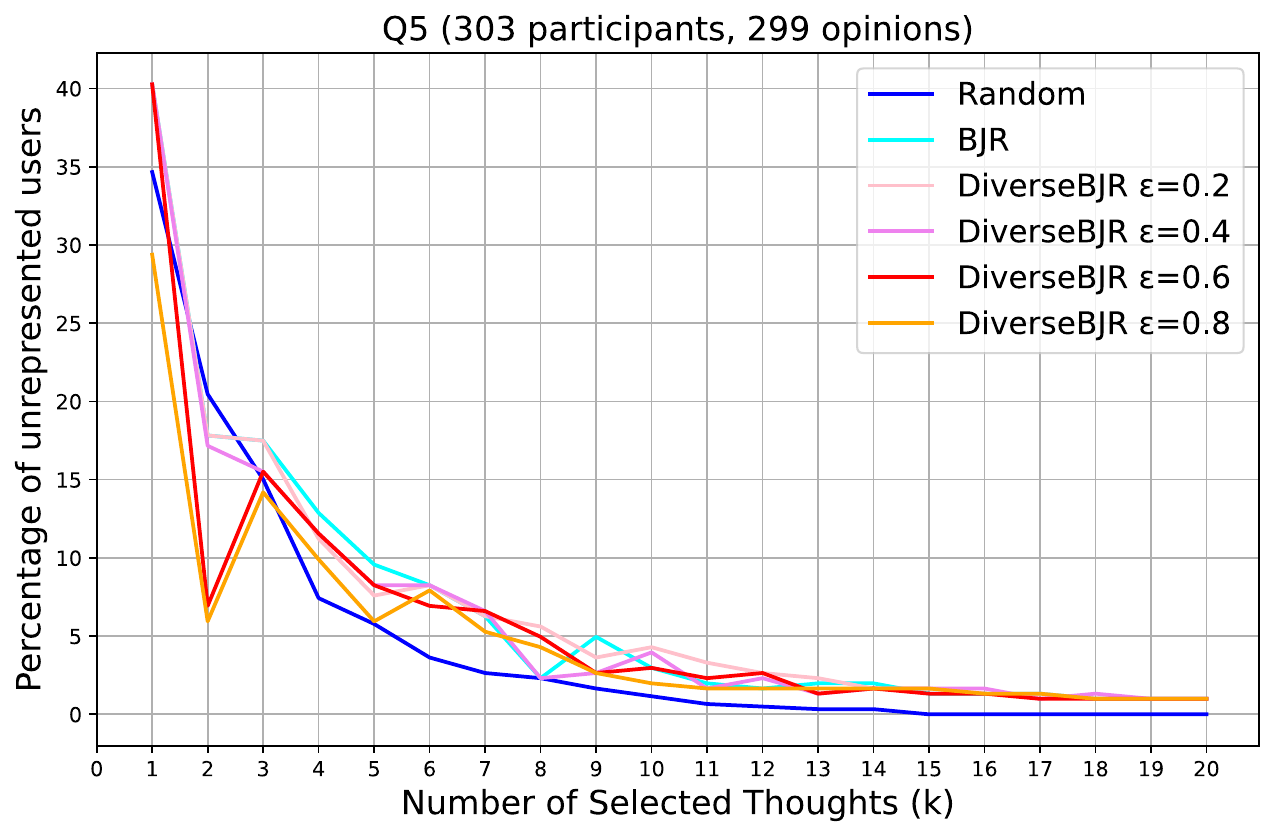}
    \end{subfigure}\hfill
  \begin{subfigure}[c]{.49\textwidth}
    \centering
    \includeinkscape[height=5cm]{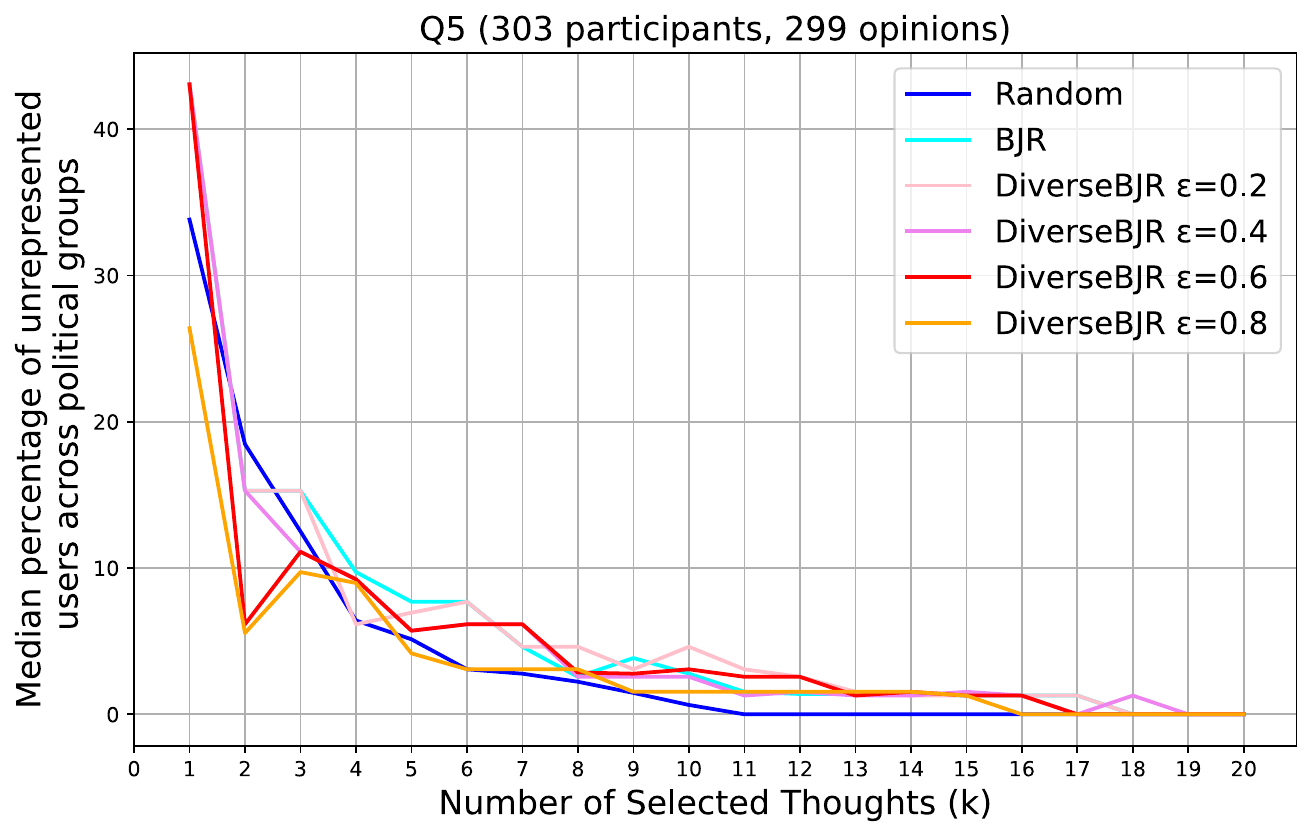}%
  \end{subfigure}\\
    \begin{subfigure}[c]{.49\textwidth}
      \centering
      \includeinkscape[height=5cm]{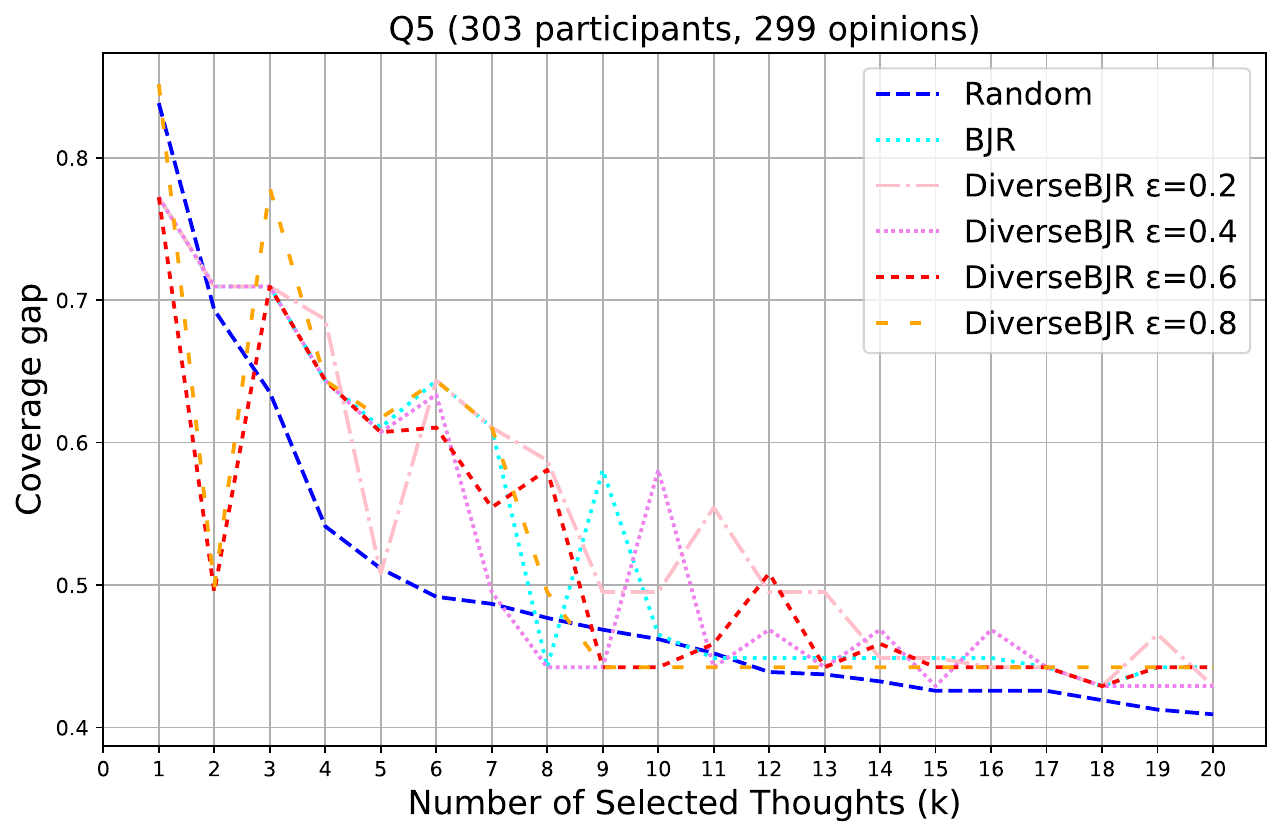}
    \end{subfigure}\hfill
  \begin{subfigure}[c]{.49\textwidth}
    \centering
    \includeinkscape[height=5.5cm]{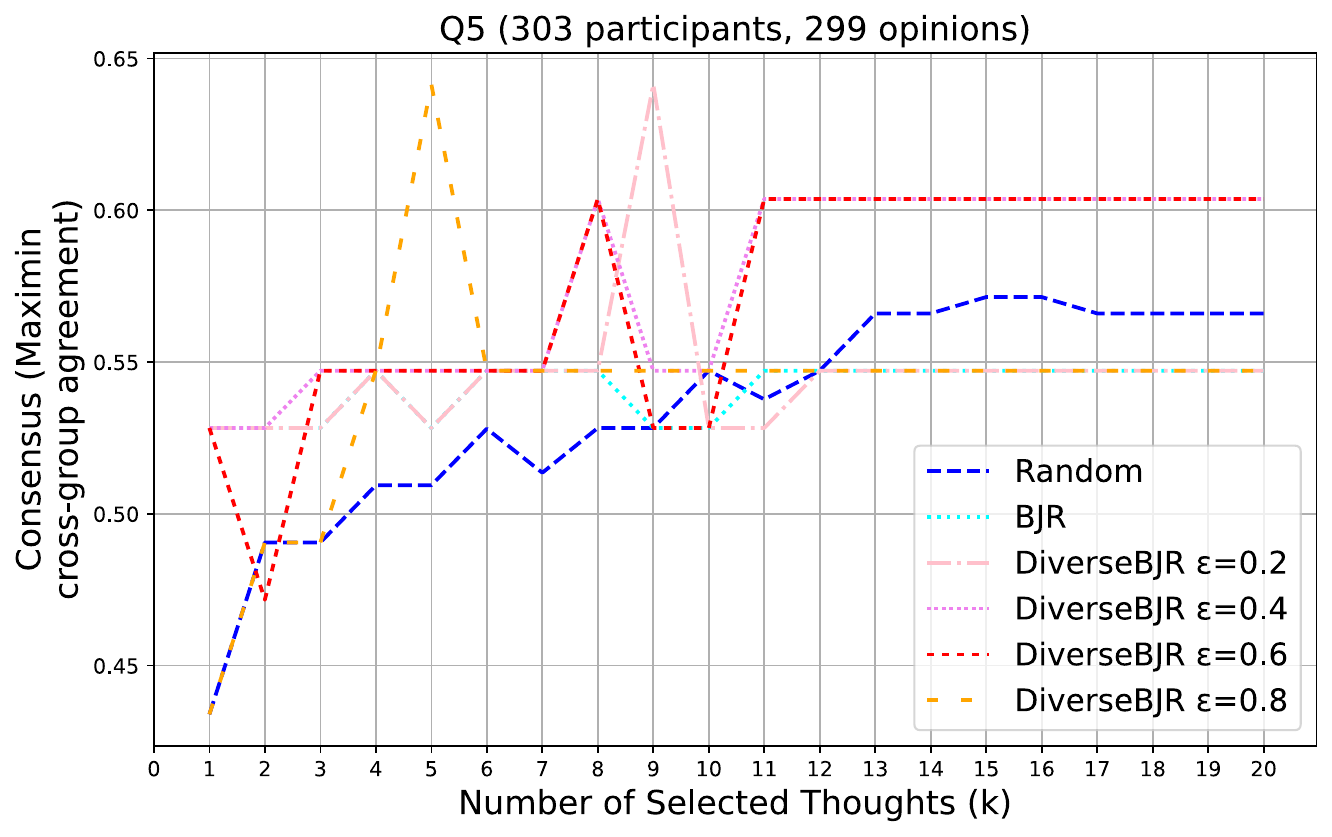}%
  \end{subfigure}
  \begin{subfigure}[c]{.49\textwidth}
    \centering
    \includeinkscape[height=5cm]{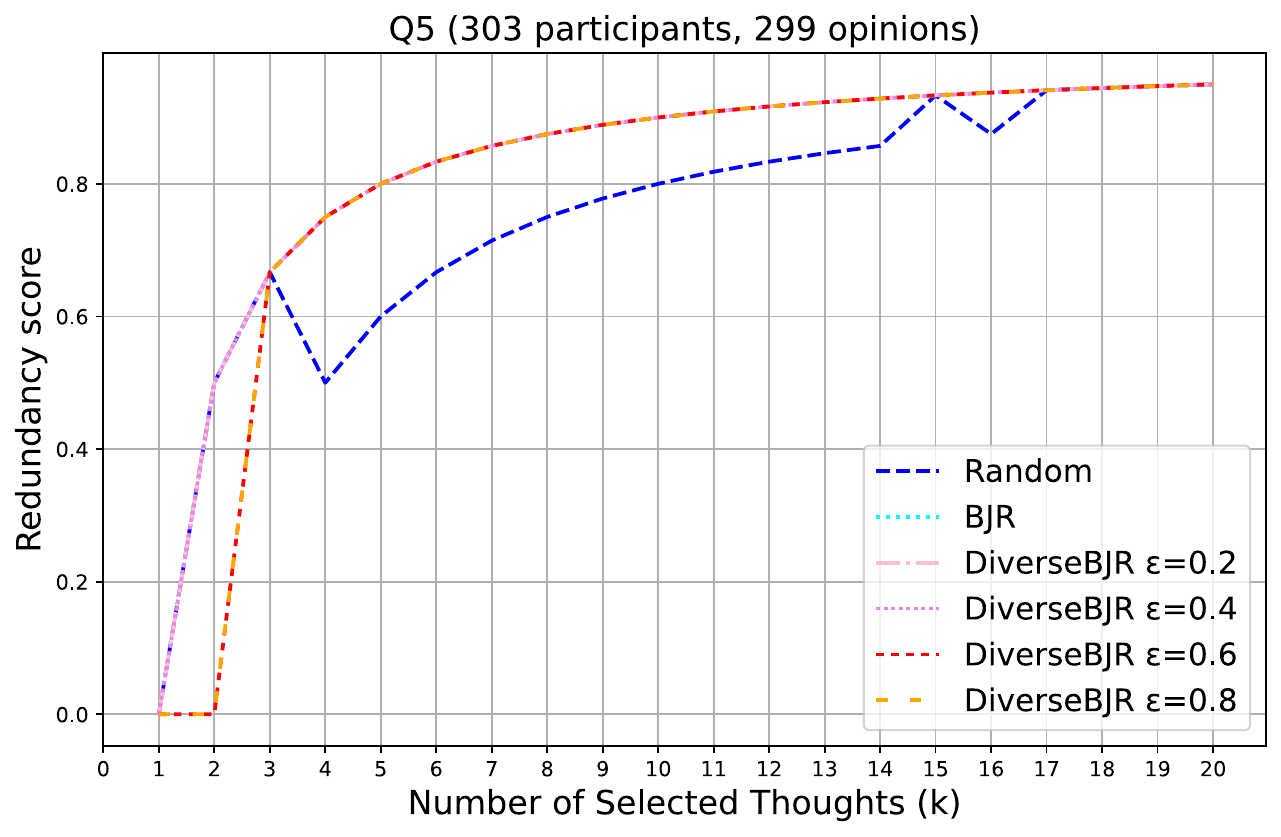}%
  \end{subfigure}\\
  \caption{Q5: "What measures (if any) could be taken to restrict or limit inappropriate protests?" (303 participants, 299 opinions)%
      \label{fig:q5-all-sensitivity-analysis}}
\end{figure}

\begin{figure}[H]%
\centering
    \begin{subfigure}[c]{.49\textwidth}
      \centering
      \includeinkscape[height=5cm]{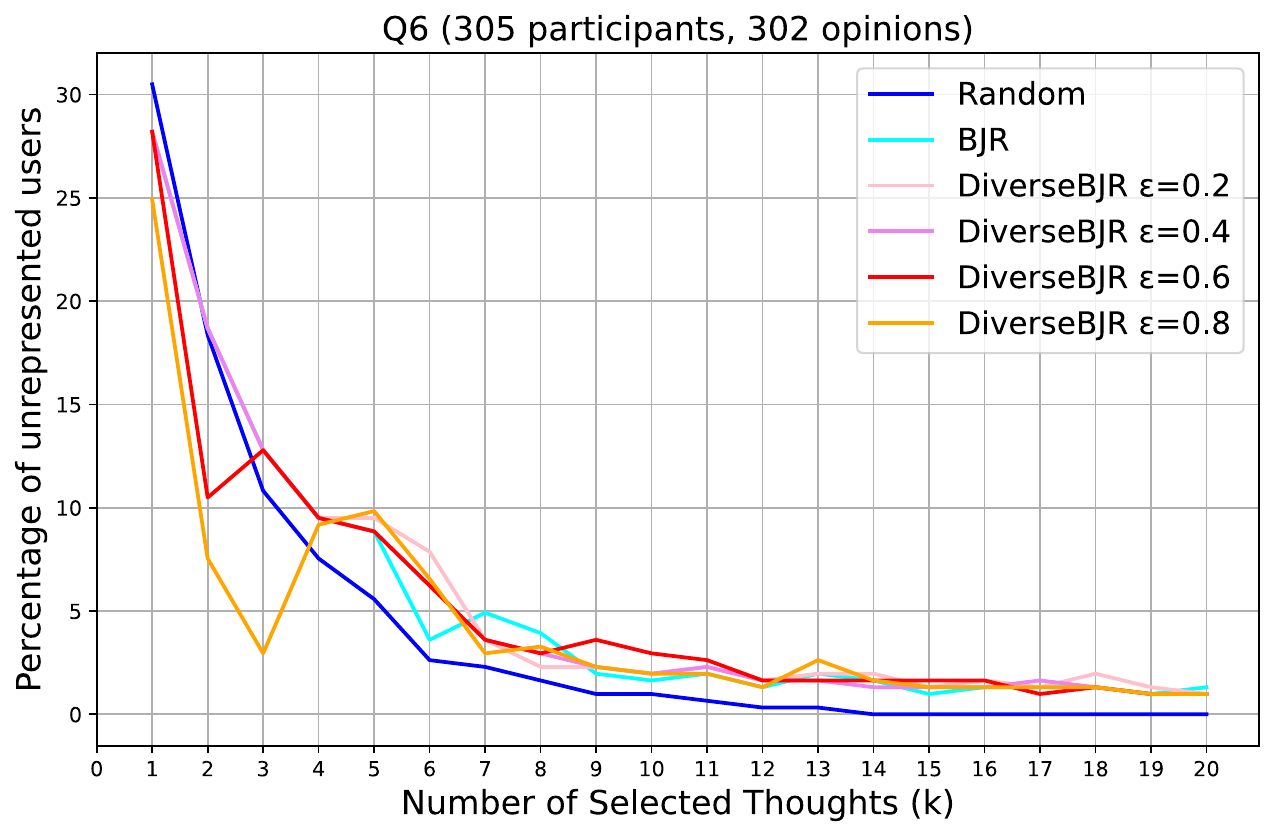}
    \end{subfigure}\hfill
  \begin{subfigure}[c]{.49\textwidth}
    \centering
    \includeinkscape[height=5cm]{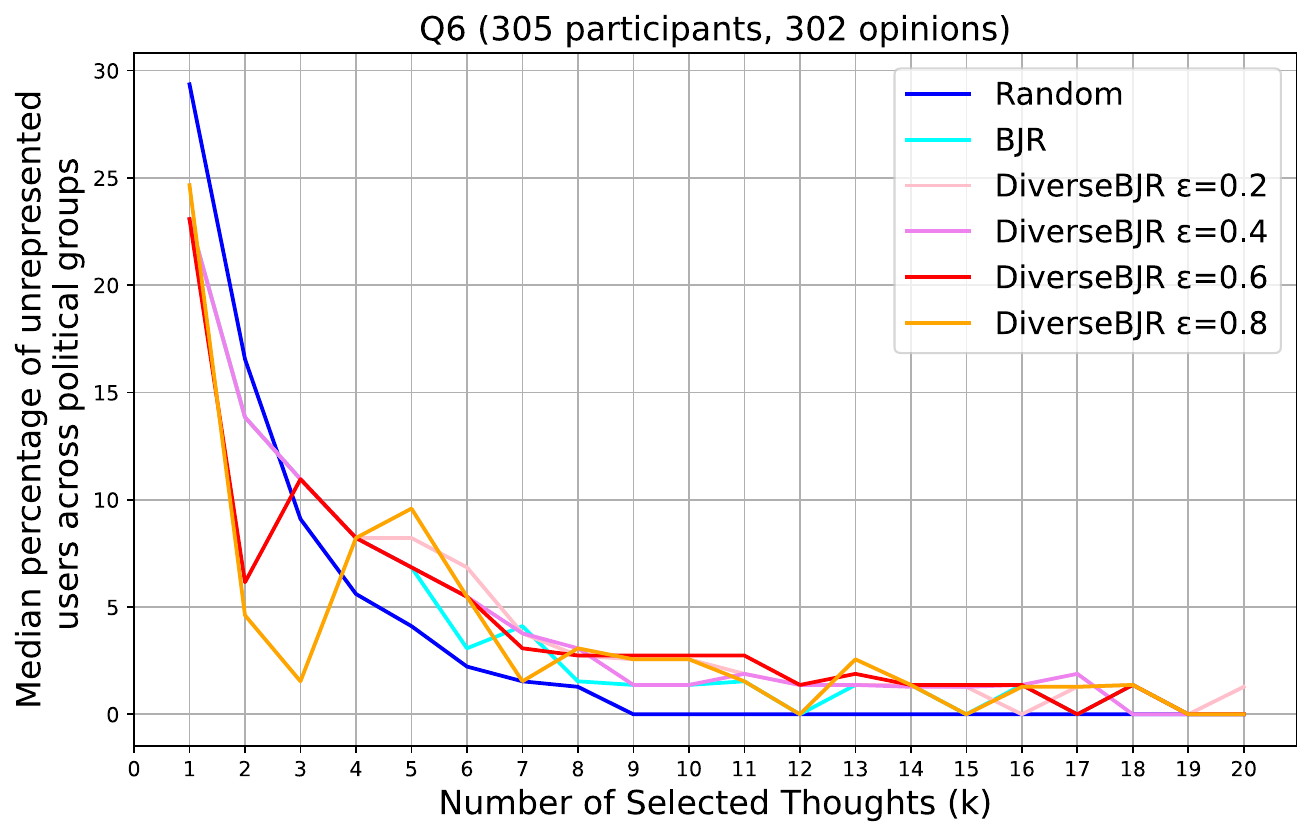}%
  \end{subfigure}\\
    \begin{subfigure}[c]{.49\textwidth}
      \centering
      \includeinkscape[height=5cm]{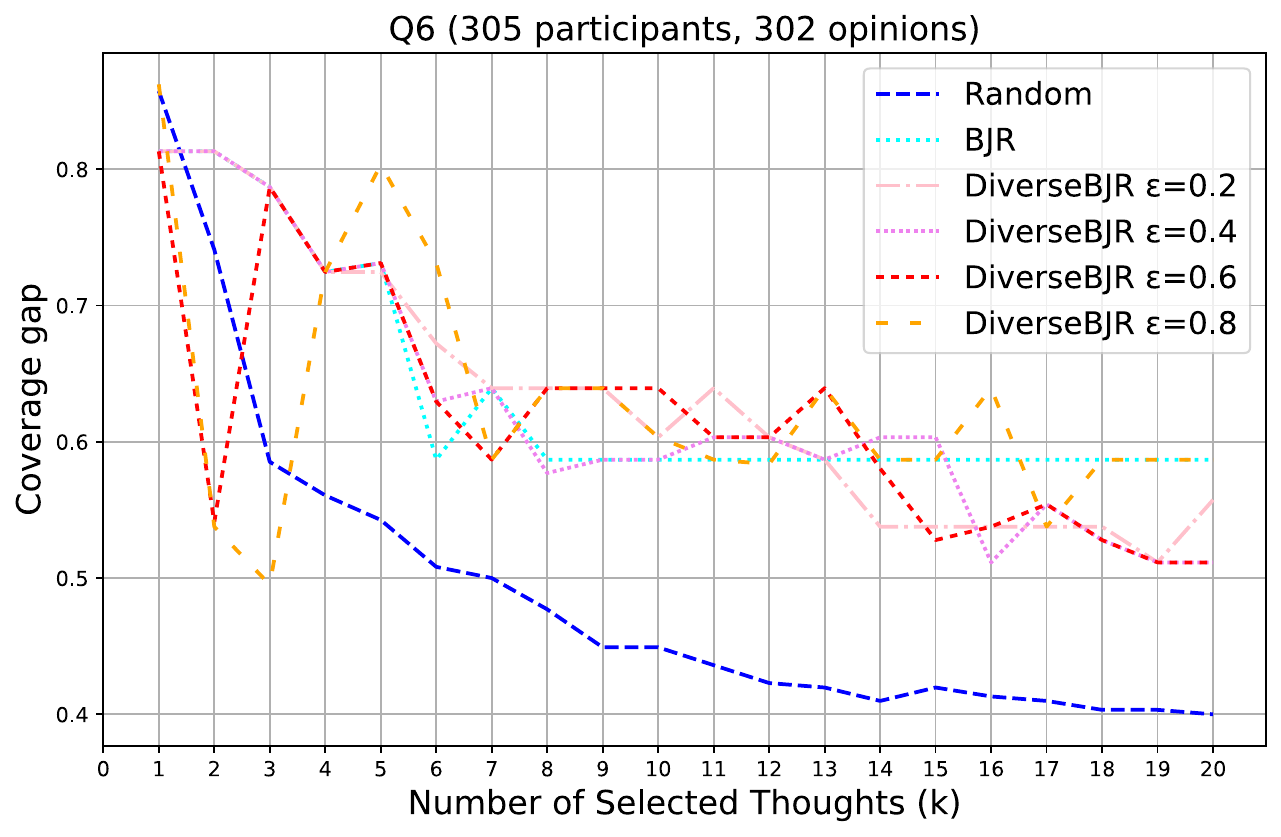}
    \end{subfigure}\hfill
  \begin{subfigure}[c]{.49\textwidth}
    \centering
    \includeinkscape[height=5.5cm]{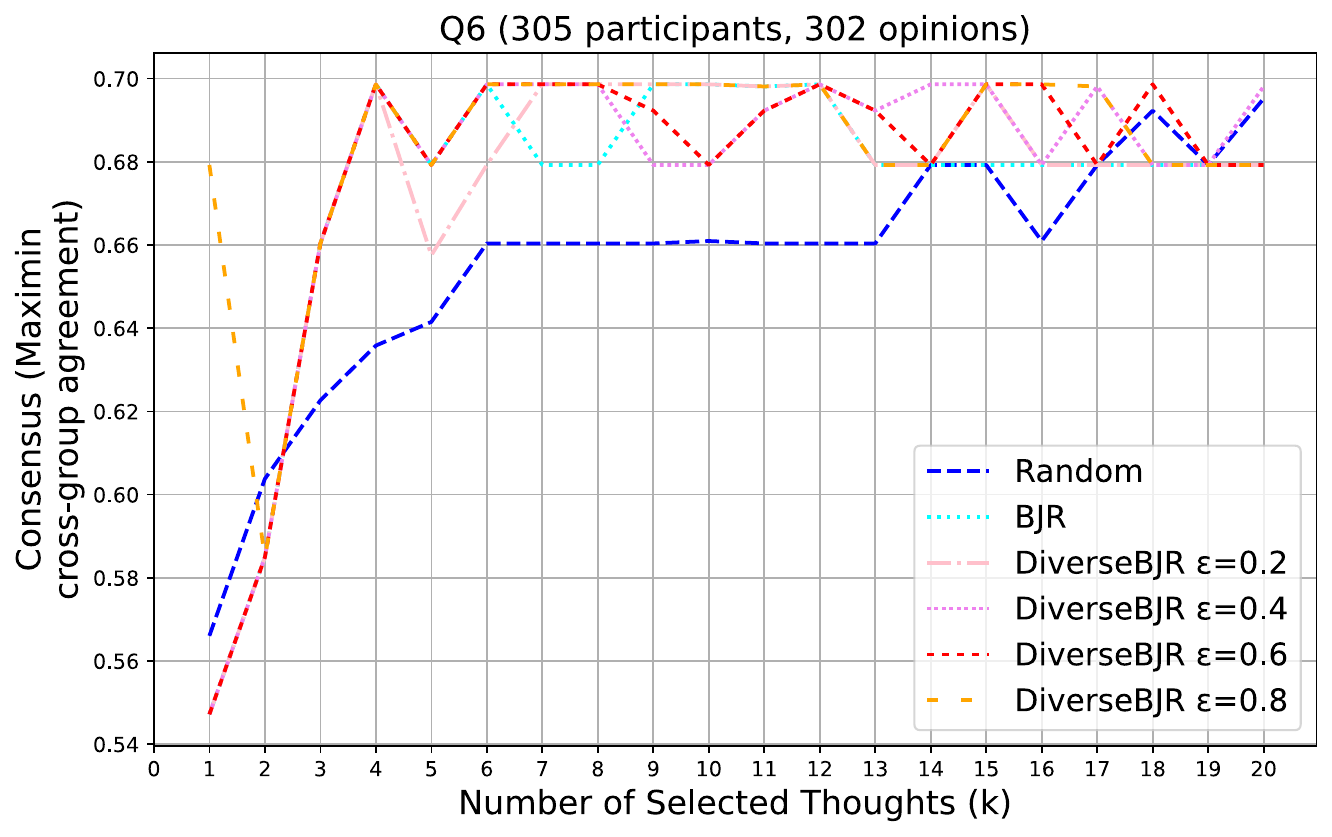}%
  \end{subfigure}
  \begin{subfigure}[c]{.49\textwidth}
    \centering
    \includeinkscape[height=5cm]{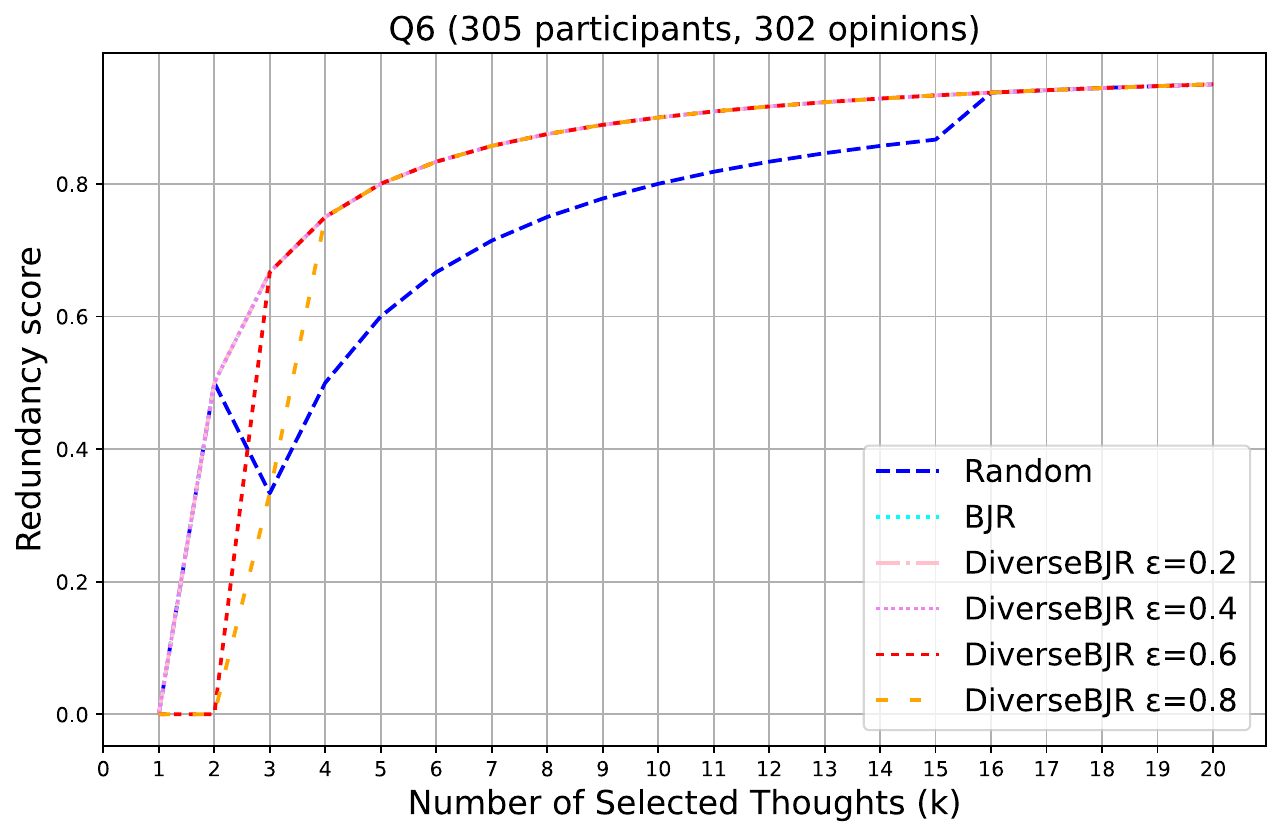}%
  \end{subfigure}\\
  \caption{Q6: "What measures (if any) could be taken to ensure appropriate protests are protected?" (305 participants, 302 opinions)%
      \label{fig:q6-all-sensitivity-analysis}}
\end{figure}

We plot the performance of DiverseBJR for various values of $\varepsilon$ and compare it against BJR (the baseline it was designed to enhance) and against random selection (sanity check). Across all data, performance enhances with higher epsilon values (see Figures \ref{fig:q1-all-sensitivity-analysis} to \ref{fig:q6-all-sensitivity-analysis}). The value of 0.8 which we set in our main experiments might still be optimizable. Crucially, DiverseBJR enhances performance over BJR for all evaluated $\varepsilon$ values, making the choice of $\varepsilon$ a case-by-case and probably data-specific problem, rather than a choice with impact on the condition the algorithm is designed to fulfill (enhancement of BJR).

\subsection{Controversial questions}

 \subsection{Consensual questions}
 
\end{document}